\documentclass[10pt,aps,prl,notitlepage,longbibliography,twocolumn]{revtex4-1}
\usepackage{float}
\usepackage{verbatim}
\usepackage{amsmath}
\usepackage{xcolor}
\usepackage{graphicx}
\usepackage{cases}
\usepackage{hyperref}
\hypersetup{ colorlinks, citecolor=green, filecolor=black,
  linkcolor=blue, urlcolor=blue}

\renewcommand{\S}{\sigma} 
\newcommand{\SSS}{\pmb{\sigma}} 

\newcommand{\s}{s} 
\newcommand{\sss}{\pmb{s}} 

\newcommand{\eH}{\Theta} 
\newcommand{\eHHH}{\pmb{\Theta}} 

\newcommand{\CCC}{\pmb{C}} 
\newcommand{\mmm}{\pmb{m}} 

\newcommand{\DDD}{\pmb{D}} 

\newcommand{\AAA}{\pmb{A}}

\newcommand{\JJJ}{\pmb{J}} 
\newcommand{\hhh}{\pmb{h}} 

\newcommand{\e}{\operatorname{e}}

\renewcommand{\th}{\operatorname{th}}

\newcommand{\ch}{\operatorname{cosh}}
\newcommand{\arth}{\operatorname{artanh}}
\newcommand{\arsh}{\operatorname{arsinh}}
\newcommand{\argmax}{\operatorname{argmax}}

\newcommand{\abs}[1]{\left\vert #1 \right\vert}
\newcommand{\mean}[1]{\left\langle #1 \right\rangle}

\newcommand{\KL}{\operatorname{KL}} 

\newcommand{\D}{\mathrm{D}} 
\newcommand{\ML}{\mathrm{ML}}

\newcommand{\MF}{\mathrm{MF}} 
\newcommand{\TAP}{\mathrm{TAP}} 
\newcommand{\BP}{\mathrm{BP}} 
\newcommand{\LR}{\mathrm{LR}} 
\newcommand{\PL}{\mathrm{PL}} 

\usepackage{hyphenat}
\makeatletter
\let\cat@comma@active\@empty
\makeatother

\usepackage{soul}
\begin{document}
\title{Inverse statistical problems: from the inverse Ising problem to data science}

\author{H. Chau Nguyen}
\affiliation{Max-Planck-Institut f\"ur Physik komplexer Systeme,
N\"othnitzer Str. 38, D-01187 Dresden, Germany}
\author{Riccardo Zecchina} 
\affiliation{Bocconi University,
via Roentgen 1, 20136 Milano, Italy
and Politecnico di Torino,
Corso Duca degli Abruzzi 24, 10129 Torino
 }
\author{Johannes Berg}
\affiliation{Institute for Theoretical Physics, University of Cologne,
  Z\"ulpicher Stra{\ss}e 77, 50937 Cologne, Germany}

\begin{abstract}
 Inverse problems in statistical physics are motivated by the
  challenges of `big data' in different fields, in particular
  high-throughput experiments in biology. In inverse problems, the usual
  procedure of statistical physics needs to be reversed: Instead of
  calculating observables on the basis of model parameters, we seek to
  infer parameters of a model based on observations.  In this review,
  we focus on the inverse Ising problem and closely related problems,
  namely how to infer the coupling strengths between spins given observed spin
  correlations, magnetisations, or other data. We review 
  applications of the inverse Ising problem, including the reconstruction of
  neural connections, protein structure determination, and the
 inference of gene regulatory networks. For the inverse Ising problem
 in equilibrium, a number of controlled and
  uncontrolled approximate solutions have been developed in the
  statistical mechanics community. A particularly strong method,
  pseudolikelihood, stems from statistics. We also review the inverse
  Ising problem in the non-equilibrium case, where the model parameters must be
  reconstructed based on non-equilibrium statistics.
\footnote{citation: Inverse statistical problems: from the inverse Ising problem to data science, 
  H.C. Nguyen, R. Zecchina and J. Berg, Advances in Physics, 66 (3), 197-261 (2017)}
\end{abstract}



\maketitle
\tableofcontents{}

\section{Introduction and applications}

The primary goal of statistical physics is to derive 
observable quantities from microscopic laws governing the constituents of a
system. In the example of the Ising model, the starting point is a
model describing interactions between 
elementary magnets (spins), the goal is to derive
observables such as spin magnetisations and correlations. 

In an inverse problem, the starting point is observations of some
system whose microscopic parameters are unknown and to be discovered. In the inverse Ising problem, the
interactions between spins are not known to us, but we want to learn
them from measurements of magnetisations, correlations, or other
observables. In general, the goal is to infer the parameters describing a
system (for instance, its Hamiltonian) from extant data. To this end,
the relationship between microscopic laws and observables needs to be
inverted. 

In the last two decades, inverse statistical problems have arisen in
different contexts, sparking interest in the statistical physics
community in taking the path from model
parameters to observables in reverse. The areas where inverse
statistical problems have arisen are characterized by (i)
microscopic scales becoming experimentally accessible and (ii) sufficient data
storage capabilities being available. 
In particular, the biological
sciences have generated several inverse statistical problems,
including the reconstruction of neural and gene regulatory networks and the
determination of the three-dimensional structure of proteins. Technological progress is likely to to open up further
fields of research to inverse statistical analysis, a development that is
currently described by the label `big data'.

In physics, inverse statistical problems also arise when we need to
design a many-body system with particular desired properties. 
Examples are finding 
the potentials that result in a particular single-particle distribution~\cite{kunkin1969inverse,chayes1984a}, 
interaction parameters in a binary alloy that yield the observed correlations~\cite{maysenholder1987determination},
the potentials between atoms that lead to specific crystal lattices~\cite{zhang2013a}, or
the parameters of a Hamiltonian that lead to a particular density matrix~\cite{changlani2015}. In the context of
soft matter, a question is how to design a many-body system that will
self-assemble into a particular spatial configuration or has
particular bulk properties~\cite{rechtsman2006a,torquato2009a}. In
biophysics, we may want to design a protein that folds into a
specified three-dimensional shape~\cite{kuhlman2003a}. For RNA, even 
molecules with more than one stable target structure
are possible~\cite{flamm2001a}. As a model of such design problems,
\cite{distasio2013a,marcotte2013a} study how to find
the parameters of Ising Hamiltonian with a prescribed ground
state.

In all these examples, `spin' variables describe microscopic degrees of
freedom particular to a given system, for instance, the states of
neurons in a neural network. The simplest description of these degrees
of freedom in terms of random binary variables then leads to
Ising-type spins. In the simplest non-trivial scenario, correlations
between the 'spins' are generated by pairwise couplings between the
spins, leading to an Ising model with unknown parameters (couplings
between the spins and magnetic fields acting on the spins). In many
cases of interest, the couplings between spins will not all be
positive, as is the case in a model of a ferromagnet. Nor will they
couplings conform to a regular lattice embedded in some finite-dimensional
space. 

For a concrete example, we look at a system of $N$ binary variables
(Ising spins) $\{s_i\}, i=1,\,\ldots,N$ with $s_i=\pm 1$. These spins
are coupled by pairwise couplings $J_{ij}$ and are subject to external
magnetic fields $h_i$. 
\begin{equation}
\label{eq:thefirst}
P(\{s_i\})=\frac{1}{Z} \exp \left[ \sum_i h_i s_i + \sum_{i < j} 
J_{ij} s_i s_j \right] \ 
\end{equation}
is the Boltzmann equilibrium distribution $P(\{s_i\})
=e^{-H(\{s_i\})}/Z$, where we have subsumed temperature into the
couplings and fields. (We will discuss this choice in section~\ref{sec:equilibrium_definition}).
The Hamiltonian 
\begin{equation}
\label{eq:thesecond}
H(\{s_i\})=-\sum_i h_i s_i - \sum_{i < j}  J_{ij} s_i s_j 
\end{equation}
specifies the energy of the spin system as a function of the
microscopic spin variables, local fields, and pairwise couplings. 
The inverse Ising problem is the determination of the couplings $J_{ij}$ and local
fields $h_i$, given a set of $M$ observed spin configurations. 
Depending on the particular nature of the system at hand,
the restriction to binary variables or pairwise interactions may need
to be lifted, or the functional form of the Hamiltonian may be
altogether different from the Ising Hamiltonian with pairwise
interactions~\eqref{eq:thesecond}. 
For non-equilibrium systems, the steady
state is not even described by a Boltzmann distribution with a known Hamiltonian. 
However,
the basic idea remains the same across different types of
inverse statistical problems: even when the frequencies of
spin configurations may be under-sampled, the data may be sufficient to infer at
least some parameters of a model.

The distribution~\eqref{eq:thefirst} is well known not only
as the equilibrium distribution of the Ising model. It is also the
form of the distribution which maximizes the (Gibbs) entropy 
\begin{equation}
\label{eq:shannon}
S[P]=-\sum_{\{s_i\}} P(\{s_i\}) \ln  P(\{s_i\})
\end{equation}
under the constraint that
$P(\{s_i\})$ is normalized and has particular first and second
moments, that is, magnetisations and correlations. We will discuss in
section~\ref{sec:maxent} how this distribution 
emerges as the `least biased distribution' of a binary random
variable with prescribed first and second moments~\cite{jaynes1957a}. The practical
problem is then again an inverse one: to find the couplings $J_{ij}$ and local fields $h_i$
such that the first and second moments observed
under the Boltzmann distribution~\eqref{eq:thefirst} 
match the mean values of $s_i$ and $s_i
s_j$ in data. In settings where third moments differ
significantly from the prediction of~\eqref{eq:thefirst} based on the
first two moments, one may need to construct 
the distribution of maximum entropy given the
first three moments, leading to three-spin
interactions in the exponent of~\eqref{eq:thefirst}.

Determining the parameters of a distribution such
as~\eqref{eq:thefirst} is always a many-body problem: changing a
single coupling $J_{ij}$ generally affects correlations between
many spin variables, and conversely a change in the correlation
between two variables can change the values of many inferred
couplings. The interplay between model parameters and observables is
captured by a statistical mechanics of inverse problems, where the
phase space consists of quantities normally considered as fixed model
parameters (couplings, fields). The observables, such as spin
correlations and magnetisations on the other hand, are taken to be
fixed, as they are specified by empirical observations. Such a
perspective is not new to statistical physics; the analysis of neural
networks in the seventies and eighties of the last century led to a
statistical mechanics of
learning~\cite{watkin1993a,hertz199a,engel2001a}, where the phase
space is defined by the set of possible rules linking the input into a
machine with an output. The set of all rules compatible with a given set
of input/output relations then defines a statistical ensemble.  In
the inverse statistical problems, however, there are generally no
explicit rules linking the input and output, but data with different
types of correlations or other observations, which are to be accounted
for in a statistical model. 

Inverse statistical problems fall into the broader realm of
statistical inference~\cite{mackay2003,bishop2006}, which
seeks to determine the properties of a probability distribution underlying
some observed data. 
The problem of inferring the parameters of a distribution such
as~\eqref{eq:thefirst} is known under different names in different
communities; also emphasis and language differ subtly across communities.
\begin{itemize}
	\item In statistics, an inverse problem is the inference of model
	parameters from data. In our case, the problem is the inference of the
	parameters of the Ising model from observed spin configurations. A
	particular subproblem is the inference of the graph formed by the
	non-zero couplings of the Ising model, termed graphical model
	selection or  reconstruction. In the specific context of statistical
	models on graphs (graphical models), the term
	\emph{inference} describes the calculation of the marginal distribution
	of one or several variables. (A marginal distribution describes
	the statistics of one or several particular variables in a
	many-variable distribution, for example, $P(x_1)=\sum_{x_2,x_3} P(x_1,x_2,x_3)$.)
	\item 
	In machine learning, a frequent task is to train an artificial neural
	network with symmetric couplings such that magnetisations and
	correlations of the artificial neurons match the corresponding values
	in the data. This is a special case of what is called Boltzmann
	machine learning; the general case also considers so-called hidden
	units, whose values are unobserved~\cite{Ackley1985a}. 
	\item 
	In statistical physics, much effort has been directed towards
	estimating the parameters of the Ising model given observed values of
	the magnetisation and two-point correlations. As we will see in
	section~\ref{sec:maxlikelihood}, this is a hard problem from an
	algorithmic point of view. Recently, threshold phenomena arising
	in inference problems have attracted much interest from the
	statistical physics community, due to the link between
	phase transitions and the boundaries separating different regimes of
	inference problems, for instance solvable and unsolvable problems, or
	easy and hard ones~\cite{mezard2009information,ZdeborovaKrzakala2015}.
\end{itemize}

Common theme across different applications and communities is the
inference of model parameters given observed data or desired properties. 
In this review we will focus on a prototype inverse statistical
problem: the inverse Ising problem and its close relatives. 
Many of the
approaches developed for this problem are also readily extended to more general
scenarios. We will start with a discussion of applications of the
inverse Ising problem and related approaches in biology,
specifically the reconstruction of neural and genetic networks, the
determination of three-dimensional protein structures, the inference
of fitness landscapes, the bacterial responses to combinations of
antibiotics, and flocking dynamics. We will find that these
applications define two distinct settings of the inverse Ising
problem; equilibrium and non-equilibrium. Part~\ref{sec:equilibrium}
of this review treats the inverse Ising problem in an equilibrium
setting, where the couplings
between spins are symmetric. Detailed balance holds and results
from equilibrium statistical physics can be used. This setting
arises naturally within the context of maximum entropy models,
which seek to describe the observed statistics of configurations with a simplified
effective model capturing, for instance, collective effects. 
We introduce the
basics of statistical inference and maximum
entropy modelling, discuss the thermodynamics of the inverse Ising
problem, and review different approaches to solve the inverse Ising
problem, pointing out their connections and comparing the resulting
parameter reconstructions. 
Part~\ref{sec:nonequilibrium} of this review
considers asymmetric coupling matrices, where in the absence of
detailed balance couplings can be reconstructed from time series, from 
data on perturbations of the system, or from detailed knowledge of the
non-equilibrium steady state.

We now turn to applications of the inverse Ising problem, which mostly
lie in the analysis of high-throughput data from biology. One aim of
inverse statistical modelling is to find the parameters of a microscopic
model to describe this data. A more ambitious aim is achieved when the
parameters of the model are informative about the
processes which produced the data, this is, when some of the mechanisms underlying
the data can be inferred. The data is large-scale measurements of the
degrees of freedom of some system. In the language of statistical
physics these describe the micro-states of a system: states of neurons,
particular sequences of DNA or proteins, or the concentration levels
of RNA. We briefly introduce some of the experimental background of
these measurements, so their potential and the limitations can be
appreciated. The models are simple models of the microscopic degrees
of freedom. In the spirit of statistical physics, these models are
simple enough so the parameters can be computed given the data, yet sufficiently complex to
reproduce some of the statistical interdependences of the observed
microscopic degrees of freedom. The simplest case, consisting of binary degrees of
freedom with unknown pairwise couplings between them, leads to the
inverse Ising problem, although we will also discuss several
extensions.

\subsection{Modelling neural firing patterns and the reconstruction of
	neural connections} 
\label{sec:neural}

Neurons can exchange information by generating discrete electrical
pulses, termed spikes, that travel down nerve fibres. Neurons can emit
these spikes at different rates, a neuron emitting spikes at a high
rate is said to be `active' or `firing', a neuron emitting spikes at a
low rate or not at all is said to be `inactive' or `silent'.
The measurement of the activity of single neurons has a long history
starting in 1953 with the development of micro-electrodes for
recording~\cite{dowben1953a}. Multi-electrodes were developed, allowing
to record multiple simultaneous neural signals
independently over long time periods~\cite{spira2013a,obien2014a}. 
Such data presents the intriguing possibility to see 
elementary brain function emerge from the interplay of a large
number of neurons.

However, even when vast quantities of data are available, the different configurations of a system are
still under-sampled in most cases. For instance, consider $N$ neurons,
each of which can be either active (firing) or inactive
(silent). Given that the firing patterns of thousands of neurons can
be recorded simultaneously~\cite{schwarz2014a}, the
number of observations $M$ will generally be far less than the total
number of possible neural configurations, $M \ll 2^N$.  For this
reason, a straightforward statistical
description that seeks to determine directly the frequency
with which each configuration appears will likely fail. 

On the other hand, a feasible starting point is a simple distribution, whose 
parameters can to be determined from the data. For a set of $N$
binary variables, this might be a
distribution with pairwise interactions between the variables.
In~\cite{Schneidman2006a}, Bialek and collaborators applied such a
statistical model to neural recordings. Dividing time into small
intervals of duration $\Delta \tau=20\text{~}\mathrm{ms}$ induces a
binary representation of neural data, where each neuron $i$ either
spikes during a given interval ($s_i=1$) or it does not ($s_i=0$). The joint
statistics observed in $40$ neurons in the retina of a salamander was modelled
by an Ising model~\eqref{eq:thefirst} with magnetic fields and pairwise symmetric
couplings.
Rather than describing the dynamics of neural spikes, 
this model describes the correlated firing of different neurons over the course of
the experiment.
The symmetric couplings $J_{ij}$
in~\eqref{eq:thefirst} describe statistical dependencies, not physical
connections. The synaptic connections between neurons, on the other hand, are generally 
not symmetric. 

In this context, the distribution~\eqref{eq:thefirst} can be viewed
as the form of the maximum entropy distribution
over neural states, given the observed one- and two-point
correlations~\cite{Schneidman2006a}.  In~\cite{Shlens2006a, Cocco2009a}, a good match was
found between the statistics of three neurons predicted
by~\eqref{eq:thefirst} and the firing patterns of the same neurons in
the data. This means that the model with pairwise couplings provides
a statistical description of the empirical data, one that can
even be used to make predictions. Similar results were obtained also from cortical cells in
cell cultures~\cite{tang2008a}. 

The mapping from the neural data to a spin model rests on dividing
time into discrete bins of duration $\Delta \tau$. A different choice
of this interval would lead to different spin configurations; in
particular changing $\Delta \tau$ affects the magnetisation of all
spins by altering the number of intervals in which a neuron fires. In~\cite{roudinirenberglatham2009a}, Roudi, Nirenberg and Latham show that
the pairwise model~\eqref{eq:thefirst} provides a good description of
the underlying spin statistics (generated by neural spike trains),
provided $N\, \Delta \tau \, \nu \ll 1$, where $\nu$ is the average
firing rate of neurons. Increasing the bin size beyond this regime
leads to an increase in bins where multiple neurons fire, as a result couplings beyond the pairwise couplings
in~\eqref{eq:thefirst} can become important.  

As a minimal model of neural correlations, the statistics~\eqref{eq:thefirst}
has been extended in several ways. Tka{\v{c}}ik et
al.~\cite{tkavcik2010} and Granot-Atedgi et al.~\cite{granot2013a} consider
stimulus-dependent magnetic fields, that is, fields which depend on the
stimulus presented to the experimental subject at a particular time of
the experiment. Ohiorhenuan et al. looks at 
stimulus-dependent couplings~\cite{ohiorhenuan2010a}. When the number of neurons increases
to $\sim100$, limitations of the pairwise model~\eqref{eq:thefirst}
become apparent, which has be addressed by adding additional terms coupling
triplets, etc. of spins in the exponent of the Boltzmann measure~\eqref{eq:thefirst}~\cite{ganmor2011a}.

The statistics~\eqref{eq:thefirst} serves as a description of the
empirical data: the couplings between spins in the
Hamiltonian~\eqref{eq:thesecond} do not 
describe physical connections between the neurons. The determination
of the network of neural connections from the observed neural
activities is thus a different question. Simoncelli and
collaborators~\cite{paninski2004a,pillow2008a} and Cocco, Leibler, and Monasson \cite{Cocco2009a}
use an integrate-and-fire model \cite{burkitt2006review} to infer how the neurons are interconnected
on the basis of time series of spikes in all neurons. In such a model, the
membrane potential of neuron $i$ obeys the dynamics
\begin{equation}
C \frac{dV_i}{dt}=\sum_{j\neq i} J_{ij} \sum_l K(t-t_{jl}) +I_i -g V_i
+\xi_i(t) \ ,
\end{equation}
where the first term on the right-hand side encodes the synaptic
connections $J_{ij} $ and a memory kernel $K$; $t_{jl}$ specifies the
time at which neuron $j$ emitted its $l^{\textrm{th}}$ spike. The remaining terms
describe a background current, voltage leakage, and white
noise. Finding the synaptic connections $J_{ij}$ that best describe a
large set of neural spike trains is a computational challenge;
\cite{paninski2006a,Cocco2009a} develop an approximation based on
maximum likelihood, see section~\ref{sec:maxlikelihood}. A related approach
based on point processes and generalized linear models (GLM) is presented in~\cite{truccolo2005point}. 
We will
discuss this problem of inferring the network of connections the context of the non-equilibrium
models in section~\ref{sec:nonequilibrium}. 

Neural recordings give the firing patterns of several neurons over
time. These neurons may have connections between them, but they also
receive signals from neural cells whose activity is not recorded~\cite{macke2011a}. 
In \cite{tyrcha2013}, the effect of
connections between neurons is disentangled from correlations arising
from shared non-stationary input. This raises the possibility that the
correlations described by the pairwise model~\eqref{eq:thefirst} in
\cite{Schneidman2006a} and related works originate
from a confounding factor (connections to a neuron other than those
whose signal is measured), rather than from connections between
recorded neurons~\cite{kulkarni2007common}.

\subsection{Reconstruction of gene regulatory networks}  
\label{sec:geneexpression}

The central dogma of molecular biology is this: Proteins are
macromolecules consisting of long chains of amino acids. The
particular sequence of a protein is encoded in DNA, a double-stranded
helix of complementary nucleotides. Specific parts of DNA, the genes,
are transcribed by polymerases, producing a single-stranded copy
called m(essenger)RNAs, which are translated by ribosomes, usually
multiple times, to produce proteins.

The process of producing protein molecules from the DNA template by
transcription and translation is called gene expression.  The
expression of a gene is tightly controlled to ensure that the right
amounts of proteins are produced at the right time.  One important
control mechanism is transcription factors, proteins which affect the
expression of a gene (or several) by binding to DNA near the
transcription start site of that gene. This part of DNA is called the
regulatory region of a gene. A target gene of a transcription factor
may in turn encode another transcription factor, leading to a cascade
of regulatory events. To add further complications, the binding of
multiple transcription factors in the regulatory region of a gene
leads to combinatorial control exerted by several transcription
factors on the expression of a gene~\cite{buchlergerlandhwa2003a}. Can
the regulatory connections between genes be inferred from data on gene
expression, that is, can we learn the identity of transcription
factors and their targets?

Over the last decades, the simultaneous measurement of expression
levels of all genes have become routine. At the centre of this
development are two distinct technological advances to measure mRNA
levels. The first is microarrays, consisting of thousands of short DNA
sequences, called probes, grafted to the surface of a small
chip. After converting the mRNA in a sample to DNA by reverse
transcription, cleaving that DNA into short segments, and
fluorescently labelling the resulting DNA segments, fluorescent DNA
can bind to its complementary sequence on the chip. (Reverse
transcription converts mRNA to DNA, a process which requires a
so-called reverse transcriptase as an enzyme.)  The amount of
fluorescent DNA bound to a particular probe depends on the amount of
mRNA originally present in the sample. The relative amount of mRNA
from a particular gene can then be inferred from the fluorescence
signal at the corresponding probes \cite{hoheisel2006a}. A limitation
of microarrays is the large amount of mRNA required: The mRNA sample
is taken from a population of cells.  As a result, cell-to-cell
fluctuations of mRNA concentrations are averaged over. To obtain time
series, populations of cells synchronized to approximately the same
stage in the cell cycle are used \cite{gasch2000a}.

The second way to measure gene expression levels is also based on reverse transcription of mRNA,
followed by high-throughput sequencing of the resulting DNA
segments. Then the relative mRNA levels follow directly from counts of
sequence reads \cite{wang2009a}. Recently, expression levels even in single
cells have been measured in this way \cite{wills2013a}. In combination
with barcoding (adding short DNA markers to identify individual cells), $10^4$ cells can have their expression
profiled individually in a single sequencing run~\cite{macosko2015a}. Such data may
allow, for instance, the analysis of the response of target genes to fluctuations in the
concentration of transcription factors. However, due to the destructive nature of
single-cell sequencing, there may never be single-cell data that give
time series of genome-wide expression levels.

Unfortunately, cellular concentrations of proteins are much harder to
measure than mRNA levels. As a result, much of the literature focuses
on mRNA levels, neglecting the regulation of translation. Advances in
protein mass-spectrometry \cite{picotti2013a} may lead to data on both
mRNA and protein concentrations. This data would pose the additional
challenge of inferring two separate levels of gene regulation: gene
transcription from DNA to mRNA and translation from mRNA to proteins.

As in the case of neural data discussed in the preceding section, gene
expression data presents two distinct challenges: (i) finding a
statistical description of the data in terms of suitable observables
and (ii) inferring the underlying regulatory connections. Both these
problems have been addressed extensively in the machine learning and
quantitative biology communities. Clustering of gene expression data
to detect sets of genes with correlated expression levels has been
used to detect regulatory relationships. A model-based approach to the
reconstruction of regulatory connections is Boolean networks.  Boolean
networks assign binary states to each gene (gene expression on/off),
and the state of a gene at a given time depends on the state of all
genes at a previous time through a set of logical functions assigned
to each gene. See \cite{Dhaeseleer2000a} for a review of clustering
and Boolean network inference and \cite{hickman2009a} for a review of
Boolean network inference.

A statistical description that has also yielded insight into
regulatory connections is Bayesian networks. A Bayesian network is a
probabilistic model describing a set of random variables (expression
levels) through conditional dependencies described by a directed
acyclic graph. Learning both the structure of the graph and the
statistical dependencies is a hard computational problem, but can
capture strong signals in the data that are often associated with a
regulatory connection. In principle, causal relationships (like the
regulatory connections) can be inferred, in particular if the
regulatory network contains no cycles.  For reviews, see
\cite{friedman2004a,koller2009a}.  Both Boolean or Bayesian networks
have been applied to measurements of the response of expression levels
to external perturbations of the regulatory network or of expression
levels, see ~\cite{ideker2000a,peer2001a}. A full review of these
methods is beyond the scope of this article, instead we focus on
approaches related to the inverse Ising problem.




For a statistical description of gene expression levels,
\cite{lezon2006} applied a model with pairwise couplings
\begin{equation}
\label{eq:Pgene_expression}
P(\{x_i\})=\exp\left[\sum_{i \leq j} J_{ij} x_i x_j + \sum_i h_i
x_i\right]/Z \ ,
\end{equation}
fitted to gene expression levels. The standard definition of 
expression levels $\{x_i\}$ is $\log_2$-values of
fluorescence signals with the mean value for each gene subtracted. Since
\eqref{eq:Pgene_expression} is a multi-variate Gaussian distribution,
the matrix of couplings $J_{ij}$ must be negative definite. 
These couplings can be inferred 
simply by inverting the matrix of variances and covariances of
expression levels. In \cite{lezon2006}, the resulting couplings
$J_{ij}$ were then used to identify hub genes which regulate many
targets. 
The same approach was used in~\cite{locasale2009} to analyse the
cellular signalling networks mediated by the phosphorylation of
specific sites on different proteins. Again, the
distribution~\eqref{eq:Pgene_expression} can be viewed
as a maximum entropy distribution for \emph{continuous} variables with
prescribed first and second moments. This approach is also linked to the
concept of partial correlations in statistics~\cite{baba2004a,krumsiek2011a}.

Again the maximum-entropy distribution~\eqref{eq:Pgene_expression}
has symmetric couplings between expression levels, whereas the network of regulatory
interactions is intrinsically asymmetric. One way to infer the
regulatory connections is time series~\cite{sima2009a}.
\cite{Bailly2010b} uses expression
levels measured at different times to infer the regulatory
connections, based on a minimal model of expression dynamics with asymmetric regulatory
connections between pairs of genes. In this model, expression levels
$x_i^t$ at successive time intervals $t$ obey 
\begin{equation}
\textrm{sign}(x^{t+1}_i)=
\begin{cases}
1    & \text{if } \sum_j J_{ij} x_j^t > \kappa\\
-1    & \text{if } \sum_j J_{ij} x_j^t <\kappa\\
\end{cases} \ ,
\end{equation} 
where $\kappa$ is a threshold. 
The regulatory connections $J_{ij} $ are taken to be discrete, with the values $-1,1,0$ denoting
repression, activation and no regulation of gene $i$ by the product of
gene $j$. The matrix of connections is then inferred based on Bayes
theorem (see section~\ref{sec:maxlikelihood}) and an
iterative algorithm for estimating marginal probabilities (message
passing, see section~\ref{sec:BP}). 

A second line of approach that can provide information on regulatory
connections is perturbations~\cite{tegner2003a}. An example is gene knockdowns, where
the expression of one or more genes is reduced, by introducing small
interfering RNA (siRNA) molecules into the cell \cite{dorsett2004a} or
by other
techniques. siRNA molecules can be introduced into cells
from the outside; after various processing steps they lead to
the cleavage of mRNA with a complementary sequence, which is then no
longer available for translation. If that mRNA translates to 
a transcription factor, all targets of that transcription factor will be
upregulated or downregulated (depending on whether the transcription factor acted as a
repressor or an activator, respectively). Knowing the responses of gene
expression levels to a sufficient number of such perturbations allows
the inference of regulatory connections.
\cite{molinelli2013a}
considers a model of gene expression dynamics based on
continuous variables $x_i$ evolving deterministically as $\partial_t x_i= a_i
\tanh (\sum_j J_{ij} x_j) - c_i x_i$. The first term
describes how the expression level of gene $j$ affects the rate of
gene expression of gene $i$ via the regulatory connection $J_{ij}$,
the second term describes mRNA degradation. The stationary points of this
model shift in response to perturbations of expression levels of
particular genes 
(for instance through knockdowns), and these changes depend on
regulatory connections. In \cite{molinelli2013a}, 
the regulatory connections are inferred from
perturbation data, again using belief propagation.

\subsection{Protein structure determination} 
\label{sec:proteinstruct}
Tremendous efforts have been made to determine the
three-dimensional structure of proteins. A linear amino acid chain
folds into a convoluted shape, the folded protein, thus bringing amino acids into close
physical proximity that are separated by a long distance along the
linear sequence.

Due to the number of proteins (several thousand per organism) and
the length of individual proteins (hundreds of amino acid residues), protein structure determination is
a vast undertaking. However, the rewards are also substantial. The
three-dimensional structure of a protein determines its physical and
chemical properties, and how it interacts with other cellular components: broadly, the shape of a protein determines many aspects of its
function. Protein structure determination relies on crystallizing
proteins and analysing the X-ray diffraction pattern of the resulting
solid. Given the experimental effort required, the determination of a
protein's structure from its sequence alone has been a key
challenge to computational biology for several
decades~\cite{dill2012a,deJuan2013a}. The computational approach
models the forces between amino acids in order to find the
low-energy structure a protein in solution will fold into. Depending
on the level of detail, this approach requires extensive computational
resources.

\begin{figure}[b!]
	\includegraphics[width = .37\textwidth]{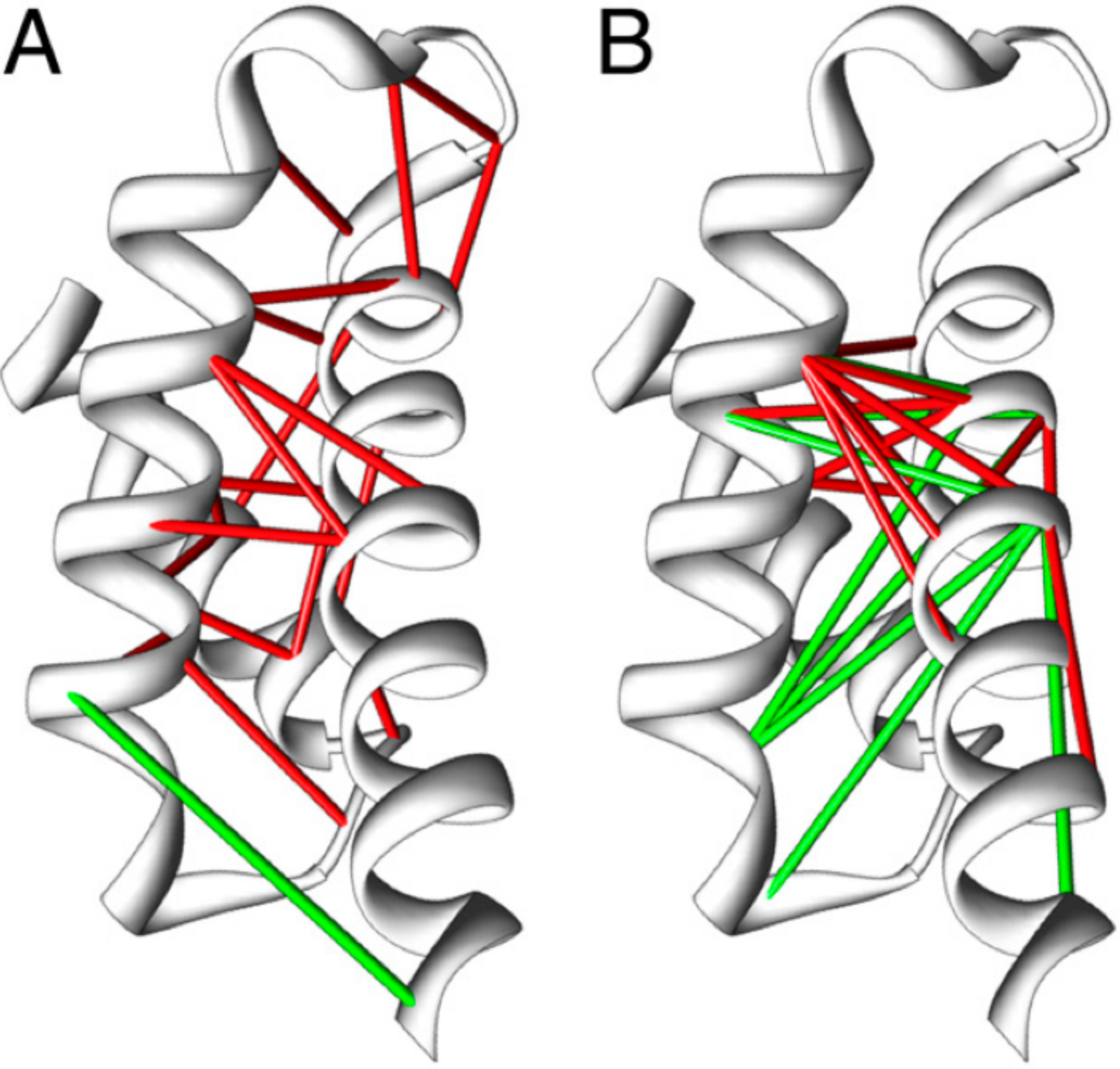}
	\caption{{\bf Correlations and couplings in protein structure
			determination.} Both figures show the three-dimensional structure
		of a particular part (region 2) of the protein SigmaE of
		\textit{E. coli}, as determined by X-ray
		diffraction. This protein, or rather a protein of similar sequence
		and presumably similar structure, occurs in many other bacterial
		species as well.
		In figure B, lines indicate
		pairs of sequence positions whose amino acids are highly correlated
		across different bacteria: for each pair of sequence positions
		at least $5$ amino acids apart, the mutual information of
		pairwise frequency counts of amino acids was calculated, and the $20$
		most correlated pairs are shown here.  
		Such pairs that also turn out to be close in the three-dimensional structure are
		shown in red, those whose distance exceeds $8{\buildrel _{\circ} \over {\mathrm{A}}}$ are
		shown in green. We see about as many highly correlated sequence
		pairs that are proximal to one another as correlated pairs that are
		further apart. By contrast, in figure A, lines show sequence pairs
		that are strongly coupled in the Potts model~\eqref{eq:protein_potts},
		whose model parameters are inferred from the correlations. The fraction of
		false contact predictions (green lines) is reduced considerably. 
		The figures are taken from~\cite{Morcos2011a}.
		\label{fig_morcos}}
\end{figure}

An attractive alternative enlists evolutionary information: Suppose
that we
have at our disposal amino acid sequences of a protein as it appears in different
related species (so-called orthologs). While the sequences are not identical across species,
they preserve to some degree the three-dimensional shape of the
protein. Suppose a specific pair of amino acids that interact strongly
with each other and bring together parts of the protein that are
distal on the linear sequence. Replacing this pair with another,
equally strongly interacting pair of amino acids would change the
sequence, but leave the structure unchanged. For this reason, we expect
sequence differences across species to reflect the structure of the
protein. Specifically, we expect correlations of amino acids in
positions that are proximal to each other in the three-dimensional
structure. In turn, the correlations observed between amino acids at
different positions might allow to infer which pairs of amino acids are
proximal to each other in three dimensions (the so-called contact
map). The use of such genomic information has recently lead to predictions of
the three-dimensional structure of many protein families inaccessible
to other methods~\cite{ovchinnikov2017}, for a review
see~\cite{coccoetal2017}. 

\begin{figure}
	\includegraphics[width = .5\textwidth]{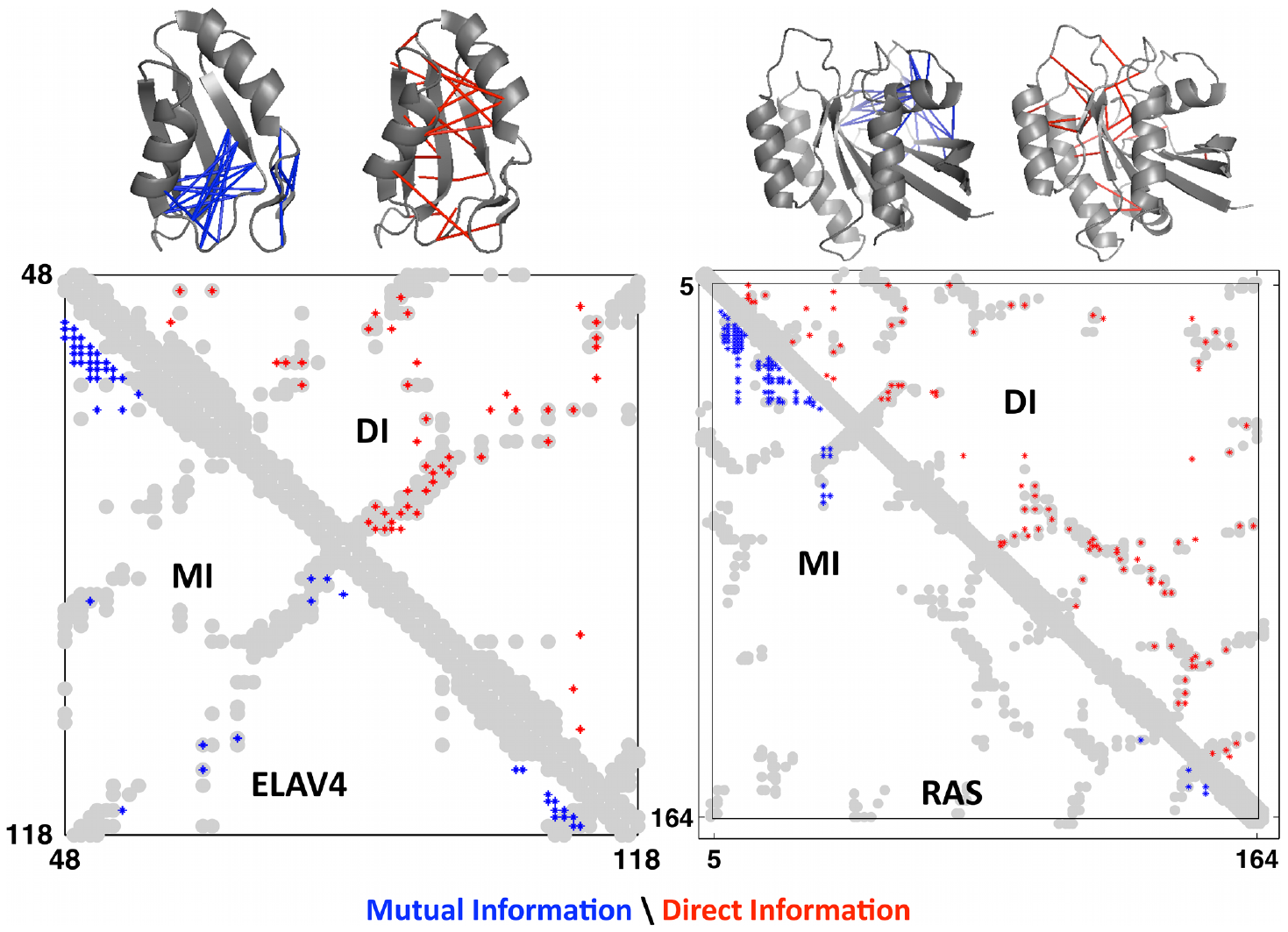}
	\caption{{\bf Protein contact maps predicted from evolutionary
			correlations.} The two figures show contact maps for the ELAV4
		protein (left) and the RAS protein (right). $x$- and $y$-axes
		correspond to sequence positions along the linear chain of
		amino acids. Pairs of sequence positions whose amino acids are in close proximity 
		in the folded protein are indicated in grey (experimental data). Pairs of sequence
		positions with highly correlated amino acids are shown in blue
		(mutual information, bottom triangle). Pairs of sequence positions
		with high direct information~\eqref{eq:direct_information} calculated
		from~\eqref{eq:protein_potts} are shown in
		red. The coincidence of red and grey points shows excellent
		agreement between predictions from direct information with the experimentally determined structure of
		the protein. The figure is taken from~\cite{marks2011a}.
		\label{fig:marks}}
\end{figure}

Early work looked at the correlations as a measure of
proximity~\cite{gobel1994,lockless1999a,socolich2005,halabi2009a}. However
correlations are transitive; if amino acids at sequence sites $i$ and $j$ are correlated
due to proximity in the folded state, and $j$ and $k$ are correlated for some reason, $i$
and $k$ will also exhibit correlations, which need not stem from
proximity. This problem is addressed by an inverse approach aimed at
finding the set of pairwise couplings that lead to the observed
correlations or
sequences~\cite{Weigt2009a,burger2010,Morcos2011a,marks2011a,sulkowska2012,
	dago2012,hopf2012a,cocco2013a,ekeberg2013}. Since
each sequence position can be taken up by one of $20$ amino acids or a
gap in the sequence alignment, there are $21^2$ correlations at each
pair of sequence positions. In~\cite{Weigt2009a,Morcos2011a} a
statistical model with pairwise interactions is formulated, based on
the Hamiltonian
\begin{equation}
\label{eq:protein_potts}
H= -\sum_{i<j} J_{ij}(s_i,s_j) - \sum_{i} h_i(s_i) \ .
\end{equation}
This Hamiltonian depends on spin variables $s_i$, one for each sequence
position $i=1,\ldots,N$. Each spin variable can take on one of $21$ values, describing the $20$
possible amino acids at that sequence position as well as the
possibility of a gap (corresponding to an extra amino acid inserted
in a particular position in the sequences of other organisms). Each pair of
amino acids $s_i,s_j$ in sequence position $i,j$ contributes $J_{ij}(s_i,s_j)$ to the energy.
The inverse problem is to find the couplings $J_{ij}(A,B)$ for each pair of sequence positions $i,j$
and pair of amino acids $A,B$, as well as field $h_i(A)$, such that the
amino acid frequencies and correlations observed
across species are reproduced. 
The sequence positions with strong pairwise couplings
are then predicted to be proximal in the protein
structure. A simple measure of the coupling between sequence positions
is the matrix norm (Frobenius norm) $\sum_{s_i,s_j} (J_{ij}(s_i,s_j))^2$. The
so-called direct information~\cite{Weigt2009a} is an alternative
measure based on information theory. A two-site model is defined with
$p_{ij}(s_i,s_j)=\exp\{ J_{ij}(s_i,s_j) + h_i(s_i) + h_j (s_j)\}/Z_{ij}$. Direct information is the mutual
information between the 
two-site model and a model without correlations between the amino acids
\begin{equation}
\label{eq:direct_information}
DI_{ij}=\sum_{s_i,s_j}p_{ij}(s_i,s_j) \ln\left(\frac{p_{ij}(s_i,s_j) }{p_{i}(s_i)   p_{j}(s_j)}\right) \ .
\end{equation}

The Boltzmann distribution resulting from~\eqref{eq:protein_potts} can
be viewed as the maximum entropy distribution with one- and two-point
correlations between amino acids in different sequence positions
determined by the data.  There is no reason to exclude higher order
terms in the Hamiltonian~\eqref{eq:protein_potts} describing
interactions between triplets of sequence positions, although the
introduction of such terms 
may lead to overfitting. 
Also, fitting the Boltzmann distribution~\eqref{eq:protein_potts} to
sequence data uses no prior information on protein structures; for this
reason it is called an \emph{unsupervised method}. Recently, neural
network models trained on sequence data \emph{and} protein structures
(\emph{supervised learning}) have
been very successful in predicting new structures~\cite{jones2015metapsicov,wang2017accurate}.  

The maximum entropy approach to structure analysis is not limited to
evolutionary data. In~\cite{zhangwolynes2015a} Zhang and Wolynes
analyse chromosome conformation capture experiments and use the
observed frequency of contacts between different parts of a chromosome
in a maximum entropy approach to predict the structure and topology of
the chromosomes.

\subsection{Fitness landscape inference} 
The concept of fitness lies at the core of evolutionary
biology. Fitness quantifies the average reproductive success (number
of offspring) of an organism with a particular genotype,
\textit{i.e.}, a particular DNA sequence. The dependence of fitness on
the genotype can be visualized as a fitness landscape in a
high-dimensional space, where fitness specifies the height of the landscape. As the
number of possible sequences grows exponentially with their length,
the fitness landscape requires in principle an exponentially large
number of parameters to specify, and in turn those parameters need an
exponentially growing amount of data to infer.

A suitable model system for the 
inference of a fitness landscape is HIV proteins, due to the large
number of sequences stored in clinical databases
and the relative ease of generating mutants and measuring
the resulting fitness. In a series of papers, Chakraborty and
co-workers proposed a fitness model the so-called Gag protein family (group-specific antigen)
of the HIV virus~\cite{dahirel2011a,ferguson2013a,shekhar2013a,mann2014a}. The model is based on pairwise interactions
between amino acids. Retaining only the information whether the
amino acid at sequence position $i$ was mutated ($s_i=1$) with respect
to a reference sequence or not ($s_i=0$), Chakraborty and
co-workers  suggest a minimal model for the fitness
landscape given by the Ising Hamiltonian~\eqref{eq:thesecond}.
Again, one can view the landscape~\eqref{eq:thesecond} as generating
the maximum entropy distribution constrained by the
observed one- and two-point correlations. 

Adding a constant
to~\eqref{eq:thesecond} in order to make fitness (expected number
of offspring) non-negative does not alter the resulting statistics.
The inverse problem is to infer the couplings $J_{ij}$ and fields
$h_i$ from frequencies of amino acids and pairs of amino acids in particular
sequence positions observed in HIV sequence data. Of course it is not clear from the outset that
a model with only pairwise interactions can describe the empirical
fitness landscape. As a test of this
approach,~\cite{ferguson2013a} compares the
prediction of~\eqref{eq:thesecond} for specific mutants to the results of
independent experimental measurements of fitness.

\begin{figure}
	\includegraphics[width = .45\textwidth]{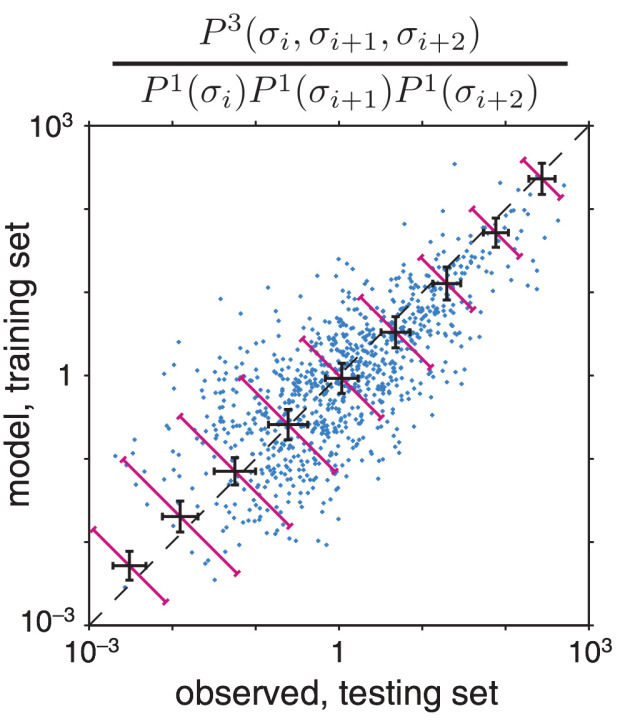} 
	\caption{{\bf Three-point correlations in an amino acid sequences and
			their prediction from a model with pairwise interactions.}  Mora \textit{et al.} look at the
		so-called D-region in the IgM
		protein (maximum length $N=8$)~\cite{mora2010a}. The D-region plays an important role in immune
		response. The frequencies at which given triplets of
		consecutive amino acids occur were compiled ($x$-axis, normalized with
		respect to the prediction of a model with independent sites). The
		results are compared to
		the prediction from a model with pairwise interactions
		like~\eqref{eq:thesecond} on the $y$-axis. 
		The figure is taken from~\cite{mora2010a}.
		\label{fig:threepoint_mora}}
\end{figure}

Statistical models of sequences described by pairwise interactions may
be useful to model a wide range of protein families with different
functions~\cite{hopf2015a}, and
have been used in other contexts as well. Santolini, Mora, and Hakim model the
statistics of sequences binding transcription factors
using~\eqref{eq:protein_potts}, with each spin taking one of four states to
characterize the nucleotides $A,C,G,T$~\cite{santolini2014a}. A similar
model is used in~\cite{mora2010a} to model the sequence diversity of
the so-called IgM protein, an antibody which plays a key role in the early immune
response. The model with pairwise interactions predicts non-trivial three-point correlations
which compare well with those found in the data, see
figure~\ref{fig:threepoint_mora}.  

\subsection{Combinatorial antibiotic treatment} Antibiotics are chemical
compounds which kill specific bacteria or inhibit their growth~\cite{walsh2000a,kohanski2010a}. 
Mutations in the bacterial DNA can lead to 
resistance against a particular antibiotic, which is a major hazard to public health~\cite{wise1998a,liu2015a}.
One strategy to slow down or eliminate the emergence of resistance is
to use a combination of antibiotics either simultaneously or in rotation~\cite{walsh2000a,kohanski2010a}.
The key problem of this approach is to find combinations of compounds
which are particularly effective against a particular strain of
bacteria. Trying out all combinations experimentally is prohibitively
expensive. Wood et al. use an inverse statistical approach to
predict the effect of combinations of several antibiotics from data on the effect of
pairs of antibiotics~\cite{wood2012a} . The available antibiotics are labelled
$i=1,\ldots,N$; in \cite{wood2012a} a distribution over continuous
variables $x_i$ is constructed, such that $\langle x_i \rangle$ gives the bacterial
growth rate when antibiotic $i$ is administered,  $\langle x_i x_j\rangle$
gives the growth rate with both $i$ and $j$ are given, etc. for
higher moments. Choosing this distribution to be a multi-variate
Gaussian $P(\{x_i\})=\exp\left[\sum_{i \le j} J_{ij} x_i x_j + \sum_i h_i
x_i\right]/Z$ results in simple relationships between the different moments,
which lead to predictions of the response to drug combinations that
are borne out well by experiment~\cite{wood2012a}. 


\subsection{Interactions between species and between individuals}
Species exist in various ecological relationships. For
instance individuals of one species hunt and eat individuals of another
species. Another example is microorganisms whose  growth can be influenced,
both positively and negatively, by the metabolic
output of other microorganisms. Such relationships form a dense web of
ecological interactions between species. Co-culturing and perturbation experiments
(for instance species removal) lead to data which may allow the
inference of
these networks~\cite{faust2012a,hekstra2013}. 

Interactions between organisms exist also at the level of individuals,
for instance when birds form a flock, or fish form
a school. This emergent collective behaviour is thought to have evolved
to minimize the exposure of individuals to
predators. In~\cite{bialek2012a,bialek2014a}, a model with pairwise
interactions between the velocities of birds in a flock is
constructed. Individual birds labelled $i=1,\ldots,N$ move with
velocity $\vec{v}_i$ in a direction specified by
$\vec{s}_i=\vec{v}_i/\vert \vec{v}_i \vert$. The statistics of the
these directions is modelled by a distribution
\begin{equation}
P(\{\vec{s}_i\})=\exp\left[\sum_{i,j} J_{ij} \vec{s}_i \cdot \vec{s}_j\right]/Z \ ,
\end{equation}
where the couplings between the spins $\vec{s}_i$ need to be inferred from the
experimentally observed correlations between normalized velocities. 
This model can be viewed as the maximum-entropy distribution
constrained by pairwise correlations between normalized velocities. From the point of view of statistical physics it describes a
disordered system of Heisenberg spins. As birds frequently change
their neighbours in flight, the couplings are not constant in time and
it makes sense to consider
couplings that depend on the distance between two
individuals~\cite{bialek2012a}. An alternative is to apply the maximum entropy
principle to entire trajectories~\cite{cavagna2014a}.

\subsection{Financial markets}

Market participants exchange commodities, shares in companies,
currencies, or other goods and services, usually for money. The change
in prices of such goods are often correlated, as the demand for
different goods can be influenced by the same
events. In~\cite{bury2013a,bury2013b}, Bury uses a spin model with
pairwise interactions to analyse stock market data. Shares in $N$
different companies are described by binary spin variables, where spin
$s_i=1$ indicates `bullish' conditions for shares in company $i$ with
prices going up at a particular time, and $s_i=-1$ implies `bearish'
conditions with decreasing prices. Couplings $J_{ij}$ 
describe how price changes in shares $i$ affect changes in the price
of $j$, or how prices are affected jointly by external events. Bury
fit stock marked data to this spin model, and found clusters in the
resulting matrix of couplings~\cite{bury2013a}. These clusters
correspond to different industries whose companies are traded on the
market. In~\cite{borysov2015a}, a similar analysis finds that heavy
tails in the distribution of inferred couplings are linked to such
clusters.  Slonim et al. identified clusters in stocks using an
information-based metric of stock prices~\cite{slonim2005a} .

\section{Equilibrium reconstruction}
\label{sec:equilibrium}

The applications discussed above can be classified according to
the symmetry of pairwise couplings: In network reconstruction,
couplings between spins are generally asymmetric, in maximum entropy
models they are symmetric. A stochastic dynamics based on symmetric
couplings entails detailed balance, leading to a steady state described by the
Boltzmann distribution~\cite{krapivsky2010kinetic}, whereas asymmetric
couplings lead to a non-equilibrium steady state. 
This distinction shapes the structure of this
review:  In this section, we discuss the inverse Ising problem in
equilibrium, in section~\ref{sec:nonequilibrium} we turn to
non-equilibrium scenarios.

\subsubsection{Definition of the problem}
\label{sec:equilibrium_definition}

We consider the Ising model with $N$ binary
spin variables $\{s_i = \pm1\},i=1,\ldots,N$. Pairwise couplings (or
coupling strengths)
$J_{ij}$ encode pairwise interactions between the spin variables, and
local magnetic fields $h_i$ act on individual spins. The energy of
a spin configuration $\sss\equiv\{s_i \}$ is specified by the Hamiltonian 
\begin{equation}
H_{\JJJ,\hhh}(\sss)= -\sum_{i<j} J_{ij} \s_i \s_j - \sum_{i} h_i \s_i
\ .
\label{eq: static sequential Hamiltonian}
\end{equation}
The equilibrium statistics of the Ising model is described by the Boltzmann distribution
\begin{equation}
p(\sss) = \frac{1}{Z}e^{-H_{\JJJ,\hhh}(\sss)} \ ,
\label{eq: static sequential Boltzmann}
\end{equation}
where we have subsumed the temperature into couplings and fields such
that $k_B T =1$: The 
statistics of spins under the Boltzmann distribution $\exp\{-\beta H\}/Z$
depends on couplings, magnetic fields, and temperature only through
the products 
$\beta J_{ij}$ and $\beta h_i$. As a result, only the products $\beta
J_{ij}$ and $\beta h_i$ can be inferred and we set $\beta$ to $1$
without loss of generality. The energy specified by the
Hamiltonian~\eqref{eq:thesecond} or its generalisation~\eqref{eq:protein_potts}
is thus a dimensionless quantity. 

$Z$ denotes the partition function
\begin{equation}
Z(\JJJ,\hhh)= \sum_{\sss} e^{-H_{\JJJ,\hhh}(\sss)} \ .
\label{eq: static sequential partition function}
\end{equation}
In such a statistical description of the Ising model, each spin is
represented by a random variable. Throughout, we denote a 
random spin variable by $\S$, and a particular realisation of
that random variable by $\s$. This distinction will become
particularly useful
in the context of non-equilibrium reconstruction in
section~\ref{sec:nonequilibrium}. 
The expectation values of spin variables and their functions then are denoted
\begin{equation}
\mean{Q(\SSS)} \equiv \sum_{\sss} p(\sss) Q(\sss) \ ,
\label{eq: static sequential def mean Q}
\end{equation}
where $Q(\sss)$ is some function mapping a spin configuration to a
number. Examples are the equilibrium magnetizations
$m_i \equiv \mean{\S_i} = \sum_{\sss} p(\sss) \s_i $ or the pair
correlations $\chi_{ij} \equiv \mean{\S_i \S_j}=\sum_{\sss} p(\sss) (\s_i \s_j) $. 
In statistics, the latter observable is  called the pair
average. We are also interested in the connected correlation
$C_{ij}=\chi_{ij}-m_i m_j$, 
which in statistics is known as the covariance.

The equilibrium statistics of the Ising problem~\eqref{eq: static
	sequential Boltzmann} is fully determined by the couplings between
spins and the magnetic fields acting on the spins. Collectively,
couplings and magnetic fields are the \emph{parameters} of the Ising
problem. The \emph{forward Ising problem} is to compute statistical
observables such as the magnetizations and correlations under the Boltzmann distribution~\eqref{eq: static sequential
	Boltzmann}; the couplings and fields are taken as given. The
\emph{inverse Ising problem} works in the reverse direction: The couplings
and fields are unknown and are to be determined from observations of
the spins.  The \emph{equilibrium inverse Ising problem} is to infer
these parameters from spin configurations sampled independently from the
Boltzmann distribution. We denote such a data set of $M$ samples by
$\D=\{\sss^{\mu}\}$ for $\mu=1,2,\ldots,M$. (This 
usage of the term `sample' appears to differ
from how it is used in the statistical mechanics of disordered systems,
where a sample often refers to a random choice of the model parameters, not
spin configurations. However, it is in line with the inverse nature of the
problem: From the point of view of the statistical mechanics of
disordered systems in an inverse statistical problem the `phase space variables' are couplings and
magnetic fields to be inferred, the `quenched disorder' is spin configurations
sampled from the Boltzmann distribution.) 

Generally, neither the values of couplings nor the graph structure
formed by non-zero couplings is known. Unlike
in many instances of the forward problem, the couplings often do not  
conform to a regular, finite-dimensional lattice; there is no sense
of spatial distance between spins. Instead, the couplings might be described by a fully
connected graph, with all pairs of spins coupling to each other,
generally all with different values of the $J_{ij}$. Alternatively,
most of the couplings might be zero, and the non-zero entries of the
coupling matrix might define a structure that is (at least locally) treelike. The graph
formed by the couplings might also be highly heterogeneous with few
highly connected nodes with many non-zero couplings and many spins
coupling only to a few other spins. These distinctions can affect how well
specific inference methods perform, a point we will revisit in
section~\ref{sec:comparison}, which compares the quality of
different methods in different situations. 

\subsection{Maximum likelihood}
\label{sec:maxlikelihood}

The inverse Ising problem is a problem of statistical
inference~\cite{mackay2003,bishop2006}. At the heart of many methods
to reconstruct the parameters of the Ising model is the maximum
likelihood framework, which we discuss here. 

Suppose a set of observations $x^1,x^2,\ldots,x^M$ drawn from a
statistical model $p(x^1,x^2,\ldots,x^M | \theta)$. In the case of the
Ising model, each observation would be a spin configuration $\sss$. 
While the
functional form of this model may be known a priori, the parameter $\theta$
is unknown to us and needs to be inferred from the observed data. Of course, with a finite
amount of data, one cannot hope to determine the parameter $\theta$ exactly. The
so-called \emph{maximum likelihood estimator}
\begin{equation}
\label{eq:maxlikelihood}
\theta^{\ML}=\argmax_{\theta} p(x_1,x_2,\ldots,x_M | \theta)
\end{equation}
has a number of attractive properties~\cite{Cramer1961a}: In the limit of a large number
of samples, $\theta^{\ML}$  converges in probability to the value
$\theta$ being estimated. This property is termed consistency. Also
for large sample sizes, there
is no consistent estimator with a smaller mean-squared error. 
For a finite number of samples, the
maximum likelihood estimator may however be biased, that is, the mean
of $\theta^{\ML}$ over many realisations of the samples does not equal $\theta$ (although the difference vanishes with the sample size).
The term \emph{likelihood} refers to $p(x_1,x_2,\ldots,x_M | \theta)$ viewed
as a function of the parameter $\theta$ at constant values of the
data $x_1,x_2,\ldots,x_M$. The same function at constant $\theta$
gives the \emph{probability} of observing the data
$x_1,x_2,\ldots,x_M$. 

The maximum likelihood estimator~\eqref{eq:maxlikelihood} can also be
derived using Bayes theorem~\cite{mackay2003,bishop2006}. In Bayesian inference, one introduces a
probability distribution $p(\theta)$ over the unknown parameter
$\theta$. This \emph{prior distribution} describes our knowledge prior to receiving the data. Upon accounting
for additional information from the data, our knowledge is described
by the \emph{posterior} distribution given by Bayes theorem 
\begin{eqnarray}
p(\theta|x_1,x_2,\ldots,x_M)
&=&\frac{p(\theta,x_1,x_2,\ldots,x_M)}{p(x_1,x_2,\ldots,x_M)}\\
&=&\frac{p(x_1,x_2,\ldots,x_M|\theta)p(\theta)}{p(x_1,x_2,\ldots,x_M)}. \nonumber
\end{eqnarray}
For the case where $\theta$ is a priori
uniformly distributed (describing a scenario where we have no prior
knowledge of the parameter value), the posterior probability distribution of the parameter
conditioned on the observations $p(\theta|x_1,x_2,\ldots,x_M)$ is proportional to 
$p(x_1,x_2,\ldots,x_M|\theta)$\footnote{This line of argument only
	works if the parameter space is bounded, so a uniform prior can be
	defined.}. Then the parameter value maximizing the probability density $p(\theta|x_1,x_2,\ldots,x_M)$ is given by the
maximum likelihood estimator~\eqref{eq:maxlikelihood}. Maximizing the logarithm of the
likelihood function, termed the log-likelihood function, leads to the same parameter estimate,
because the logarithm is a strictly monotonic function. As the
likelihood scales exponentially with the number of samples, the 
the log-likelihood is more convenient to use. (This is simply the
convenience of not having to deal with very small numbers: the logarithm is
not linked to the quenched average considered in the statistical
mechanics of disordered systems; there is no average involved and the likelihood
depends on both the model parameters and the data.) 

We now apply the principle of maximum likelihood to the inverse Ising
problem. Assuming that the configurations in the dataset were sampled
independently from the Boltzmann distribution~\eqref{eq:thefirst}, the
log-likelihood of the model parameters given the observed
configurations $\D=\{\sss^{\mu}\}$ is derived
easily.
\begin{eqnarray}
\label{eq: static sequential log-likelihood}
L_{\D}(\JJJ,\hhh) &=& \frac{1}{M} \ln p(\D \vert \JJJ,\hhh)  \\ 
&=& \sum_{i<j} J_{ij} \frac{1}{M} \sum_{\mu} \s_i^{\mu} \s_j^{\mu} + \sum_{i} h_i \frac{1}{M} \sum_{i} \s_i^{\mu} -  \ln Z(\JJJ,\hhh) \nonumber \\
&=& \sum_{i<j} J_{ij} \mean{\S_i \S_j}^\D + \sum_{i} h_i \mean{\S_i}^\D  -  \ln Z(\JJJ,\hhh),\nonumber
\end{eqnarray}
gives the log-likelihood per sample, a quantity of order zero
in $M$ since the likelihood scales
exponentially with the number of samples.
The sample averages of spin variables and their functions are defined by
\begin{equation}
\mean{Q}^\D = \frac{1}{M} \sum_{\mu} Q(\sss^{\mu}) \ .
\end{equation}
Beyond the parameters of the Ising model, the log-likelihood
\eqref{eq: static sequential log-likelihood} depends only on the correlations
between pairs of spins observed in the data $\mean{\S_i \S_j}^\D$ and the
magnetizations $\mean{\S_i}^\D$. To determine the maximum-likelihood
estimates of the model parameters we thus only need the pair
correlations and magnetizations observed in the sample (sample averages); at least in
principle, further observables are superfluous.  In the language of
statistics, these sets of sample averages provide \emph{sufficient
	statistics} to determine the model parameters. 

The log-likelihood~\eqref{eq: static sequential log-likelihood} has a
physical interpretation: The first two terms are the sample average of
the (negative of the) energy, the second term adds the free energy. Thus the
log-likelihood is the (negative of the) entropy of the Ising system, based on
the sample estimate of the energy. We will further
discuss this connection in section \ref{sec:thermodynamics}.

A second interpretation of the log-likelihood is based on the difference
between the Boltzmann distribution~\eqref{eq: static sequential
	Boltzmann} and the empirical distribution of the data in the sample 
$\D$, denoted $p^{\D}(\sss)\equiv \frac{1}{M}\sum_{\mu}
\delta_{\sss^{\mu},\sss}$.
The difference between two probability
distributions $p(\sss)$ and $q(\sss)$ can be quantified by 
the Kullback--Leibler (KL) divergence
\begin{equation}
\label{eq:def_kullbachleibler}
\KL (p \vert q) = \sum_{\sss} p(\sss) \ln
\frac{p(\sss)}{q(\sss)} \ ,
\end{equation}
which is non-negative 
and reaches zero only when the two distributions are
identical~\cite{Cover2006a}. The KL divergence between the empirical
distribution and the Boltzmann distribution is 
\begin{eqnarray}
\label{eq:ml_kl}
\KL (p^{\D}\vert p) &=& \sum_{\sss} p^{\D}(\sss) \ln
\frac{p^{\D}(\sss)}{p(\sss)}\\
&=&-L_{\D}(\JJJ,\hhh) + \sum_{\sss} p^{\D}(\sss) \ln p^{\D}(\sss). \nonumber
\end{eqnarray}
The second term (the negative empirical entropy) is independent of the model parameters; the best match
between the Boltzmann distribution and the empirical distribution
(minimal KL divergence) is thus achieved when the likelihood~\eqref{eq:
	static sequential log-likelihood} is maximal. 

Above, we derived the principle of maximum likelihood~\eqref{eq:maxlikelihood} from 
Bayes theorem under the assumption that the model parameter $\theta$
is sampled from a uniform prior distribution. Suppose we had the prior information that
the parameter $\theta$ was taken from some non-uniform distribution, the
posterior distribution would then acquire an additional dependence on the
parameter. In the case of the inverse Ising problem, prior information
might for example describe the sparsity of the coupling matrix, with a suitable
prior  $p_{\JJJ} \sim \exp\left[-\gamma
\sum_{i<j} |J_{ij}|\right]$ that assigns small probabilities to large entries in the
coupling matrix. 
The resulting (log) posterior is
\begin{equation}
\label{eq:posteriorwithregularizer}
\ln p(\JJJ,\hhh|\D) =M L_{\D}(\JJJ,\hhh) -\gamma \sum_{i<j} |J_{ij}| 
\end{equation}
up to terms that do not depend on the model parameters. The maximum of
the posterior is now no longer achieved by maximizing the likelihood,
but involves a  second term that
penalizes coupling matrices with large entries. Maximizing the
posterior with respect to the parameters no longer makes the Boltzmann
distribution as similar to the empirical distribution as possible, but
strikes a balance between making these distributions similar while 
avoiding large values of the couplings. 
In the context
of inference, the second term is called a regularisation term. Different
regularisation terms have been used, including the absolute-value term in~\eqref{eq:posteriorwithregularizer} 
as well as a penalty on the square values of couplings $\sum_{i<j}
J_{ij}^2$ (called $\ell_1$- and $\ell_2$-regularisers, respectively). One
standard way to determine the value of the regularisation coefficient $\gamma$
is to cross-validate with a part of the data that is initially
withheld, that is to probe (as a function of $\gamma$) how well the model can predict
aspects of the data not yet used to infer the model parameters~\cite{Hastie2009a}. 



\subsubsection{Exact maximization of the likelihood}

The maximum likelihood estimate of couplings and magnetic fields
\begin{equation}
\{\JJJ^{\ML},\hhh^{\ML}\} = \argmax L_{\D}(\JJJ,\hhh)
\label{eq: static sequential optimisation}
\end{equation}
has a simple interpretation. Since $\ln Z(\JJJ,\hhh)$ serves as a
generating function for expectation values under the Boltzmann
distribution, we have
\begin{eqnarray}
\label{eq:Boltzmannlearning}
\frac{\partial L_{\D}}{\partial h_i} (\JJJ,\hhh)&=& \mean{\S_i}^{\D}-  \mean{\S_i}  \\
\frac{\partial L_{\D}}{\partial J_{ij}} (\JJJ,\hhh)&=& \mean{\S_i\S_j}^{\D} -
\mean{\S_i \S_j} \ . \nonumber
\end{eqnarray} 
At the maximum of the log-likelihood these derivatives are zero; 
the maximum-likelihood estimate of the parameters is reached when the 
expectation values of pair correlations and magnetizations under
the Boltzmann statistics 
match their sample averages
\begin{eqnarray}
\label{eq: static sequential matching m and C}
\mean{\S_i} &=& \mean{\S_i}^\D \\
\mean{\S_{i}\S_{j}} &=& \mean{\S_i \S_{j}}^\D \ . \nonumber
\end{eqnarray}

The log-likelihood~\eqref{eq: static sequential
	log-likelihood} turns out to be a concave function of the
model parameters, see~\ref{sec:uniquenessofmaximum}. Thus, in principle, it can be
maximized by a convex optimization algorithm.
One particular way to reach the maximum of the likelihood is a
gradient-descent algorithm called Boltzmann machine
learning~\cite{Ackley1985a}. 
At each step of the algorithm
fields and couplings are updated according to 
\begin{align}
\label{eq:Boltzmannlearning_algo}
h_i^{n+1} &=  h_i^{n}
+ \eta \frac{\partial L_{\D}}{\partial h_i} (\JJJ^n,\hhh^n) \\
J_{ij}^{n+1} &=  J_{ij}^{n}+ \eta \frac{\partial L_{\D}}{\partial
	J_{ij}} (\JJJ^n,\hhh^n) \ .
\end{align}
The parameter $\eta$ is the learning rate of the algorithm, which has
\eqref{eq: static sequential matching m and C} as its fixed point. 

In order to
calculate the expectation values $\mean{\S_i}$ and $\mean{\S_{i}\S_{j}}$
on the left-hand side of these equations, one
needs to perform thermal averages of the form~\eqref{eq: static
	sequential def mean Q} over all $2^N$ configurations, which is
generally infeasible for all but the smallest
system sizes. Analogously, when maximizing the
log-likelihood~\eqref{eq: static sequential log-likelihood} directly, the
partition function is the sum over $2^N$ terms.
Moreover, the expectation values or the partition function need to be evaluated many
times during an iterative search for the solution of~\eqref{eq:
	static sequential matching m and C} or the maximum of the likelihood. As a result, also numerical sampling techniques such
as Monte Carlo sampling
are cumbersome, but have been used for moderate
system sizes~\cite{broderick2007faster}. Habeck proposes a Monte
Carlo sampler that draws model parameters from the posterior
distribution~\cite{habeck2014}. A recent algorithm uses information contained in
the shape of the likelihood maximum to speed up the
convergence~\cite{ferrari2015a}. An important development in machine
learning has led to the so-called restricted Boltzmann machines,
where couplings form a symmetric and bipartite graph. Variables fall
into two classes, termed `visible' and `hidden', with couplings
never linking variables of the same
class. This allows fast learning algorithms~\cite{fischer2012introduction} at the
expense of additional hidden variables. 

We stress that the difficulty of maximizing the
likelihood is associated with the restriction of our input to the first two moments (magnetisations and
correlations) of the data. On the one hand, this restriction is natural, as the
likelihood only depends on these two moments. On the other hand, 
\emph{computationally efficient methods} have been developed that effectively
use correlations in the data beyond the first two moments. An
important example is pseudolikelihood, which we will discuss in
section~\ref{sec:pseudo-likelihood}. Other learning techniques that
sidestep the computation of the partition function include
score matching~\cite{hyvarinen2005estimation} and minimum probability
flow~\cite{sohl2011new}. Also, when the number of samples is small (compared to the number of 
spins), the likelihood need no longer be the best quantity to
optimize. 


\subsubsection{Uniqueness of the solution}

\label{sec:uniquenessofmaximum}
We will show that the log-likelihood~\eqref{eq: static sequential
	log-likelihood} is a strictly concave function of the model parameters
(couplings and magnetic fields). As the space of parameters is convex, the maximum of the
log-likelihood is unique. 

We use the shorthands
$\pmb{\lambda}= \{\JJJ,\hhh\}$ and $\{Q_k(\sss)\}=\{s_i s_j, s_i\}$ for the model
parameters and the functions coupling to them and
write the Boltzmann distribution as
\begin{equation}
p (\sss) = \frac{1}{Z(\pmb{\lambda})} e^{\sum_{k} \lambda_k Q_k
	(\sss)} \ .
\end{equation}
For such a general class of exponential distributions~\cite{Hastie2009a}, the second
derivatives of the log-likelihood $L_\D$ with respect to a parameters obey
\begin{equation}
\label{eq:d2Ldlambda2}
-\frac{\partial^2 L_{\D}}{\partial \lambda_i \partial \lambda_j}
(\pmb{\lambda})= \mean{Q_i Q_j} - \mean{Q_i} \mean{Q_j} \ .
\end{equation}
This matrix of second derivatives is
non-negative (has no negative eigenvalues) since $\sum_{ij}  (\mean{Q_i Q_j} -
\mean{Q_i} \mean{Q_j}) x_i x_j= \mean{\left[ \sum_{k} \left(x_k Q_k
	- \mean{x_k Q_k} \right) \right]^2} \ge 0$ for all
$x_i$. If no non-trivial linear combination of the observables $Q_k$ has vanishing fluctuations,
the Hessian matrix is even positive-definite.
For the inverse Ising problem, there are indeed no non-trivial linear
combinations of the spin variables $\S_i$ and pairs of spins variables
$\S_i \S_j$ that do not fluctuate under the Boltzmann measure, unless
some of the couplings or fields are infinite. As a result, the maximum
of the likelihood, if it exists, is unique. However, it can happen
that the maximum lies at infinite values of some of the
parameters (for instance when the samples contain only positive values
of a particular spin, the maximum likelihood value of the
corresponding magnetic field is infinite). These divergences can be
avoided with the introduction of a regularisation term, see section~\ref{sec:maxlikelihood}.


%
\subsubsection{Maximum entropy modelling}
\label{sec:maxent}

The Boltzmann distribution in general and the Ising
Hamiltonian~\eqref{eq:thesecond} in particular can be derived from information
theory and the principle of maximum entropy. This principle has been
invoked in neural modelling~\cite{Schneidman2006a}, protein
structure determination~\cite{Weigt2009a}, and DNA sequence analysis~\cite{mora2010a}. 
In this section, we discuss the statistical basis of 
Shannon's entropy, the principle of maximum entropy, and their application
to inverse statistical modelling. 

Consider $M$ distinguishable balls, each to be placed in a box with $R$
compartments. The number of ways of placing the balls such that 
$n_r$ balls are in the $r$th compartment ($r \in
1,\ldots,R$) is 
\begin{equation}
\label{eq:entropycombinatorics1}
W=\frac{M!}{\prod_{r=1}^{R}n_r!}
\end{equation}
with $\sum_{r=1}^{R}n_r=M$. For large $M$, we write $n_r=M q_r$ and exploit
Stirling's formula $n_r! \approx e^{n_r}n_r^{n_r}$, yielding the Gibbs entropy
\begin{equation}
\label{eq:entropycombinatorics2}
\ln W/M \approx - \sum_{r=1}^{R} q_r \ln q_r \ .
\end{equation}
This combinatorial result forms the basis of equilibrium statistical
physics in the classic treatment due to Gibbs and can be found in
standard textbooks. In the context of statistical physics,
each of the 
$R$ compartments corresponds to a microstate of a system, 
and each microstate $r$ is associated with energy $E_r$. The $M$
balls in the compartments describe a set of copies of the system, or
a so-called \textit{ensemble of replicas}. The replicas may exchange
energy with each other, 
while the ensemble of replicas itself is isolated and has a fixed total energy $M E$ (and possibly other conserved
quantities). In this way, the replicas can be thought
of as providing a heat-bath for each other. If we assume that each state of the
ensemble of replicas with a given total energy is equally likely, the
statistics of $q_r$ is dominated by a sharp maximum of $W$ as a
function of the $q_r$, subject to the constraint $\sum_r q_r=1$ and
$\sum_r E_r q_r=E$. Using Lagrange multipliers to
maximize~\eqref{eq:entropycombinatorics2} subject to these constraints
yields the Boltzmann distribution~\cite{jaynes1957a}.

This seminal line of argument can also be
used to derive Shannon's information entropy~\eqref{eq:shannon}. The argument is due to Wallis
and is recounted in~\cite{Jaynes1989a}. Suppose we want to find a
probability distribution $p_r$ compatible with a certain constraint, for
instance a specific expectation value $\sum_r p_r E_r=E$ to
within some small margin of error. Consider $M$ independent casts of a fair
die with $R$ faces. We denote the number of times outcome $r$ is
realized in these throws as $n_r$. The probability of a particular set
$\{n_r\}$ is 
\begin{equation}
\label{eq:entropycombinatorics3}
\frac{M!}{\prod_{r=1}^{R}n_r!} \prod_{r=1}^{R} (1/R)^{n_r} \ .
\end{equation}
In the limit of large $M$, the logarithm of this probability is
$-M\sum_r q_r \ln q_r-M\ln R$ with $q_r=n_r/M$. 

Each set of $M$ casts defines one
instance of the $\{n_r\}$. In most instances, the 
constraint will not be realized. For those (potentially rare) instances obeying
the constraint, we can ask what are the most likely values of $n_r$, and
correspondingly $q_r$. Maximising the Shannon's information entropy $-\sum_r q_r \ln
q_r$ subject to the constraint and the normalisation $\sum_r q_r=1$
gives the so-called \textit{maximum-entropy estimate} of $p_r$. If the underlying set of
probabilities (the die with $R$ faces) differs from the uniform
distribution, so outcome $r$ occurs with probability $q^0_r$, it
is not the entropy but the relative entropy
$-\sum_r q_r \ln \frac{q_r}{q^0_r}$ that is to be maximised. Up to a
sign, this is the Kullback-Leibler
divergence~\eqref{eq:def_kullbachleibler} between $q_r$ and $q_r^0$.

The maximum-entropy estimate can be used to approximate an unknown
probability distribution $q_r$ that is under-sampled. Suppose 
data is sampled several times from some unknown probability
distribution. With a sufficient number of samples $M$, the
distribution $q_r$ can be easily determined from frequency counts 
$q_r=n_r/M$. Often this is not feasible; if $q_rM \ll 1$, $n_r$ fluctuates strongly from one set of samples to the
next. This situation appears naturally when the number of possible
outcomes $R$ grows exponentially with the size of the system, see \textit{e.g.}
section~\ref{sec:neural}. Nevertheless the data may be sufficient to pin down
one or several expectation values. The maximum-entropy estimate has
been proposed as the most unbiased
estimate of the unknown probability distribution compatible with the
observed expectation values~\cite{Jaynes1989a}. For a discussion of
the different ways to justify the maximum entropy principle, and
derivations based on robust estimates see~\cite{tikochinsky1984alternative}. 

Many applications of the maximum-
entropy estimate are in image analysis and spectral analysis~\cite{gull1984a},
for reviews in physics and biology see~\cite{banavar2010a,bialek2007rediscovering,Mora2011b}, and for critical discussion see~\cite{aurell2016maximum,vannimwegen2016inferring}. 

The connection between maximum entropy and the inverse Ising problem
is simple: For a set of $N$ binary variables, the distribution with given first
and second moments maximizing the information entropy
is the Boltzmann distribution~\eqref{eq:thefirst} with the Ising
Hamiltonian~\eqref{eq:thesecond}. We use Lagrange multipliers to maximize the information entropy~\eqref{eq:shannon}
subject to the normalization condition and the constraints on the first and second moments (magnetisations and
pair correlations) of $p(\sss)$ to be $\mmm$ and $\pmb{\chi}$. Setting the derivatives of 
\begin{align}
&\sum_{\sss} -p(\sss) \ln  p(\sss) + \eta [1-\sum_{\sss} p(\sss)]  +  \nonumber \\ &\sum_i h_i [m_i -  \sum_{\sss} p(\sss) s_i] +\sum_{i<j} J_{ij} [\chi_{ij} - \sum_{\sss} p(\sss) s_i s_j] 
\end{align}
with respect to $p(\sss)$ to zero yields the Ising model~\eqref{eq:thefirst}. 
The Lagrange multipliers $\hhh$ and $\JJJ$ need to be chosen to reproduce the first and second moments (magnetisations and
correlations) of the data and can be interpreted as couplings between spins
and magnetic fields. 

While this principle appears to provide a
statistical foundation to the model~\eqref{eq:thefirst}, there is no a
priori reason to disregard empirical data beyond the first two
moments.  Instead, the pairwise couplings result from the
particular choice of making the probability distribution match the first two moments of the
data. The reasons for this step may be different in different applications.
\begin{itemize}
	\item Moments beyond the first and second may be poorly determined by
	the data. Conversely, with an increasing number of samples, the
	determination of higher
	order correlations and hence interactions between triplets of spin variables \textit{etc.}
	becomes viable. 
	\item The data may actually be generated by an equilibrium model with
	(at most) pairwise interactions between spin variables. This need not be
	obvious from observed correlations of any order, but can be
	tested by comparing
	three-point correlations predicted by a model with
	pairwise couplings to the corresponding correlations in the data. Examples are found in sequence
	analysis, where population dynamics leads to an equilibrium steady
	state~\cite{BergWillmannLaessig2004,SellaHirsh2005} and the energy
	can often be approximated by pairwise
	couplings~\cite{mora2010a, santolini2014a}. For a review
	see~\cite{stein2015a}. Surprisingly, also in neural data (not
	generated by an equilibrium model), three-point correlations are predicted well
	by a model with pairwise interactions~\cite{tkacik2009spin,tkavcik2014a}.
	\item A model of binary variables interacting via a high-order
	coupling terms $J_{ijk\ldots}s_is_js_k\ldots$ can sometimes be
	approximated surprisingly well by pairwise interactions. 
	This seems to be the case when the couplings
	are dense, so that each variable appears in several coupling
	terms~\cite{merchan2015a}. 
	\item Often one seeks to describe a subset of $n$ variables
	$s_1,s_2,\ldots,s_n$ from 
	a larger set of $N$ variables, for instance when only the
	variables in the subset can be observed. The subset of variables is
	characterized by effective interactions which stem from interactions
	between variables in the subset, and from interactions with the other
	variables. If the subset is sufficiently small, the resulting
	statistics is often described by a model with pairwise couplings~\cite{roudinirenberglatham2009a}. 
	\item The true probability distribution underlying some data may be
	too complicated to calculate in practice. A more modest goal then is to
	describe the data using an effective statistical model such
	as~\eqref{eq:thefirst}, which is tractable and allows the derivation
	of bounds on the entropy or the free energy. Examples are the description
	of neural data and gene expression data using the Ising
	model with symmetric couplings (see~\ref{sec:neural}
	and~\ref{sec:geneexpression}).
	\item There are also useful models which are computationally tractable but do not
	maximize the entropy. An example is Gaussian models to generate
	artificial spike trains with prescribed pair correlations~\cite{amari2003synchronous,macke2009generating}.
\end{itemize}


\subsubsection{Information theoretic bounds on graphical model reconstruction}
\label{sec:infortheo}

A particular facet of the inverse Ising problem is graphical
model selection. 
Consider the Ising problem on a graph. A graph is a set
of nodes and edges connecting these nodes, and nodes associated with
spin variables. Couplings between node
pairs connected by an edge are non-zero, couplings between unconnected node
pairs are zero. The graphical model selection problem
is to recover the underlying graph (and usually also the values of the
couplings) from data sampled independently from the Boltzmann distribution. 
Given a particular number of
samples, one can ask with which probability a given method can
reconstruct the graph correctly (the reconstruction fluctuates between
different realisations of the samples). 
Notably, there are also \emph{universal}
limits on graphical model selection that are independent of a
particular method. 

In~\cite{santhanam2012information}, Santhanam and Wainwright derive
information-theoretic limits to graphical model selection. Key result
is the dependence of the required number of samples on the smallest
and on the largest coupling 
\begin{equation}
\alpha= \min_{i<j} | J_{ij}| \ , \ \ \beta= \max_{i<j} | J_{ij}|
\end{equation}
and on the maximum node connectivity (number of neighbours on the
graph) $d$. Reconstruction of the graph, by any method, is impossible
if fewer than
\begin{equation}
\max\left\{
\frac{\ln N}{2\alpha \tanh\alpha}, 
\frac{e^{\beta d} \ln(Nd/4 - 1)}{4d\alpha e^{\alpha}},
\frac{d}{8} \ln \left(\frac{N}{8d}\right)
\right\}
\end{equation}
samples are available (the precise statement is of a probabilistic nature, see~\cite{santhanam2012information}). If the maximum connectivity $d$ grows with the system size,
this result implies that at least $c \max\{d^2,\alpha^{-2}\} \ln N$
samples are required (with some constant $c$)~\cite{santhanam2012information}.  
The derivation of this and other results is based on Fano's inequality
(Fano's lemma)~\cite{Cover2006a}, which gives a lower bound for the
probability of error of a classification function (such as the mapping
from samples to the graph underlying these samples).

\subsubsection{Thermodynamics of the inverse Ising problem}
\label{sec:thermodynamics}
Calculations in statistical physics
are greatly simplified by introducing thermodynamic potentials. In
this section, we will discuss the method of thermodynamic potentials
for the inverse Ising problem. It turns out that the maximum likelihood
estimation of the fields and couplings is simply a transformation of
the thermodynamic potentials.

Recall that the thermodynamic potential most useful for the forward problem, where
couplings and magnetic fields are given, is the
Helmholtz free energy $ F(\JJJ,\hhh)=-\ln
Z(\JJJ,\hhh)$. Derivatives of this free energy give the magnetizations,
correlations, and other observables. The thermodynamic potential most useful
for the inverse problem, where the pair correlations $\pmb{\chi}$ and magnetizations $\mmm$ are given, is the Legendre transform of
the Helmholtz free energy with respect to both couplings and fields~\cite{Sessak2009a,Cocco2011a,Cocco2012a}
\begin{equation}
\label{eq:entropy_by_legendre}
S(\pmb{\chi},\mmm)= \min_{\JJJ,\hhh} \left[-\sum_{i} h_i m_i - \sum_{i<j} J_{ij} \chi_{ij} - F(\JJJ,\hhh) \right].
\end{equation}  
This thermodynamic potential is readily recognised as the entropy
function; up to a sign, it gives the maximum
likelihood \eqref{eq: static sequential log-likelihood} of the model
parameters. The transformation~\eqref{eq:entropy_by_legendre} thus provides a link between the
inference via maximum likelihood and the statistical physics of
the Ising model as described by its Helmholtz free energy.
The couplings and the fields are found by differentiation, 
\begin{eqnarray}
\label{eq:entropy_reconstruction}
J_{ij}&=& -\frac{\partial S}{\partial \chi_{ij}}  (\pmb{\chi},\mmm) \\
h_i &=& -\frac{\partial S}{\partial m_i} (\pmb{\chi},\mmm) \nonumber \ ,
\end{eqnarray} 
where the derivatives are evaluated at the sample 
correlations and magnetizations. These relationships follow from
the inverse transformation of~\eqref{eq:entropy_by_legendre}
\begin{equation}
F(\JJJ,\hhh)= \min_{\pmb{\chi},\mmm} \left[ -\sum_{i} h_i m_i - \sum_{i<j} J_{ij} \chi_{ij} - S(\pmb{\chi},\mmm) \right],
\end{equation}
by setting derivatives of the term in square brackets with
respect to $\pmb{\chi}$ and $\mmm$ to zero. 

In practice, performing the Legendre transformation of both
$\hhh$ and $\JJJ$ is often not necessary; 
derivatives of the Helmholtz free energy $F(\JJJ,\hhh)$ with respect to $\JJJ$ can also be generated by differentiating with respect to $\hhh$, \textit{e.g.},
\begin{equation}
\frac{\partial F}{\partial J_{ij}} (\JJJ,\hhh)= \frac{\partial^2 F}{\partial h_{i} \partial h_{j}}(\JJJ,\hhh)-\frac{\partial F}{\partial h_{i}} (\JJJ,\hhh) \frac{\partial F}{\partial h_{j}} (\JJJ,\hhh).
\end{equation}
The thermodynamics of the inverse problem can thus be reduced to a
single Legendre transform of the Helmholtz free energy, yielding the
Gibbs free energy
\begin{equation}
\label{eq:gibbs_free}
G(\JJJ,\mmm) = \max_{\hhh} \left[\sum_{i} h_i m_i  +F(\JJJ,\hhh) \right] \ .
\end{equation}
The magnetic fields are given by the first derivative of the Gibbs
free energy
\begin{equation}
\label{eq:dGdm}
h_i = \frac{\partial G}{\partial m_i} (\JJJ,\mmm) \ .
\end{equation}
To infer the couplings, we consider the second derivatives of Gibbs potential, which give
\begin{eqnarray}
\label{eq:d2Gdm2}
\frac{\partial^2 G}{\partial m_j \partial m_i} (\JJJ,\mmm) =(\CCC^{-1})_{ij}  \ ,
\end{eqnarray} 
where $\CCC$ is the matrix of connected correlations $C_{ij}\equiv
\chi_{ij}-m_i m_j$. \eqref{eq:d2Gdm2} follows from the inverse function theorem,
\begin{eqnarray}
\left[ \frac{\partial(h_1, \ldots,h_N)}{\partial(m_1, \ldots, m_N)} \right]_{ij} =\left[ \left({\frac{\partial(m_1, \ldots, m_N)}{\partial(h_1, \ldots,h_N)}}\right)^{-1} \right]_{ij},
\end{eqnarray} 
and linear response theory
\begin{equation}
\label{eq:linearresponse}
C_{ij}=\frac{\partial m_j}{\partial h_i}(\JJJ,\hhh)=-\frac{\partial^2
	F}{\partial h_i \partial h_j} (\JJJ,\hhh) \ ,
\end{equation}
which links the susceptibility of the magnetization to a small change in
the magnetic field with the connected correlation~\cite{stanley1987introduction}.

The result~\eqref{eq:d2Gdm2} turns out to be central to many methods for the inverse
Ising problem. The left-hand side of this expression is a
function of $J_{ij}$. If the Gibbs free
energy \eqref{eq:gibbs_free} can be evaluated or
approximated,~\eqref{eq:d2Gdm2} can be solved to yield the couplings. 
Similarly~\eqref{eq:dGdm} with the estimated couplings and the sample magnetisations
gives the magnetic fields, completing the reconstruction of 
the parameters of the Ising model.

\subsubsection{Variational principles}

For most systems, neither the free energy $F(\JJJ,\hhh)$ nor other
thermodynamic potentials can be evaluated. However, there are many
approximation schemes for $F(\JJJ,\hhh)$~\cite{AMF}, which lead to approximations
for the entropy and the Gibbs free energy. Direct approximation schemes for
$S(\pmb{\chi},\mmm)$ and $G(\JJJ,\mmm)$ have also be formulated within
the context of the inverse Ising problem. The key idea behind most of
these approximations is the variational principle.

The variational principle for the free energy is
\begin{equation}
F(\JJJ,\hhh)= \min_{q}\left\{U[q] - S[q]\right\} \equiv \min_{q} F[q]\ ,
\label{eq: variational F}
\end{equation}
where $q$ denotes a probability distribution over spin
configurations. $U[q] \equiv \mean{H}_{q}$ and $S[q]\equiv \mean{\ln q}_{q}$  and the
minimisation is taken over all distributions $q$. This principle finds
its origin in information theory. Take an arbitrary trial distribution
$q(\sss)$, the Kullback-Leibler
divergence~\eqref{eq:def_kullbachleibler} quantifies the difference between $q$ and the Boltzmann distribution $p$
is positive and vanishes if and only if $q=p$~\cite{Cover2006a}. One then arrives directly at~\eqref{eq: variational F} when rewriting  $\KL (q \vert p)= U[q] - S[q] - F(\JJJ,\hhh)$. 

We will refer to $F[q] \equiv U[q]-S[q]$ as the \emph{functional} Helmholtz free energy,  also  
called the non-equilibrium free energy in the context of non-equilibrium statistical physics~\cite{parrondo2015a}. 
Another term in use is `Gibbs free energy'~\cite{AMF}, which we
have reserved for the thermodynamic potential~\eqref{eq:gibbs_free}.

So far, nothing has been gained as the minimum is
over all possible distributions $q$, including the Boltzmann
distribution itself. A practical approximation arises when 
a constraint is put on $q$, leading to a family of \emph{trial distributions} $q$. Often the minimisation can then be carried 
out over that family, yielding an upper bound to the Helmholtz free energy~\cite{AMF}.

In the context of the inverse problem, it is useful to derive the
variational principles for other thermodynamic potentials as well. 
Using the definition of Gibbs potential~\eqref{eq:gibbs_free} and the
variational principle for the Helmholtz potential~\eqref{eq:
	variational F} we obtain
\begin{equation}
G(\JJJ,\mmm)= \max_{\hhh} \left\{ \sum_{i} h_i m_i + \min_{q} \left\{ U[q] - S[q] \right\} \right\}.
\end{equation}
By means of Lagrange multipliers it is easy to show that this
double extremum can be obtained by a single conditional minimisation,
\begin{equation}
G(\JJJ,\mmm)= \min_{q \in \mathcal{G}} \left\{ -\sum_{i<j} J_{ij}
\mean{\S_i \S_j}_q - S[q]\right\},
\label{eq: variational G}
\end{equation}
where the set $\mathcal{G}$ denotes all distributions $q$ with a given
$\mean{\S_i}_q= m _i$~\cite{AMF}. 
We will refer to the functional $G[q]=\left\{ -\sum_{i<j} J_{ij}
\mean{\S_i \S_j}_q-S[q]\right\}$ 
as the functional Gibbs free energy defined on $\mathcal{G}$. 

Similarly, the variational principle can be applied to the entropy function
$S(\pmb{\chi},\mmm)$, leading once again to a close relationship between
statistical modelling and thermodynamics. The entropy \eqref{eq:entropy_by_legendre}
is found to be
\begin{equation}
S(\pmb{\chi},\mmm)= \max_{q \in \mathcal{S}} \left\{S[q]\right\} \ ,
\end{equation}
where $\mathcal{S}$ denotes distributions with $\mean{\S_i}_q= m_i$
and $\mean{\S_i \S_j}_q= \chi_{ij}$.  This is nothing but the 
maximum entropy principle~\cite{Jaynes1989a}: the variational
principle identifies the distribution with the maximum information
entropy subject to the constraints on magnetisations and spin
correlations, which are set equal to their sample averages (see the
section on maximum entropy modelling above).

\subsubsection{Mean-field theory}
\label{sec:meanfield}

As a first demonstration of the variational principle, we derive the mean-field theory for the inverse Ising problem. 
The starting point is an ansatz for the Boltzmann
distribution~\eqref{eq: static sequential Boltzmann} which factorises
in the sites~\cite{stanley1987introduction,AMF,tanaka2000a}
\begin{equation}
\label{mf_distribution}
p^{\MF} (\sss)= \prod_{i} \frac{1+ \tilde{m}_i \s_i}{2} \ ,
\end{equation}
thus making the different spin variables statistically independent of one another.
The parameters $\tilde{m}_i$ of this ansatz describe the spin
magnetizations; each spin has a magnetisation resulting from the effective magnetic field acting on that
spin.  This effective field arises from its local magnetic field $h_i$, as well as from
couplings with other spin. The mean field giving its name to
mean-field theory is the average over typical configurations of the effective field. 

Using the mean-field ansatz, we now estimate the Gibbs free energy. Within the mean-field ansatz, the minimisation of the variational 
Gibbs potential \eqref{eq: variational G} is trivial: there is only a single mean-field distribution \eqref{mf_distribution} that
satisfies the constraint $\mathcal{G}$ that spins have magnetisations $\mmm$, namely $\tilde{\mmm}=\mmm$. We 
can thus directly write down the mean-field Gibbs free energy 
\begin{eqnarray}
&&G^{\MF}(\mmm,\JJJ) = -\sum_{i<j} J_{ij} m_i m_j  \nonumber \\
&&\quad  +\sum_{i}\left[ \frac{1+m_i}{2} \ln
\frac{1+m_i}{2} + \frac{1-m_i}{2} \ln
\frac{1-m_i}{2} \right]. 
\label{eq: mf G}
\end{eqnarray}
The equation for the couplings $\JJJ$ follows from the second order
derivative of $G(\mmm,\JJJ)$, \textit{cf.} equation~\eqref{eq:d2Gdm2} 
\begin{equation}
(\CCC^{-1})_{ij}= -J_{ij}^{\MF}, (i \ne j) \ .
\label{eq: mean field J}
\end{equation}
Similarly, the reconstruction of the magnetic field follows from the derivative
of $G(\mmm,\JJJ)$ with respect to $m_i$, \textit{cf.} equation~\eqref{eq:dGdm}, 
\begin{equation}
h_i^{\MF} = -\sum_{j\ne i} J_{ij}^{\MF} m_j + \arth m_i \ .
\label{eq: mean field h}
\end{equation} 
This result establishes a simple relationship between the observed
connected correlations and the couplings between spins in terms of the inverse
of the correlation matrix. The matrix inverse in~\eqref{eq: mean field J} is of course much simpler to
compute than the maximum over the likelihood~\eqref{eq: static sequential
	log-likelihood} and takes only a
polynomial number of steps: Gau{\ss}-Jordan elimination for
inverting an $N\times N$ matrix
requires ${\mathcal O}(N^3)$ operations, compared to the exponentially
large number of steps to compute a partition function or its derivatives. 

The standard route to mean-field reconstruction
proceeds somewhat differently, namely via the Helmholtz free energy
rather than the Gibbs free energy. Early work to address the inverse
Ising problem using the mean-field approximation was performed by
Peterson and Anderson~\cite{peterson1987mean}. In~\cite{Kappen1998b}, Kappen
and Rodr\'{\i}guez construct the 
Helmholtz functional free energy $F^{\MF}_{\tilde{\mmm}}(\JJJ,\hhh)$
given by \eqref{eq: variational F}
under the mean-field ansatz~\eqref{mf_distribution}. $F^{\MF}_{\tilde{\mmm}}(\JJJ,\hhh)$ is then minimised with respect to
the parameters of the mean-field ansatz $\tilde{\mmm}$ by
setting its derivatives with respect to $\tilde{\mmm}$ to zero. This yields equations
which determine the values $m_i^{\MF}$ of the magnetization parameters that minimize the
KL divergence between the mean-field ansatz and the Boltzmann
distribution, namely the well-known self-consistent equations
\begin{equation}
\label{eq:c3 mean field mi} 
m_i^{\MF} =  \th(h_i+\sum_{j \ne i} J_{ij} m_j^{\MF}) \ . 
\end{equation}
Using $m^{\MF}_i$ as an approximation for the equilibrium magnetisations $m_i$ one can derive
the so-called linear-response approximation for the connected correlation function
$C^{\MF-\LR}_{ij}\equiv \frac{\partial m^{\MF}_i}{\partial h_j}
(\hhh)$. Taking 
derivatives of the self-consistent 
equations~\eqref{eq:c3 mean field mi} with respect to local fields gives
\begin{equation}
\sum_{j} \left(\frac{\delta_{ij}}{1-m_i^2}-J_{ij} \right)
C^{\MF-\LR}_{jk} = \delta_{ik} \ ,
\label{eq: static sequential MF LR}
\end{equation}
where we have used the fact that diagonal terms of the coupling matrix
are zero. Identifying the result for the connected correlations
$C^{\MF-\LR}_{jk}$ with the sample correlations $C_{jk}$ leads to a
system of linear equations to be solved for the
couplings~\cite{Kappen1998b}. However, to obtain~\eqref{eq: static
	sequential MF LR}, we have used that the diagonal elements $J_{ii}$
are zero. With these constraints, the system of equations~\eqref{eq:
	static sequential MF LR} becomes over-determined and in general
there is no solution. Although different procedures have been
suggested to overcome this problem~\cite{Kappen1998b,Ricci2012a,JaquinRancon2016},
there seems to be no canonical way out of this
dilemma. The most
common approach is to ignore the constraints on the diagonal elements
altogether and invert equation~\eqref{eq: static sequential MF LR} to
get
\begin{equation}
\label{eq:meanfield_reconstruction}
J^{\MF-\LR}_{ij}=\frac{\delta_{ij}}{1-m_i^2}-(\CCC^{-1})_{ij} \ .
\end{equation} 
This result agrees with the reconstruction via the Gibbs free
energy except for the non-zero diagonal couplings, which bear no
physical meaning and are to be ignored. No diagonal couplings arise in the
approach based on the Gibbs free energy since
equation~\eqref{eq:d2Gdm2} with $j=i$ does not involve any
unknown couplings $J_{ij}$.

\subsubsection{The Onsager term and TAP reconstruction}

The variational estimate of the Gibbs free energy \eqref{eq: mf G} can be improved
further. In 1977, Thouless, Anderson, and Palmer (TAP) advocated
adding a term to the Gibbs free energy
\begin{equation}
\label{eq:TAPfreeenergy}
G^{\TAP}(\JJJ,\mmm)= G^{\MF} (\JJJ,\mmm) - \frac{1}{2}\sum_{i<j} J_{ij}^2
(1-m_i^2) (1-m_j^2) \ .
\end{equation} 
This term can be interpreted as describing the effect of fluctuations of a spin variable on the
magnetisation of that spin via their impact on neighbouring
spins~\cite{Thouless1977a}. It is called the Onsager
term, which we will derive in section~\ref{sec:plefka} 
in the context of a systematic expansion around the mean-field ansatz. 
For the forward problem, adding this term modifies the self-consistent
equation~\eqref{eq:c3 mean field mi} to the so-called
TAP equation
\begin{equation}
\label{eq:TAPeq} 
m_i^{\TAP} =  \th\left( h_i+ \sum_{j \ne i} J_{ij} m_j^{\TAP} 
-m_i^{\TAP}  \sum_j J_{ij}^2 (1-(m_j^{\TAP})^2)\right) \ . 
\end{equation}
In the inverse problem, the TAP free energy\eqref{eq:TAPfreeenergy}
gives an the equation for the couplings based on~\eqref{eq:d2Gdm2} 
\begin{equation}
(\CCC^{-1})_{ij}= - J_{ij}^{\TAP} - 2 (J_{ij}^{\TAP})^2 m_i m_j \ .
\end{equation}
Solving this quadratic equation gives the TAP reconstruction~\cite{Kappen1998b,tanaka1998mean}
\begin{equation}
\label{eq:TAPreconstruction}
J_{ij}^{\TAP}= \frac{-2 (\CCC^{-1})_{ij}}{1+\sqrt{1-8(\CCC^{-1})_{ij}m_i
		m_j}} \ ,
\end{equation}
where we have chosen the solution that coincides with the mean-field
reconstruction when the magnetisations are zero.
The magnetic fields can again be found by differentiating the Gibbs free energy
\begin{equation}
\label{eq: TAP h}
h_i= \arth (m_i) - \sum_{j \ne i} J_{ij}^{\TAP} m_j + m_i \sum_{j \ne
	i} (J^{\TAP}_{ij})^2 (1-m_j^2) \ .
\end{equation}

\subsubsection{Couplings without a loop: mapping to the minimum spanning tree
	problem}
\label{sec:mst}

The computational hardness of implementing Boltzmann machine
learning~\eqref{eq:Boltzmannlearning_algo} comes from the difficulty
of computing correlations under the Boltzmann measure, which can
require a computational time that scales exponentially with the system
size.  This scaling originates from the presence of loops in the graph
of couplings between the spins. Graphs for which correlations can be
computed efficiently are the acyclic graphs or trees, so it comes as
no surprise that the first efficient method to solve the inverse Ising
problem was developed for trees already in 1968. This was done by
Chow and Liu~\cite{Chow1968a} in the context of a product
approximation to a multi-variate probability distribution. While the
method itself can be used as a crude approximation for models with
loops or as reference point for more advanced methods, the exact
result by Chow and Liu is of basic interest in itself as it provides a
mapping of the inverse Ising problem for couplings forming a tree onto
a minimum spanning tree (MST) problem.  MST is a core problem in
computational complexity theory, for which there are many efficient
algorithms. This section on Chow and Liu's result also provides some
of the background needed in section~\ref{sec:bethe} on the Bethe
ansatz and section~\ref{sec:BP} on belief propagation.

We consider an Ising model whose pairwise couplings form a tree. The
graph of couplings may consist of several parts that are not connected
to each other (in any number of steps along connected node pairs), or
it may form one single connected tree, but it contains no loops. We
denote the set of nodes (vertices) associated with a spin of the tree $T$ by $V_T$ and the
set of edges (couplings between nodes) by $E_T$. It is straightforward
to show that in this case, the
Boltzmann distribution for the Ising model can be written in a
pairwise factorised form
\begin{align}
\label{eq:P-tree} 
p_{T}(\sss) &=\prod_{i \in V_{T}} p_i (\s_i) \prod_{(i,j) \in
	E_{T}}\frac{p_{ij}
	(\s_i,\s_j)}{p_i(\s_i) p_j(\s_j)} \\
&=\prod_{(ij)\in
	E_{T}}p_{ij}(s_{i},s_{j})\prod_{i \in V_{T}}p_{i}(s_{i})^{1-|\partial i|}\
.\nonumber 
\end{align}
$\partial i$ denotes the set of neighbours of node $i$, so 
$|\partial i|$ is the number of nodes $i$ couples to. 
The distributions $p_i$ and $p_{ij}$ denote the one-point and
two-point marginals of $p_{T}$. 

The KL divergence~\eqref{eq:ml_kl} between the empirical distribution
$p^\D(\sss)$ and $p_{T}$ is given by 
\begin{align}
\label{eq:mst_KL}
\KL(p^\D \vert p_{T})= &
\sum_{\sss}p^\D(\sss)\ln\frac{p^\D(\sss)}{\prod_{i \in V_{T}} p_i (\s_i) \prod_{(i,j) \in
		E_{T}}\frac{p_{ij} (\s_i,\s_j)}{p_i(\s_i) p_j(\s_j)}}  \ . 
\end{align}
For a given tree, it is straightforward to show that the KL divergence
is minimized when the marginals $p_i$ and $p_{ij}$ equal the empirical
marginals $p^\D_i$ and $p^\D_{ij}$. This gives
\begin{align}
\min_{\{p_i, p_{ij}\}}\KL(p^\D \vert p_{T})=  -H +\sum_{i\in V_{T}}H_i -\sum_{(ij)\in E_{T}}I_{ij} \ , \label{eq:KL-tree}
\end{align} 
where $H =-\sum_{\{\sss\}} p^\D (\sss)\ln p^\D (\sss)$
is the entropy of the empirical distribution $p^\D$, $H_i=-\sum_{s_i} p^\D_i(s_{i})\ln p^\D_i (s_{i})$
is the single site entropy and $I_{ij}$ is the mutual information
between a pair of spins
\begin{equation}
I_{ij}=\sum_{s_{i},s_{j}}p^\D_{ij}(s_{i},s_{j})\ln \frac{p^\D_{ij}(s_{i},s_{j})}{p^\D_{i}(s_{i})p^\D_{j}(s_{j})}.\label{eq:mutual}
\end{equation}

Assuming the graph of couplings is an (unknown) tree and that
empirical estimates of all pairs of mutual informations are available, the inverse Ising
problem can then be solved by minimizing the KL divergence~\eqref{eq:mst_KL}
over all possible $N^{N-2}$ trees $T$. 
The first two terms in eq.~(\ref{eq:KL-tree})
do not depend on the choice of the tree, so only the last term needs
to be minimized over, which is a sum over local terms on the
graph. The optimal tree topology $T_\text{opt} $ is thus
given by
\[
E_{T_\text{opt}}=\text{{argmin}}_{\{E_{T}\}}\left(-\sum_{(ij)\in
	E_{T}}I_{ij}\right) \ .
\]
This is where the mapping onto MST problem emerges: $T_\text{opt}$
connects all vertices of the graph and its edges are such that the
total sum of their weights is minimal.  In our case, each edge weight
is the (negative) pairwise mutual information $I_{ij}$ between spin
variables. Finding the MST does not require an infeasible exploration
of the space of all possible trees. On the contrary, it can be found
in a number of steps bounded by $\mathcal{O}(|V|^{2}\ln|V|)$ by greedy
iterative procedures which identify the optimal edges to be added at
each step ($V$ is the set of nodes in the data).  The most famous
algorithms for the MST problem date back to the 1950s and are known
under the names of Prim's algorithm, Kruskal's algorithms and
Boruvka's algorithm (see e.g.~\cite{Moore2011a}).  In practice, one has
to compute the empirical estimates of the mutual information from
samples and then proceed with one of the above algorithms. An
interesting observation which makes the Chow--Liu approach even easier
to apply is that one may use as edge weights also the connected
correlations between spins \cite{Chow1968a}.

Once the optimal tree $T_\text{opt}$ has been identified, we still need to find the
optimal values of the couplings $J_{ij}$ of the Ising model. This
is, however, an easy task: the factorised form of the probability measure
over the tree (\ref{eq:P-tree}) allows one to compute the couplings using
the independent-pair approximation, see subsection~\ref{sec:ace}.

\subsubsection{The Bethe--Peierls ansatz}
\label{sec:bethe}

The factorising probability distribution \eqref{eq:P-tree} can also be
used as an ansatz in situations where the graph of couplings is not a
tree. In this context,
\begin{equation}
p^{\BP}(\{\s_i\})=\prod_{i} p_i (\s_i) \prod_{(i,j) \in E}\frac{p_{ij} (\s_i,\s_j)}{p_i(\s_i) p_j(\s_j)}
\label{eq: static sequential Bethe ansatz}
\end{equation}
is called the Bethe--Peierls ansatz~\cite{bethe1935,peierls1936}.
$(i,j) \in E$ runs over pairs of interacting spins, or equivalently
over edges in the
graph of couplings. One can parameterize the marginal distribution
$p_i$ and $p_{ij}$ using the 
magnetisation parameters $\tilde{m}_i$ and the (connected) correlation parameters $\tilde{C}_{ij}$,
\begin{eqnarray}
p_{i} (\s_{i}) &=& \frac{1+\tilde{m}_i \s_i}{2} \\
p_{ij} (\s_{i},s_{j}) &=& \frac{(1+\tilde{m}_i \s_i)(1+\tilde{m}_j
	\s_j) + \tilde{C}_{ij} \s_i \s_j}{4} \nonumber
\end{eqnarray}
subject to the constraints 
\begin{eqnarray}
&-1 \le \tilde{m}_i \le  1 &  \label{eq: static sequential BP constraint m}\\
&-1+\abs{\tilde{m}_i+\tilde{m}_i} \le \tilde{C}_{ij} +\tilde{m}_i
\tilde{m}_j \le +1 - \abs{\tilde{m}_i -\tilde{m}_j} \ & \label{eq:
	static sequential BP constraint C}\ . \nonumber 
\end{eqnarray}
The Bethe--Peierls ansatz can be
compared to the 
mean-field ansatz~\eqref{mf_distribution}, which assigns a magnetisation (or an
effective field) to each spin. The
Bethe--Peierls ansatz goes one step
further; it assigns to each \emph{coupled pair of spins} correlations
as well as magnetisations. These correlations and magnetisations are
then determined self-consistently. 
An important feature of the Bethe--Peierls ansatz \eqref{eq: static sequential Bethe ansatz} is that its Shannon entropy~\eqref{eq:shannon} can be decomposed into spin pairs
\begin{equation}
S[p^{\BP}]= \sum_{i} S[p_i] +  \sum_{(i,j)} \left( S[p_{ij}] - S[p_i] - S[p_j] \right) .
\end{equation}
For graphs containing loops, the entropy generally contains terms involving
larger sets of spins than pairs, a situation we will discuss in section~\ref{sec:ace}.

The Bethe--Peierls ansatz is well defined, and indeed exact, when the
couplings form a tree. When the graph of couplings contains loops, the
probability distribution~\eqref{eq: static sequential Bethe ansatz} is
not normalized and the ansatz is not well defined. In that case, the
Bethe--Peierls ansatz is an uncontrolled approximation, although
recently progress has been made in the control of the resulting
error~\cite{Chertkov2006a,Parisi2006a}. We start with the assumption
that there are no loops.

To address the inverse Ising problem, we use the Bethe--Peierls
ansatz~\eqref{eq: static sequential Bethe ansatz} as a variational
ansatz to minimise the functional 
Gibbs free energy~\eqref{eq: variational G}. Again the constraint $\mathcal{G}$ in~\eqref{eq: variational G}
implies $\tilde{\mmm}=\mmm$. The remaining minimisation is over the correlation parameters $\tilde{\CCC}$,
\begin{equation}
G^{\BP}(\JJJ,\mmm)=  \min_{\tilde{\CCC}} \left\{ -\sum_{(i,j)} J_{ij}
\mean{\S_i \S_j}_{p^{\BP}}-S[p^{\BP}]\right\} \ 
\label{eq: Bethe G minimisation}
\end{equation}
and yields
\begin{eqnarray}
&&G^{\BP}(\JJJ,\mmm)=  -\sum_{(ij)} J_{ij} (C^{\BP}_{ij}+ m_i m_j) \\
&&\qquad\quad + \sum_{i} (1-z_i) \sum_{\s_i} \frac{1+ m_i \s_i}{2} \ln \frac{1+ m_i \s_i}{2}  \nonumber \\
&&\qquad\quad + \sum_{(i,j)} \sum_{\s_i,\s_j} \frac{(1+ m_i \s_i)(1+ m_j \s_j) + C^{\BP}_{ij} \s_i \s_j}{4}  \nonumber \\ 
&&\qquad\qquad\quad \times \ln \frac{(1+m_i
	\s_i)(1+m_j \s_j) + C^{\BP}_{ij} \s_i \s_j}{4}\nonumber  ,
\label{eq: static sequential Gibbs BP} 
\end{eqnarray}
where $\CCC^{\BP}$ ist the optimal value of $\tilde{\CCC}$ and satisfies
\begin{equation}
J_{ij} =\sum_{\s_i, \s_j} \frac{\s_i \s_j}{4} \ln \frac{(1+ m_i \s_i)(1+ m_j s_j)+C^{\BP}_{ij} \s_i \s_j}{4}.
\label{eq: IP BP}
\end{equation}
$z_i$ denotes the number of neighbours (interaction partners with non-zero couplings) of node $i$. 
From~\eqref{eq:d2Gdm2}, the equation for the couplings can be found again by
equating the second derivative of the Gibbs free energy with the
inverse of the correlation matrix, that is,
\begin{equation}
(\CCC^{-1})_{ij}=
\frac{C^{\BP}_{ij}}{(C^{\BP}_{ij})^2-(1-m_i^2)(1-m_j^2)}\ \ (j \ne i)
\ .
\end{equation}
This quadratic equation can be solved for $C^{\BP}_{ij}$; inserting
the solution 
\begin{equation}
\label{eq:BPcouplingsimplicit}
C^{\BP}_{ij}= \frac{1}{2}\left\{ \frac{1}{(\CCC^{-1})_{ij}}-\sqrt{\frac{1}{(\CCC^{-1})_{ij}^2}-4(1-m_i^2)^2(1-m_j^2)^2}\right\}
\end{equation}
for $(\CCC^{-1})_{ij} \ne 0$ and $C^{\BP}_{ij}= 0$ for
$(\CCC^{-1})_{ij}=0$ into~\eqref{eq: IP BP} gives the couplings of the
Bethe--Peierls reconstruction~\cite{Nguyen2012a,Ricci2012a}.
For the special case $m_i=0$, one obtains a particularly simple result
\begin{equation}
J^{\BP}_{ij}= -\frac{1}{2}\arsh [ 2 (\CCC^{-1})_{ij}],\ (j \ne i).
\end{equation}
In graph theory, this formula can be related to the expression for the
distance in a tree whose links carry weights specified by
pair correlations between spin pairs~\cite{Bapat2005a}. 
Correspondingly, the magnetic fields follow from the first
derivative of the Gibbs free energy as in~\eqref{eq:dGdm} giving
\begin{eqnarray}
\label{eq:BPfields}
&&h^{\BP}_i = (1-z_i) \arth{m_i} - \sum_{j \in \partial i}
J^{\BP}_{ij} m_j \\
&& +\sum_{j \in \partial i}  \sum_{\s_i, \s_j} \frac{\s_i + m_j s_i
	s_j}{4}  \ln \frac{(1+m_i \s_i)(1+m_j \s_j) + C^{\BP}_{ij} \s_i
	\s_j}{4}\nonumber \ .
\end{eqnarray}

In section~\ref{sec:plefka}, we will show that the Bethe--Peierls
ansatz, and hence the resulting reconstruction, is exact for 
couplings forming a tree. However, the reconstruction of couplings and magnetic
fields based on the Bethe--Peierls ansatz can also be applied to cases
where the couplings do not form a tree. Although the results then
arise from an uncontrolled approximation, the quality of the
reconstruction can still be rather good. For a comparison of the different approaches, see
section~\ref{sec:comparison}.

\subsubsection{Belief propagation and susceptibility propagation}
\label{sec:BP}

Belief propagation is a distributed algorithm to compute marginal
distributions of statistical models, 
such as $p_i(s_i)$ and $p_{ij}(s_i,s_j)$ of the preceding sections. Again, it is exact on trees,
but also gives a good approximation when the graph of
couplings is only locally treelike (so any loops present are long). The term belief propagation is used in
the machine learning~\cite{pearl1988probabilistic} and statistical
physics communities, in coding theory the approach is known as the sum-product
algorithm~\cite{gallager1962}. Belief propagation shares a deep
conceptual link with the Bethe--Peierls ansatz; 
Yedidia \textit{et al.} \cite{Yedidia2005a} showed that 
belief propagation is a numerical scheme to efficiently compute the
parameters of Bethe--Peierls ansatz. 

A detailed exposition and many applications of belief propagation can be found in the
textbook by M\'ezard and Montanari~\cite{mezard2009information}. Here,
we briefly introduce the basics and discuss applications to the inverse Ising
problem. We start by considering the ferromagnetic Ising model in 1D, that is, a
linear chain of $N$ spins. The textbook solution of this problem considers the partitions function of the
system when the last spin is constrained to take on the values
$s_N=\pm1$, $Z_N(+1)$ and $Z_N(-1)$. The corresponding partition
functions for the linear chain with $N+1$ spins are linked to the
former via the so-called transfer matrix~\cite{baxter1982exactly}.
The partition function for a system of any size can be computed iteratively starting from a
single spin and extending the 1D lattice with each multiplication of
the transfer matrix. In fact the transfer matrix can also be used to
solve the Ising model on a tree~\cite{Eggarter1974a}.
Belief propagation is similar in spirit, but can be extended (as an
approximation) also to graphs which are not trees. The best
book-keeping device is again a restricted partition function, namely
$Z_{i \to j}(s_i)$. It is defined as the partition function of the part of the system containing $i$ when the coupling
present between spins $i$ and $j$ has been deleted from the tree and
spin $i$ is constrained to take on $s_i$. 
(Deleting the edge
$(i,j)$ splits the tree containing $i$ and $j$ into two disconnected
parts.) 
On a tree we obtain the recursion relation relation
\begin{equation}
\label{eq:cavity recursion}
Z_{i \to j}(s_i) = e^{h_i s_i} \prod_{k \in \partial i \setminus j}
\left(\sum_{s_k} Z_{k \to i}(s_k) e^{J_{ik} s_i s_k}\right) \ ,
\end{equation}
which can be computed recursively starting from leaves of the tree
(nodes connected to a single edge only). In statistical physics,
\eqref{eq:cavity recursion} is called the cavity recursion for
partition functions, since deleting a link can be thought of as
leaving a `cavity' in the original system. 
The partition function for the entire tree with spin $i$ constrained
to $s_i$ is then
\begin{equation}
Z_i(s_i) =  e^{h_i s_i} \prod_{k \in \partial i}  \left(\sum_{s_k} Z_{k \to
	i}(s_k) e^{J_{ik} s_i s_k}\right) \ .
\end{equation}
The marginal distribution $p_i(s_i)$ can be calculated by normalizing $p_i(s_i) \propto Z_i (s_i)$ and the marginal distribution $p_{ij}(s_i,s_j)$ can be calculated by normalizing  $p_{ij} (s_i, s_j) \propto Z_{i \to j} (s_i) e^{J_{ij} s_i s_j} Z_{j \to i} (s_j)$.

The power of the cavity recursion~\eqref{eq:cavity recursion_probs} lies in its application to graphs
which are not trees. It is particularly effective when the graph of
couplings is at least locally treelike, so any loops present are
long. To extend the cavity recursion as an approximation to graphs which are not
trees, it makes sense to consider normalized quantities and define
$\pi_{i \to j}(s_i) = Z_{i \to j}(s_i)/\sum_s Z_{i \to j}(s)$ with the
recursion
\begin{equation}
\label{eq:cavity recursion_probs}
\pi_{i \to j}(s_i) \propto e^{h_i s_i} \prod_{k \in \partial i}
\left(\sum_{s_k} \pi_{k \to i}(s_k) e^{J_{ik} s_i s_k}\right)
\end{equation}
leading to the estimate of the two-point marginals
$\pi_{ij}(s_i,s_j) \propto e^{J_{ij} s_i s_j} \pi_{i \to j} (s_i) \pi_{j \to
	i} (s_j)$. (This step is necessary, as deleting the link $(i,j)$ on a tree
leads to two disjoint subtrees with separate partition functions and
zero connected correlations. On a locally treelike graph,
correlations between $s_i$ and $s_j$ can still be small once the link
$(i,j)$ has been cut, but the tree does not split into disjoint parts
with separate partition functions.) Associating each edge $(i,j)$ in the graph with particular
values of $\pi_{i \to j}(s_i)$ and $\pi_{j \to i}(s_j)$, one can update
these values in an iterative scheme which replaces them with the right-hand side of \eqref{eq:cavity recursion_probs} at each step. The fixed point
of this procedure obeys \eqref{eq:cavity recursion_probs}. In this context,
$\pi_{i \to j}(s_i)$ is termed a `message' that is being `passed'
between nodes. Belief propagation is an example of a message passing
algorithm.

Belief propagation has been used to solve the inverse Ising problem in
two different ways. In the first, belief propagation is used to
approximately calculate the magnetisations and correlations given some
parameters of the Ising model, and then fields and couplings are
updated according to the Boltzmann learning
rule~\eqref{eq:Boltzmannlearning_algo}. To estimate the correlations, 
one has to define additional messages also for the susceptibilities 
of each spin together with their update rules. This approach, termed
susceptibility propagation, was developed by Welling and
Teh~\cite{Welling2003a,Welling2004a} for the forward problem.
Susceptibility propagation thus solves the `forward' problem, that is, it offers a
computationally efficient approximation to the correlations and
magnetisations, which are then used for Boltzmann machine learning. 

A sophisticated variant of susceptibility propagation was
developed by M\'ezard and Mora~\cite{Mezard2009a, Aurell2010a}
specifically for the inverse Ising problem, where also the couplings are updated at each
step. This approach gives the same reconstruction as the Bethe--Peierls
reconstruction from section~\ref{sec:bethe}. However, the iterative
equations can fail to converge, even when the analytical
approach~\eqref{eq:BPcouplingsimplicit}--\eqref{eq:BPfields} gives
valid approximate solutions~\cite{Nguyen2012a}.

%

\subsubsection{The independent-pair approximation and the Cocco--Monasson adaptive-cluster expansion}
\label{sec:ace}

We expect the Bethe--Peierls ansatz to work well when the couplings
are locally treelike and loops are long. Conversely, we expect the
ansatz to break down when the couplings generate many short
loops. Cocco and Monasson developed an iterative procedure to identify
clusters of spins whose couplings form short loops and evaluate their
contribution to the
entropy~\eqref{eq:entropy_by_legendre}~\cite{Cocco2011a,Cocco2012a}.
In the context of disordered systems, the expansion of the entropy in
terms of clusters of connected spins is known as Kikuchi's cluster
variational method~\cite{kikuchi1951,yedidia2001bethe}.

The Cocco--Monasson
adaptive-cluster expansion directly approximates the entropy
potential~\eqref{eq:entropy_by_legendre}. We start by considering the statistics of a single spin variable
described by its magnetisation $m_i$, 
with the entropy
\begin{equation}
S_{(1)} (m_i) = \sum_{s_i} \frac{1+m_i s_i}{2} \ln \frac{1+m_i s_i}{2}
\ .
\end{equation} 
The simplest entropy involving coupled spins is the two-spin
entropy
\begin{eqnarray}
&&S_{(2)} (m_i,m_j,\chi_{ij})= \sum_{s_i, s_j} \frac{1+m_i s_i + m_j s_j + \chi_{ij} s_i s_j}{4} \times \nonumber \\ 
&& \qquad\qquad \ln \frac{1+m_i s_i + m_j s_j + \chi_{ij} s_i s_j}{4} \ .
\end{eqnarray}
When the two spins are statistically independent, we obtain $S_{(2)}
(m_i,m_j,\chi_{ij})=S_{(1)}(m_i) + S_{(2)} (m_j)$. Hence the residual
entropy which accounts for the correlation between the two spins is
$\Delta S_{2} (m_i,m_j,\chi_{ij})\equiv S_{(2)} (m_i,m_j,\chi_{ij}) -
S_{(1)}(m_i) - S_{(1)} (m_j)$. To make the notation uniform, we also
define $\Delta S_{(1)} (m_i)\equiv S_{(1)} (m_i)$. A very simple
approximation is based on the assumption that the $N$-spin entropy
~\eqref{eq:entropy_by_legendre} is described by pairwise terms
\begin{equation}
S(\pmb{\chi},\mmm) \approx \sum_{i} \Delta S_{(1)} (m_i) + \sum_{(i,j)}\Delta S_{(2)} (m_i,m_j,\chi_{ij}),
\label{eq:independent_pair_entropy}
\end{equation}
where the pair $(i,j)$ denotes a cluster of two distinct spins. The
couplings and fields can then be obtained 
via differentiation as in~\eqref{eq:entropy_reconstruction},
\begin{eqnarray}
\label{eq:independent_pair_reconstruction}
J_{ij} &=& \sum_{\s_i, \s_j} \frac{\s_i \s_j}{4} \ln \frac{1+ m_i \s_i+ m_j s_j+\chi_{ij} \s_i \s_j}{4} \ , \\
h_i &=& \frac{2-N}{2} \ln \frac{1+m_i}{1-m_i} + \nonumber \\ && \sum_{j \ne i} \sum_{s_i,s_j}\frac{s_i}{4}\ln \frac{1+m_i s_i + m_j s_j + \chi_{ij} s_i s_j}{4} \ . \nonumber 
\end{eqnarray}
This result
is called the
independent-pair approximation for the
couplings, see~\cite{roudi2009isingquality}
and~\cite{Roudi2009b}. (Expressions (8a) and (8b) in~\cite{Roudi2009b}
differ slightly from~\eqref{eq:independent_pair_reconstruction} due to a typo.) When the
topology of couplings is known and forms a tree, 
equation~\eqref{eq:independent_pair_entropy} gives the exact
entropy of the system when the second sum is restricted to pairs of
interacting spins (see sections~\ref{sec:mst} and~\ref{sec:bethe}). In this
case,~\eqref{eq:independent_pair_reconstruction} gives the exact
couplings.

However, in most cases the topology is not known, so second sum
in~\eqref{eq:independent_pair_entropy} runs over all pairs of
spins. In this case, equation~\eqref{eq:independent_pair_entropy} is
only a (rather bad) approximation to the
entropy. However, the independent-pair approximation~\eqref{eq:independent_pair_entropy} can serve as
starting point for an expansion that includes clusters of spins of 
increasing size
\begin{eqnarray}
S(\pmb{\chi},\mmm) = && \sum_{i} \Delta S_{(1)} (m_i) + \\
&&\sum_{(i,j)} \Delta S_{(2)} (\chi_{ij},m_i,m_j) + \nonumber  \\
&& \sum_{(i,j,k)} \Delta S_{(3)}
(\chi_{ij},\chi_{jk},\chi_{kl},m_i,m_j,m_k) + \cdots \ .\nonumber
\end{eqnarray}
In this expansion, the contribution from clusters consisting of $3$
spins is 
\begin{eqnarray}
&&\Delta S_{(3)}= S_{(3)}(\chi_{ij},\chi_{jk},\chi_{kl},m_i,m_j,m_k) - \\
&& \qquad S_{(2)} (\chi_{ij},m_i,m_j)- S_{(2)} (\chi_{jk},m_j,m_k) - \nonumber \\
&& \qquad S_{(2)} (\chi_{ki},m_k,m_i) + S_{(1)} (m_i) + \nonumber \\
&& \qquad S_{(1)} (m_j) + S_{(1)} (m_k)  \nonumber \ ,
\end{eqnarray}
where we have dropped the argument of $\Delta S_{(3)}$ to simplify the notation. The residual entropies of higher clusters are defined analogously.  

The evaluation of $S_{(3)}$ is not as straightforward as $S_{(2)}$ and
$S_{(1)}$, but still tractable. In general, the evaluation of
$\Delta S_{(k)}$ requires the computation of order $2^k$ steps and therefore
becomes quickly intractable. Cocco and Monasson argue that the
contribution of $\Delta S_{k}$ decreases with $k$
and can be neglected from a certain order on~\cite{Cocco2011a}. This inspires an
adaptive procedure to build up the library of all clusters which
contribute significantly to the total entropy of the system. Starting
from $1$-clusters, one constructs all $2$-clusters 
by merging pairs of $1$-clusters. If the entropy contribution of the new
cluster is larger than some threshold, the new cluster is added to a
list of clusters. One then constructs the $3$-clusters,
$4$-clusters in the same way until no new cluster gives a contribution
to the entropy exceeding the threshold. The total entropy and the reconstructed
couplings and fields are updated each time a new cluster is added. 
This procedure fares well
in situations when there are many short loops, like in a regular lattice
in more than one dimension or when the graph of couplings is highly
heterogeneous and some subsets of spins are strongly connected among
each other.

The approaches introduced so far are each built on an ansatz for
the Boltzmann distribution such as the mean-field
ansatz~\eqref{mf_distribution}
or the Bethe--Peierls ansatz~\eqref{eq: static sequential Bethe
	ansatz}: They are all uncontrolled approximations. In the next sections, we discuss controlled approximations based
on either an expansion in small couplings or small
correlations.

\subsubsection{The Plefka expansion}
\label{sec:plefka}

In 1982, Plefka gave a systematic expansion of the Gibbs free energy of the
Sherrington--Kirkpatrick model~\cite{sherringtonkirkpatrick1975} in the couplings between
spins~\cite{Plefka1982a}. Already 8 years earlier, Bogolyubov Jr. \textit{et
	al.} had used similar ideas in the context of the ferromagnetic Ising
model on a lattice~\cite{bogolyubov1976high}. The resulting estimate
of the partition function can also be used
to derive new solutions of the inverse Ising problem.  Zeroth- and
first-order terms of Plefka's expansion turn out to yield mean-field
theory~\eqref{eq: mf G}, the second-order term gives the
TAP free energy~\eqref{eq:TAPfreeenergy}, and correspondingly the
reconstructions of couplings~\eqref{eq: mean field J}
and~\eqref{eq:TAPreconstruction}.

Plefka's expansion is a Legendre transformation 
of the cumulant expansion of 
the Helmholtz free energy $F(\JJJ,\hhh)$. Plefka introduced a perturbation parameter $\lambda$ to the interacting part of the Hamiltonian
\begin{equation}
H(\sss) = H_0 (\sss) + \lambda V (\sss),
\end{equation} 
where $H_0 (\sss) = -\sum_{i} h_i \s_i$ and $V (\sss)= -\sum_{i<j}
J_{ij} \s_i \s_j$. The perturbation parameter $\lambda$ serves to
distinguish the different orders in the strength of couplings and will
be set to one at the end of the calculation. 
The standard cumulant expansion of the Helmholtz free energy $F^{\lambda} (\JJJ,\hhh)$ then reads
\begin{equation}
F^{\lambda}(\JJJ,\hhh)= F^{(0)} (\hhh) - \lambda \mean{V}_0 + \frac{\lambda^2}{2} \left[ \mean{V^2}_0 - \mean{V}^2_0 \right] + \cdots \ ,
\label{eq: cumulant expansion}
\end{equation}
where $F^{(0)} (\hhh)=\sum_{i} 2 \ch h_i$ and $\mean{}_0$ denotes the
average with respect to the Boltzmann distribution corresponding to
the non-interacting part $H_0$ of the Hamiltonian.
Next we perform the Legendre transformation of $F^{\lambda}(\JJJ,\hhh)$ with respect to $\hhh$
to obtain the perturbative series for the Gibbs free energy
\begin{equation}
G^{\lambda}= G^{(0)} + \lambda G^{(1)} + \lambda^2 G^{(2)} + \cdots,
\end{equation}
where we suppressed the dependence of $G$ on $\JJJ$ and $\mmm$ to simplify the notation.
Using the definition of the Gibbs free energy~\eqref{eq:gibbs_free}
we solve
\begin{equation}
G^{\lambda} (\JJJ,\mmm) = \sum_{i} h_i m_i +F^{\lambda}(\JJJ,\hhh) 
\end{equation}
with the local fields satisfying 
\begin{equation}
m_i = -\frac{\partial F^{\lambda}(\JJJ,\hhh)}{\partial h_i}.
\label{eq: Plefka m_i}
\end{equation}
Plugging the perturbative series 
\begin{equation}
\hhh^{\lambda}= \hhh^{(0)} + \lambda \hhh^{(1)} + \lambda^2 \hhh^{(2)} + \cdots
\end{equation}
into \eqref{eq: Plefka m_i} and re-expanding the right-hand side in
$\lambda$, one finds successive orders of $\hhh$. 
In particular, the lowest two orders are found easily as
\begin{eqnarray}
h^{(0)} &=& \arth m_i \\ 
h^{(1)} &=& -\sum_{j} J_{ij} m_j. \nonumber
\end{eqnarray}
This gives the Gibbs free energy up to first order in $\lambda$,
\begin{eqnarray}
\label{eq:plefka_gibbs_zero_one}
G^{(0)}(\mmm) &=& \sum_{i} \left\{\frac{1+m_i}{2} \ln \frac{1+m_i}{2} + \frac{1-m_i}{2} \ln \frac{1-m_i}{2}\right\} \nonumber \\
G^{(1)}(\mmm) &=& -\sum_{i<j} J_{ij} m_i m_j. 
\end{eqnarray}
Continuing to the second order, one obtains the Onsager term~\eqref{eq:TAPfreeenergy}
\begin{equation}
\label{eq:plefka_gibbs_zero_two}
G^{(2)}(\mmm) = -\frac{1}{2} \sum_{i<j} J_{ij}^2 (1-m_i^2) (1-m_j^2).
\end{equation}
A systematic way to perform this expansion has been developed by 
Georges and Yedidia~\cite{Georges1991a} leading to
\begin{eqnarray}
\label{eq:plefka_gibbs_three_four}
&&G^{(3)}(\mmm)= -\frac{2}{3} \sum_{(i,j)} J_{ij}^3 m_i (1- m_i^2) m_j (1-m_j^2) \\
&&\quad - \sum_{(i,j,k)} J_{ij} J_{jk} J_{ki} (1- m_i^2) (1-m_j^2)(1- m_k^2) \nonumber \\
&&G^{(4)}(\mmm)= \frac{1}{12} \sum_{(i,j)} J^4_{ij} (1- m_i^2)(1-m_j^2) \nonumber \\
&&\quad \times (1+3 m_i^2+3 m_j^2-15 m_i^2 m_j^2) \nonumber \\
&&-\sum_{(i,j,k,l)} J_{ij} J_{jk} J_{kl} J_{li} (1-m_i^2)(1- m_j^2)(1- m_k^2)(1- m_l^2) \nonumber \\
&&-2 \sum_{(i,j,k)} J_{ij}^2 J_{jk} J_{ki} m_i
(1-m_i^2) m_j (1-m_j^2) (1- m_k^2). \nonumber
\end{eqnarray}
The mean-field and TAP
reconstructions of couplings~\eqref{eq:meanfield_reconstruction}
and~\eqref{eq:TAPreconstruction} then follow from~\eqref{eq:d2Gdm2} and 
the first derivatives of $G$ with respect to the magnetizations gives
the corresponding reconstructions of the magnetic fields~\eqref{eq: mean field h} and~\eqref{eq: TAP h}. 
Higher order terms of the Plefka expansion are discussed by Georges
and Yedidia~\cite{Georges1991a,Yedidia2001a}, but have not yet been
applied to the inverse Ising problem. For a
fully-connected ferromagnetic model with all couplings set to $J/N$ (to
ensure an extensive Hamiltonian) one finds that already the
second-order term $G^{(2)}$ vanishes relative to the first in the
thermodynamic limit. For the Sherrington-Kirkpatrick model
of a spin glass~\cite{sherringtonkirkpatrick1975}, couplings are of the
order of $N^{-1/2}$, again to make the energy extensive. In that case, $G^{(0)}$, $G^{(1)}$,
and $G^{(2)}$ turn out to be extensive, but the third order vanishes
in the thermodynamic limit. In specific instances (not independently
and identically distributed couplings), higher order terms of the Plefka
expansion may turn out to be important. 

The terms in $J_{ij}$, $J_{ij}^2$, $J_{ij}^3$, \textit{etc.} appearing
in~\eqref{eq:plefka_gibbs_zero_one}-\eqref{eq:plefka_gibbs_three_four}
resum to the results of the Bethe--Peierls
ansatz~\cite{Georges1991a}. If the couplings form a tree, these
terms are the only ones contributing to the Gibbs free energy; terms
such as $J_{ij}J_{jk}J_{ki}$ in~\eqref{eq:plefka_gibbs_three_four}
quantify the coupling strengths around a closed loop of spins and are zero 
if couplings form a tree. This finally shows that the Bethe--Peierls
ansatz is exact on a tree. However, when the couplings contain loops, these terms contribute to the
Gibbs free energy. Beyond evaluating
the Plefka expansion at different orders, one can also resum the
contributions from specific types of loops to infinite order. We will
discuss such an approach in the next section.

\subsubsection{The Sessak--Monasson small-correlation expansion}

The Sessak--Monasson expansion is a perturbative expansion of the
entropy $S(\pmb{\chi},\mmm)$ in terms of the \emph{connected
	correlation} $C_{ij}\equiv\chi_{ij}-m_i m_j$~\cite{Sessak2009a}. For zero
correlations $C_{ij}$, the couplings $J_{ij}$ should also be
zero. This motivates an expansion of the free energy
$F(\JJJ,\hhh)$ in terms of the connected correlations around a
non-interacting system; the Legendre
transformation~\eqref{eq:entropy_by_legendre} then yields
the perturbative series of the entropy function
$S(\pmb{\chi},\mmm)$. Equivalently, this is an expansion of 
equations~\eqref{eq: static
	sequential matching m and C} for the fields and the couplings in the connected correlation $\CCC$.
To make the perturbation explicit, we replace $\CCC$ by $\lambda
\CCC$, where the perturbation parameter $\lambda$ is to be set to
one at the end. The couplings and fields are the solutions of
\begin{eqnarray}
\label{eq:sm1}
m_i &=& -\frac{\partial F (\JJJ,\hhh)}{\partial h_i}  \\
\lambda C_{ij}+m_i m_j &=& -\frac{\partial F (\JJJ,\hhh)}{\partial J_{ij}}, \nonumber
\end{eqnarray}
where the latter can be replaced by 
\begin{equation}
\label{eq:sm2}
\lambda C_{ij}= -\frac{\partial^2 F(\JJJ,\hhh)}{\partial h_i \partial h_j}  \ .
\end{equation}
We then expand the solution $\hhh$ and $\JJJ$ in a power series around the uncorrelated
case $\lambda=0$
\begin{eqnarray}
\hhh^\lambda &=& \hhh^{(0)} + \lambda \hhh^{(1)} + \frac{\lambda^2}{2} \hhh^{(2)} + \cdots \\
\JJJ^\lambda &=& \JJJ^{(0)} + \lambda \JJJ^{(1)} + \frac{\lambda^2}{2} \JJJ^{(2)} + \cdots \nonumber \ .
\end{eqnarray}
At the zeroth-order in the connected correlations, spins are uncorrelated, so $\JJJ^{(0)}=0$ and $h^{(0)}= \arth (m_i)$.
The expansion of model parameters induces an expansion 
of the Hamiltonian 
\begin{eqnarray}
H^{\lambda}= H^{(0)}+ \lambda H^{(1)} + \frac{\lambda^2}{2} H^{(2)} + \cdots \ ,
\end{eqnarray}
where 
\begin{equation}
H^{(k)} (\sss)= -\sum_{i<j} J^{(k)}_{ij} \s_i \s_j - \sum_{i} h_i^{(k)} \s_i.
\end{equation}

Likewise, the free energy $F(\JJJ,\hhh)=-\ln Z(\JJJ,\hhh)$ (or the cumulant generator) can be expanded in $\lambda$,
\begin{eqnarray}
&&F^{\lambda} = F^{(0)} + \lambda F^{(1)} + \lambda^2 F^{(2)} + \cdots \\
&&\quad = F^{(0)} - \lambda \mean{H^{(1)}}_0 \nonumber  \\
&& \quad + \frac{\lambda^2}{2} \left[-\mean{H^{(2)}}_0 +
\mean{(H^{(1)})^2}_0 - \mean{H^{(1)}}_0^2 \right] + \cdots \ , \nonumber
\end{eqnarray}
where the subscript $\mean{}_0$ refers to the thermal average under the zeroth-order Hamiltonian $H^{(0)} (\sss)$. For example, the first-order term is
\begin{eqnarray}
\label{eq:sessakmonasson_Ffirstorder}
F^{(1)} &=& - \mean{H^{(1)}}_0 \\
&=& \sum_{i<j} J_{ij}^{(1)} \th (h^0_i) \th (h^0_j) + \sum_{i}
h_i^{(1)} \th (h^0_i) \nonumber \ .
\end{eqnarray}

Equations~\eqref{eq:sm1} and~\eqref{eq:sm2} then read 
\begin{eqnarray}
m_i &=& -\frac{\partial F^{\lambda}}{\partial h^{(0)}_i}=-\frac{\partial F^{(0)}}{\partial h^{(0)}_i}
-\lambda \frac{\partial F^{(1)}}{\partial h^{(0)}_i} 
-\lambda^2 \frac{\partial F^{(2)}}{\partial h^{(0)}_i}
-\ldots  \\
\lambda C_{ij} &=& -\frac{\partial^2 F^{\lambda}}{\partial h^{(0)}_{i}\partial h^{(0)}_{j}}=
-\frac{\partial^2 F^{(0)}}{\partial h^{(0)}_{i}\partial h^{(0)}_{j}}
-\lambda \frac{\partial^2 F^{(1)}}{\partial h^{(0)}_{i}\partial h^{(0)}_{j}}
-\ldots \nonumber \ .
\end{eqnarray}
Evaluating these equations successively gives solutions for the
different orders of $h_i^{(k)}$ and $J^{(k)}_{ij}$, which in turn
yield expressions for reconstructed fields and couplings when the
magnetisations $\mmm$ and the connected 
correlations $\CCC$ are identified with their empirical values. 
Recalling the zeroth-order solution $h^{(0)}_i = \arth (m_i)$, 
the first-order contribution of the free energy \eqref{eq:sessakmonasson_Ffirstorder} leads to
\begin{eqnarray}
J^{(1)}_{ij} &=& \frac{C_{ij}}{[1-(m_i)^2][1-(m_j)^2]} \\
h^{(1)}_i &=& -\sum_{j \ne i} J_{ij}^{(1)} m_j \ .   \nonumber
\end{eqnarray}
Higher-order terms are given in~\cite{Sessak2009a}. 
Also in~\cite{Sessak2009a}, Sessak and Monasson give a diagrammatic framework suitable for the derivation of
higher order terms in the couplings and sum specific terms to infinite order. Roudi \textit{et
	al.}~\cite{roudi2009isingquality} simplified the results yielding
the reconstruction
\begin{equation}
J_{ij}= -(\CCC^{-1})_{ij} + J_{ij}^{\mathrm{IP}}  -
\frac{C_{ij}}{(1-m_i^2)(1-m_j^2)-(C_{ij})^2} \
\end{equation}
for couplings with $i \ne j$, where $J^{\mathrm{IP}}_{ij}$ is the
independent pair
approximation~\eqref{eq:independent_pair_reconstruction}. This result
can be interpreted as follows~\cite{roudi2009isingquality}: 
The first term is the mean-field reconstruction, which is of a
sub-series of the Sessak--Monasson expansion. One then adds the
sub-series that constitutes the independent pair
approximation~\eqref{eq:independent_pair_reconstruction}. The last
term is the overlap of the two series: the mean-field reconstruction
of two spins considered independently, which needs to be
subtracted.
The resummation of other sub-series for special cases is also
possible~\cite{Sessak2009a}; in~\cite{JaquinRancon2016} loop diagrams
are resummed to obtain a reconstruction that is particularly robust
against sampling noise.



\subsection{Logistic regression and pseudolikelihood}
\label{sec:pseudo-likelihood}

Pseudolikelihood is an alternative to the likelihood
function~\eqref{eq: static sequential log-likelihood} and leads to
the exact inference of model parameters in the limit of an infinite
number of
samples~\cite{arnoldstrauss1991a,hyvarinen2006consistency,mozeika2014}. 
The computational complexity of this approach scales 
polynomially with the number of spin variables $N$, but also with the
number of samples $M$. 
In practice, this is usually much faster than exact maximisation of the
likelihood function, 
whose computational complexity is exponential in the system size. 
Pseudolikelihood inference was developed by Julian Besag in 1974
in the context of statistical inference of data with
spatial dependencies~\cite{Besag1974a}. It is closely related to logistic regression. 
While pseudolikelihood and regression have been used widely in statistical
inference~\cite{besag1986,kalbfleisch1986,Ravikumar2010a},
this approach was not well-known in the physics community until quite
recently~\cite{Aurell2012b,mozeika2014,ekeberg2013,Decelle2014a}. 

Our derivation focuses on the physical character of regression
analysis, which we then link to Besag's pseudolikelihood. Key observation is that although the likelihood
function~\eqref{eq: static sequential log-likelihood} depends on the model parameters in a complicated way, one can
simplify this dependence
by separating different groups of parameters. 
Let us consider how the statistics of a particular spin variable $\S_i$ depends on the
configuration of all other spins. We split the Hamiltonian
into two parts
\begin{eqnarray}
\label{eq:pseudolikelihood_splithamiltonian}
H(\sss)&=&H_{i}(\sss)+H_{\setminus i}(\sss\setminus\s_i) \\
&=& - h_i \s_i - \s_i \sum_{j \ne i} J_{ij} \s_j + H_{\setminus
	i}(\sss\setminus\s_i) \ , \nonumber
\end{eqnarray}
such that the first part $H_{i}(\sss)$ depends on the magnetic field $h_i$ and the couplings of spin $i$ to other spins $J_{ij}$,
while the part $H_{\setminus i}(\sss\setminus\s_i)$ does
not. $\sss\setminus\s_i$ denotes all spin variables except spin $s_i$.
This splitting of variables is possible since the statistics of $\S_i$ conditioned on the remaining spins
$\{\s_j\}_{j \ne i}$ is fully captured by $h_i$ and $J_{ij}, (j \in 1,\ldots,N)$. 

The expectation values of $\S_i$ can be
computed based on the partition function
\begin{equation}
Z(\JJJ,\hhh)= \sum_{\sss \setminus \s_i} 2 \ch \left( h_i + \sum_{j
	\ne i } J_{ij} \s_j\right) \e^ {-H_{\setminus i}(\sss \setminus
	\s_i)} \ ,
\label{eq: static sequential partial partition function}
\end{equation}
where only spin $i$ has been summed over.  
Differentiating the partition function in this form with respect to
$h_i$ and $J_{ij}$ yields the expectation values
\begin{eqnarray}
\mean{\S_i} &=& \mean{\th \left( h_i + \sum_{k \ne i} J_{ik} \S_k \right)}
\label{eq: static sequential Callen m} \\
\mean{\S_i \S_j} &=& \mean{\S_j \th \left( h_i + \sum_{k \ne i} J_{ik}
	\S_k \right)} \ , 
\label{eq: static sequential Callen C}
\end{eqnarray}
where on both sides the thermal average is over the Boltzmann measure
$e^{-H(\sss)}/Z$. The first equation follows from 
\begin{align}
\mean{\S_i} &=
\frac{1}{Z} \sum_{\sss \setminus\s_i} e^{-H_{\setminus i}} \sum_{s_i} s_i
e^{-H_i}
=\frac{1}{Z} \sum_{\sss \setminus\s_i} e^{-H_{\setminus i}} 2 \sinh
\Theta_i \nonumber\\
&=\frac{1}{Z} \sum_{\sss \setminus\s_i} e^{-H_{\setminus i}} \tanh \Theta_i
2 \cosh \Theta_i \nonumber \\
&=\frac{1}{Z} \sum_{\sss \setminus\s_i} e^{-H_{\setminus i}} \tanh \Theta_i
\sum_{s_i} e^{-H_i}
=\mean{\tanh \Theta_i} 
\end{align}
with the shorthand $\Theta_i = h_i + \sum_{k \ne i} J_{ik} \S_k$. 
The second equation follows analogously.

In statistical physics, equations~\eqref{eq: static sequential Callen
  m} and~\eqref{eq: static sequential Callen C} 
are known as Callen's
identities~\cite{Callen1963a} and have been used to compute coupling parameters from
observables in
Monte Carlo simulations for the numerical calculation of
renormalisation-group trajectories~\cite{Swendsen1984a}.
While these equations are exact, the expectation values on the right
hand sides still contain the average over the $N-1$ spins other than $i$. 

The crucial step and the only approximation involved is to replace the
remaining 
averages in~\eqref{eq: static sequential Callen C} 
with the sample means
\begin{eqnarray}
\mean{\S_i}^\D &=& \mean{\th \left( h^{\PL}_i + \sum_{k \ne i} J^{\PL}_{ik} \S_k \right)}^\D \label{eq: static sequential pseudo-likelihood m} \\
\mean{\S_i \S_j}^\D &=& \mean{\S_j \th \left( h^{\PL}_i + \sum_{k \ne
		i} J^{\PL}_{ik} \S_k \right)}^\D \ . \nonumber
\end{eqnarray}
We now have a system of non-linear equations to be solved for
$h_i$ and $J_{ij}$ for fixed $i$ and various $j$, $j \ne i$.  Standard methods
to solve such equations are Newton--Raphson or conjugated gradient
descent~\cite{Press2002a,Hastie2009a}.

With these steps, we have broken down the problem of estimating the magnetic fields
and the full coupling matrix to $N$ separate problems of
estimating one magnetic field and a single row of the coupling
matrix for a specific spin $i$. Crucially, the
Boltzmann average over $2^{N-1}$ states has been replaced with an average
over all configurations of the samples. As a result, the
computation of~\eqref{eq: static sequential pseudo-likelihood m} 
uses not only the sample magnetizations and correlations (sufficient statistics), but all spin
configurations that have been sampled.
The time to evaluate~\eqref{eq: static sequential pseudo-likelihood m}  is thus linear in the
number of samples $M$. In general, the coupling matrix
inferred in this way will be asymmetric, $J_{ij}
\ne J_{ji}$ due to sampling noise. A practical solution is to use the average $\frac{1}{2}(J_{ij}+ J_{ji})$
as estimate of the (symmetric) coupling matrix. 

The set of equations~\eqref{eq: static sequential pseudo-likelihood m}
can be viewed as solving a gradient-descent problem of a logistic regression~\cite{Ravikumar2010a}.  The statistics of $\S_i$ conditioned on the values of the 
remaining spins $\{\s_j\}_{j\ne i}$ can be written as 
\begin{align}
\label{eq:conditional_statistics_single_spin}
p (s_i \vert \{s_j\}_{j \ne i}) & = \frac{1}{2}\left[ 1+ \s_i \th \left(
h_i + \sum_{j\ne i} J_{ij} \s_j \right) \right] \\ 
&= \frac{1}{1+ e^{- 2 s_i(h_i + \sum_{j \ne i} J_{ij} s_j)}}. \nonumber
\end{align}
From this expression for the conditional probability, the $i$-th row
of couplings $J_{i\ast}$ and the field $h_i$ are simply the
coefficients and the intercept in the multi-variate logistic regression
of the response variable $\S_i$ on the 
variables $\{\s_j\}_{j\ne i}$. The log-likelihood for the $i$-th row
of couplings $J_{i\ast}$ and the magnetic field $h_i$ of this regression problem is 
\begin{align}
\label{eq:static sequential pseudo-likelihood} 
L^{i}_{\D} (J_{i\ast},h_i) &=  \frac{1}{M}\sum_{\mu} \ln p (s_i^{\mu} \vert \{s_j^{\mu}\}_{j \ne i})\\
&=\frac{1}{M}\sum_{\mu} \ln
\frac{1}{2}\left[ 1+ \s^{\mu}_i \th \left( h_i + \sum_{j\ne i} J_{ij}
\s^{\mu}_j \right) \right] \ . \nonumber
\end{align}
Setting the derivatives of this likelihood function with respect to the magnetic
field $h_i$ and the entries of 
$J_{i\ast}$ to zero recovers~\eqref{eq: static sequential pseudo-likelihood m}. 

Consider all couplings and fields together, one can sum~\eqref{eq:static sequential pseudo-likelihood}
over all rows of couplings to obtain the so-called (log)-\emph{pseudolikelihood}~\cite{besag1986}
\begin{equation}
L^{\PL} (\JJJ,\hhh) = \sum_{i} L^{i}_{\D} (J_{i\ast},h_i) \ ,
\end{equation} 
which can be maximized with respect to all rows of the coupling
matrix, yielding an asymmetric matrix as discussed above. The pseudolikelihood can also be maximized within the space of symmetric
matrices, although the maximisation problem is
harder~\cite{Aurell2012b}; rather than solving $N$ independent
gradient-descent problems in $N$ variables we have a single problem in
$N(N+1)/2$ variables. In practice, maximising the pseudolikelihood
without the symmetry constraint on the coupling matrix is 
preferred because of its simplicity and efficiency. 

The reconstruction based on maximizing the pseudolikelihood is of a
different nature than the previous methods, which approximated the
Boltzmann measure. The Boltzmann measure specifies the probability of
observing a particular spin configuration $\sss$ in equilibrium. On the
other hand, the conditional probability $\ln p (s_i^{\mu} \vert
\{s_j^{\mu}\}_{j \ne i})$ only gives the probability of observing a
given spin $s_i$ conditioned on the remaining spin variables and cannot be used to generate
the configurations of all spins. Yet, as a function of the model
parameters, the log-pseudolikelihood~\eqref{eq:static sequential pseudo-likelihood} 
has the same maximum as the likelihood~\eqref{eq: static sequential
	log-likelihood} in the limit of $M\to \infty$, when the sample
average in~\eqref{eq: static sequential pseudo-likelihood m} coincides
with the corresponding expectation values under the Boltzmann
distribution. 
%


Curiously, the pseudolikelihood can also be used to obtain a variant of the 
mean-field and TAP reconstructions. We follow Roudi and
Hertz~\cite{Roudi2011a} and replace expressions such as $\mean{\th \left( h_i + \sum_{j\ne i} J_{ij} \S_j \right)}$
by $\th \left( h_i + \sum_{j\ne i} J_{ij} \mean{\S_j} \right)$, thus replacing the effective local field
by its mean. The resulting approximation of Callen's identity
\eqref{eq: static sequential Callen m} is 
\begin{equation}
m^{\MF}_i= \th \left(h_i + \sum_{j \ne i} J_{ij} m^{\MF}_j  \right)\ ,
\end{equation} 
which is the mean-field equation~\eqref{eq:c3 mean field mi}  for the magnetizations. Then, replacing $\S_i$ by $\mean{\S_i}
+ (\S_i-\mean{\S_i})$ in~\eqref{eq: static sequential Callen C} and expanding
the equation to the lowest orders in $(\S_i-\mean{\S_i})$ gives
\begin{eqnarray}
C^{\PL-\MF}_{ij}= \frac{1}{1-(m^{\MF}_i)^2} \sum_{k \ne i} J_{ik}
C^{\PL-\MF}_{kj}\ .
\end{eqnarray}
Identifying the pseudolikelihood mean-field magnetisations $m^{\MF}_i$
and connected correlations
$C^{\PL-\MF}_{ij}$ with the sample magnetisations $m_i$ and sampled correlations
$C_{ij}$, this linear equation can be solved for
$J_{ik}^{\PL-\MF}$ for a fixed $i$, giving
\begin{equation}
J_{ik}^{\PL-\MF}= [{1-m_i^2}] \sum_{j \ne i}  C_{ij} \times
[(\CCC_{\setminus i})^{-1}]_{jk}\ ,
\end{equation} 
where
$\CCC_{\setminus i}$ is the submatrix of the correlation matrix with row and
column $i$ removed. This reconstruction is closely related, but not identical to
the mean-field reconstruction~\eqref{eq:meanfield_reconstruction}.  In
particular, the diagonal terms are naturally excluded. Numerical
experiments show that this reconstruction gives a good correlation
between the reconstructed and true couplings, but the magnitude of the
couplings is systematically
underestimated.
Continuing the expansion to second order in $(\S_i-\mean{\S_i})$  leads to
a variant of the TAP reconstruction~\eqref{eq:TAPreconstruction}. 

\subsection{Comparison of the different approaches}
\label{sec:comparison}


	\begin{center}
\begin{table}

	{\begin{tabular}{llll}
			\toprule
			\parbox[t]{1.5cm}{associated \\ potential} & \parbox[t]{2cm}{exact} & \parbox[t]{2cm}{variational \\ approximation} & \parbox[t]{2cm}{perturbative \\ expansion}\\
			\colrule
			$F(\JJJ,\hhh)$ & \parbox[t]{2cm}{convex \\ nonlinear \\ optimisation} & & \\
			& & & \\
			$G(\JJJ,\mmm)$ & & \parbox[t]{2cm}{mean-field, \\ Bethe--Peierls} & \parbox[t]{2cm}{Plefka, \\ mean-field, \\ TAP, \\ Bethe--Peierls} \\
			& & & \\
			$S(\pmb{\chi},\mmm)$ & & \parbox[t]{2cm}{independent-pair \\ approximation, \\  Cocco--Monasson} & \parbox[t]{2cm}{Sessak--Monasson} \\
			& & & \\
			\botrule
		\end{tabular}}
		\label{tab:comparison}
				\caption{\textbf{Classification of reconstruction methods based on the
						thermodynamic potentials used and the
						approximations employed to evaluate them.} Pseudolikelihood
					maximisation falls outside this classification scheme, as it is not
					based on an approximation of the Boltzmann measure.}
	\end{table}

	\end{center}

Table~\ref{tab:comparison} gives a classification of the
different reconstruction methods based on the thermodynamic potentials they
employ and the approximations used to evaluate them.  
Some of these approximations are exact in particular limits, and fail
in others. This has consequences for how well a 
reconstruction method works in a given regime.  
For instance, the TAP equations become exact for fully
connected systems (with couplings between all spin pairs drawn from a
distribution with variance $1/N$) in
the thermodynamic limit. Hence, we expect the TAP
reconstruction to perform well when couplings are
uniformly distributed across spin pairs, and couplings are
sufficiently weak so there is no ergodicity breaking
(see subsection~\ref{sec:nonergodic}).
Similarly, the Bethe--Peierls
approximation is exact when the non-zero couplings between spin pairs form a tree. Hence, we
expect the Bethe--Peierls reconstruction to work perfectly in this
case, and to work well when the graph of couplings
is locally treelike (so there are no short loops). The adaptive
cluster expansion, on the other hand, is expected to work well even
when there are short loops. 

\begin{figure}[!tbh]
	\includegraphics[width=0.5\textwidth]{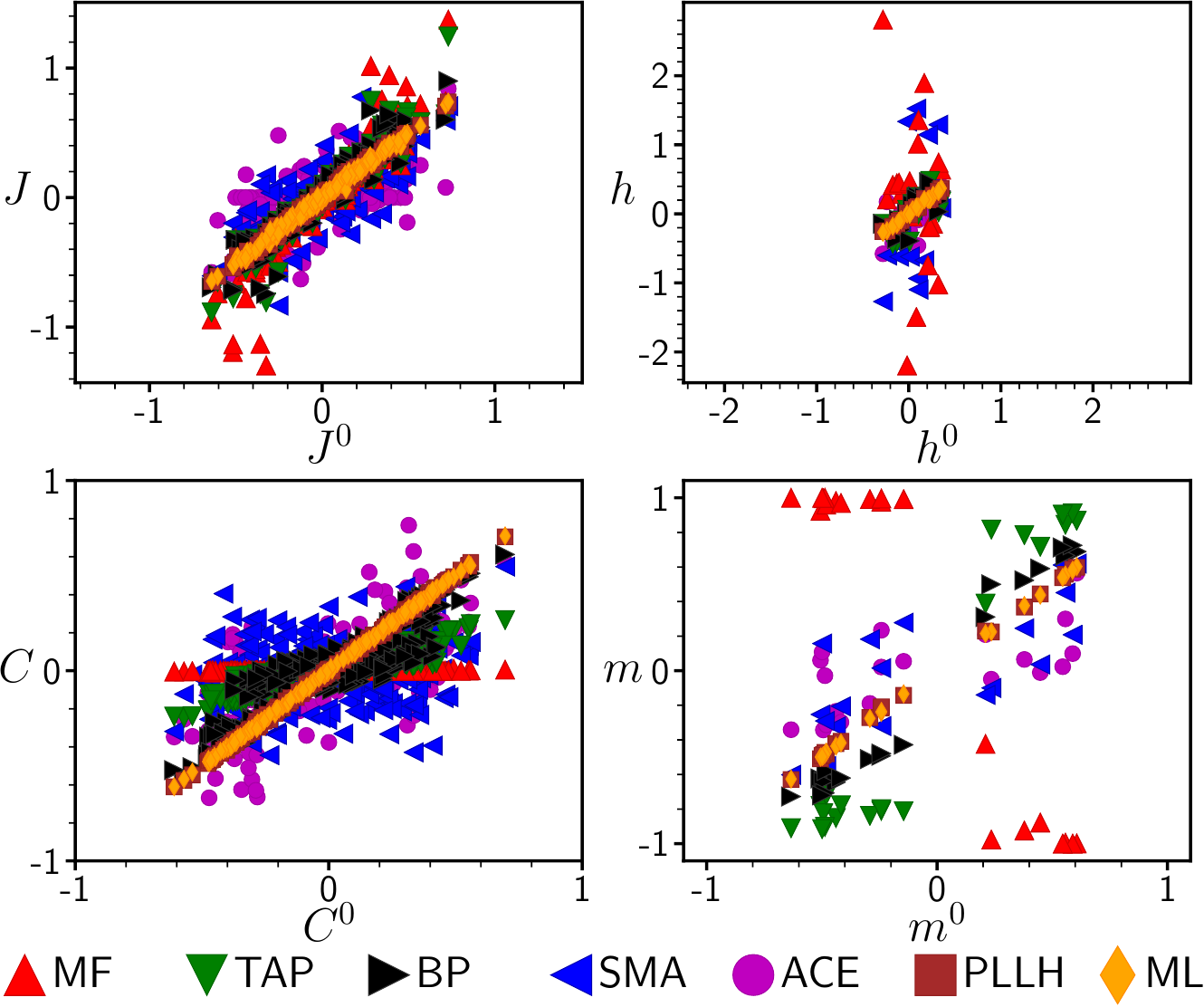}
	\caption{{\bf Reconstructing a fully connected Ising model with different
			methods.} 
		The scatter plots are  generated from a particular realization of the
		Sherrington-Kirkpatrick model with $N = 20$, $\beta = 1.3$
		and $M = 15000$ samples (configurations drawn from the Boltzmann
		measure, see text). 
		The colour legend indicates the
		different reconstruction methods: mean-field (MF), TAP, Bethe--Peierls (BP), Sessak--Monasson (SM),
		the adaptive cluster expansion (ACE), maximum pseudolikelihood (MPL), and maximum likelihood (ML). 
		The top plots show the reconstructed couplings/fields against the
		couplings/fields of the original model. 
		The bottom plots show the
		connected correlations and magnetisations calculated from $M$
		samples generated using the underlying model parameters (on the
		$x$-axis) and the same quantities generated using the inferred
		parameters ($y$-axis).
		The Sessak--Monasson expansion was
		computed as described in~\cite{Sessak2009a} with the series in the
		magnetic fields
		truncated at the third order excluding the loop terms. The adaptive
		cluster expansion was carried out using the
		ACE-package~\cite{Cocco2011a,Barton2016a} with a maximum cluster size
		$k=6$ and default parameters. The numerical maximisation of the
		likelihood and pseudolikelihood was done using the Eigen~3
		wrappers~\cite{eigenweb} for Minpack~\cite{minpack}.
		\label{fig:comparison-scatter}}
\end{figure}

\begin{figure}[!tbh]
	\includegraphics[width=0.43\textwidth]{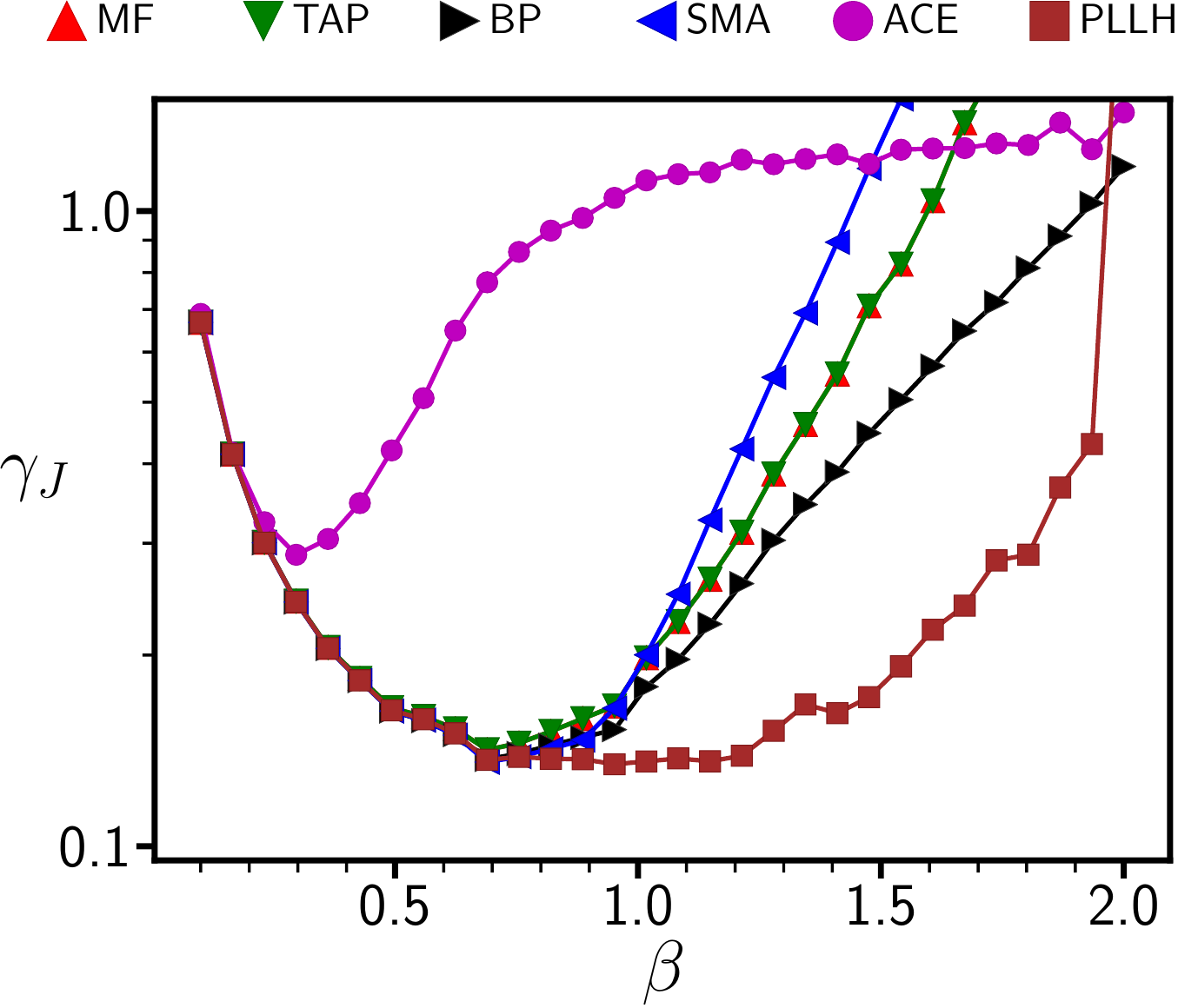} \\
	\vspace{3pt}
	\includegraphics[width=0.43\textwidth]{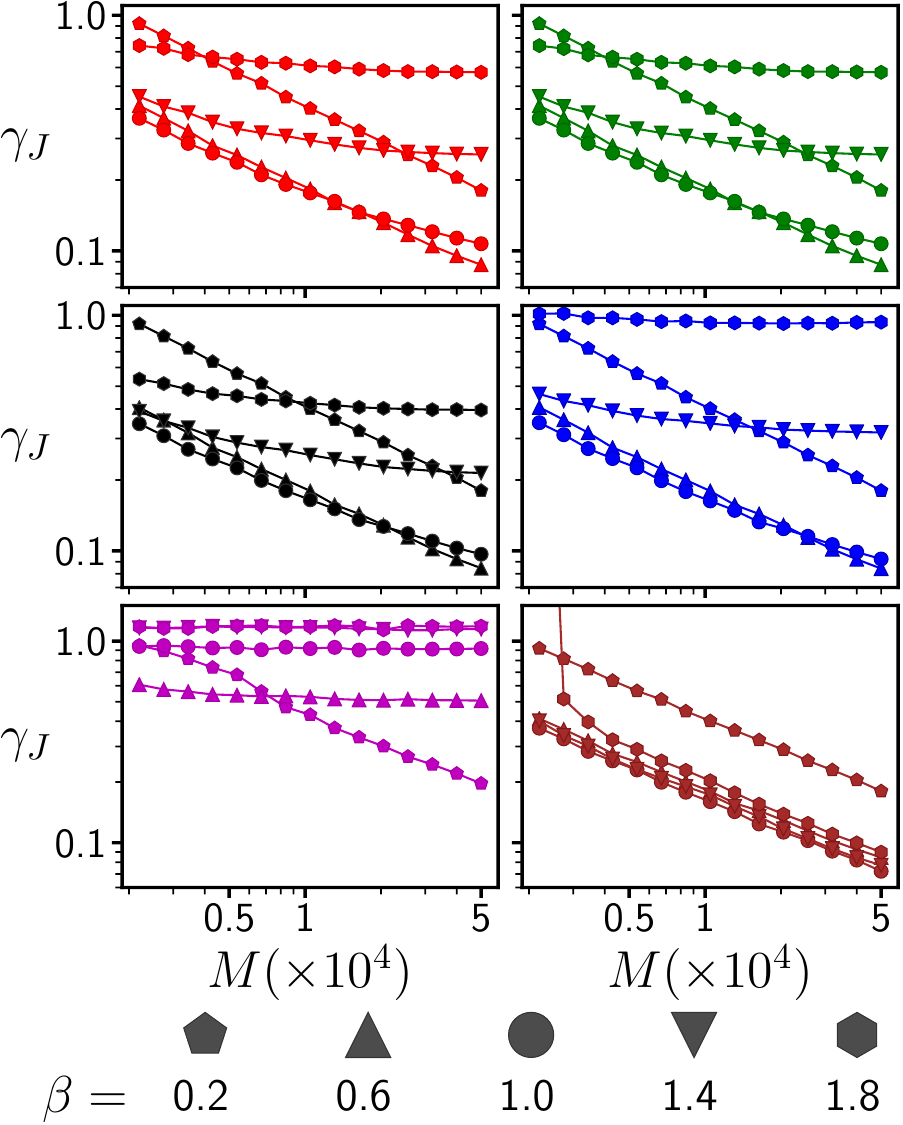} 
	\caption{\textbf{Reconstruction of a fully connected Ising model as a function of the coupling
			strength $\beta$ and the number of samples $M$.} 
		The top panel shows the reconstruction error $\gamma_J$ defined
		by~\eqref{eq:reconstruction_error} as a function of $\beta$ at a 
		constant $M=15000$ samples. The smaller panels show $\gamma_J$ as a function
		of the number of samples $M$ at different coupling strengths $\beta$
		and for different methods (mean-field, TAP, Bethe--Peierls,
		Sessak--Monasson, ACE, and maximum pseudolikelihood). For these plots a larger system
		size $N=64$ was used, so evaluating the likelihood is not feasible; at
		lower system sizes we found pseudolikelihood performs as well
		as the exact maximization of the likelihood in the parameter range
		considered.}
	\label{fig:comparison-SK}
\end{figure}

In this section, we compare the reconstruction methods discussed so
far. We consider the reconstruction problem of an Ising
model~\eqref{eq:thesecond} with couplings $J^0_{ij}$ between pairs of
spins defined on a certain graph. We explore two aspects of the model
parameters: the type of the interaction graph and the coupling
strength (temperature). We consider three different graphs: the fully
connected graph, a random graph with fixed degree as a
representative of graphs with long loops, and the $2$D square lattice
as a representative of graphs with short loops. In this section, we denote the couplings of the
model underlying the data by the superscript `$0$' to distinguish them from
the inferred couplings. For each graph, every
edge is assigned a coupling $J^0_{ij}$ drawn from a certain
distribution. We use the Gaussian distribution with zero mean and
standard deviation $\beta/\sqrt{N}$ for couplings on the fully
connected graph, leading to the Sherrington--Kirkpatrick (SK) model.  For the
tree and square lattice, the uniform distribution on
the interval $[-\beta,\beta]$ is
used for the couplings.  By tuning $\beta$, we effectively
change the coupling strength.  For the SK-model, external magnetic
fields field $h_i^0$ are drawn uniformly from the interval $[-0.3\beta,0.3 \beta]$.

At each value of 
$\beta$, we generate $M$ samples (spin configurations) drawn from the
Boltzmann distribution. Each sample is obtained by
simulating $10^4 N$ Monte Carlo steps 
using the Metropolis transition rule starting from a random
configuration. Although this is not always sufficient to guarantee that the
system has reached equilibrium, the results are not
sensitive to increasing this breaking-in time. 

Figure~\ref{fig:comparison-scatter} top left compares the
reconstructed couplings with the couplings of the original model at
$\beta=1.3$. We chose this value as it results in couplings which are
sufficiently large so that the different methods perform differently
(we will see that at low $\beta$ all methods perform similarly).  On
the fully connected graph, the adaptive cluster expansion performs
poorly as it sets too many couplings to zero. Mean-field
reconstruction somewhat overestimates large
couplings~\cite{barton2014a}. This error also affects the estimate of
the magnetic fields.  The TAP and Bethe--Peierls
reconstructions correct this overestimate quite effectively.  The reconstruction by
pseudolikelihood stands out by providing an accurate reconstruction of
the model parameters. One consequence is that in
figure~\ref{fig:comparison-scatter} the symbols indicating the results from pseudolikelihood
are largely obscured by those from the exact maximization of the
likelihood~\eqref{eq: static sequential log-likelihood}. We also
compare the correlations and magnetizations based on the reconstructed
parameters with those of the original model. The correlations and
magnetisations are sampled as described above. We find
significant bias in the results from all methods except for
likelihood maximization and pseudolikelihood maximization (bottom
panels of figure~\ref{fig:comparison-scatter}).


Next, we explore how the reconstruction quality depends on the coupling
strength $\beta$ and the number of samples $M$.  
To this end, we define the relative reconstruction error 
\begin{equation}
\label{eq:reconstruction_error}
\gamma_J= \sqrt{\frac{\sum_{i<j} (J_{ij} - J^0_{ij})^2 }{\sum_{i<j}
		(J^0_{ij})^2}} \ ,
\end{equation}
which compares the reconstructed couplings $\JJJ$ to the couplings of
the original model underlying the data
$\JJJ^0$. Figure~\ref{fig:comparison-SK}, large panel, shows the
relative reconstruction error $\gamma_J$ as a function of coupling
strength $\beta$. At low 
$\beta$, the connected correlations are small, so all reconstruction
methods are equally limited by sampling noise: The
relative reconstruction error increases with decreasing $\beta$ as the
couplings become small relative to the sampling error.  This is also
compatible with the result that at weak couplings the relative errors
of all methods decrease with an increasing number of samples, as seen
in the $6$ small panels of figure~\ref{fig:comparison-SK}.

\onecolumngrid
\begin{center}
	\begin{figure}[tb]
		\begin{tabular}{ccc}
			A & \hspace{30pt} & B \\
			\includegraphics[width=0.43\textwidth]{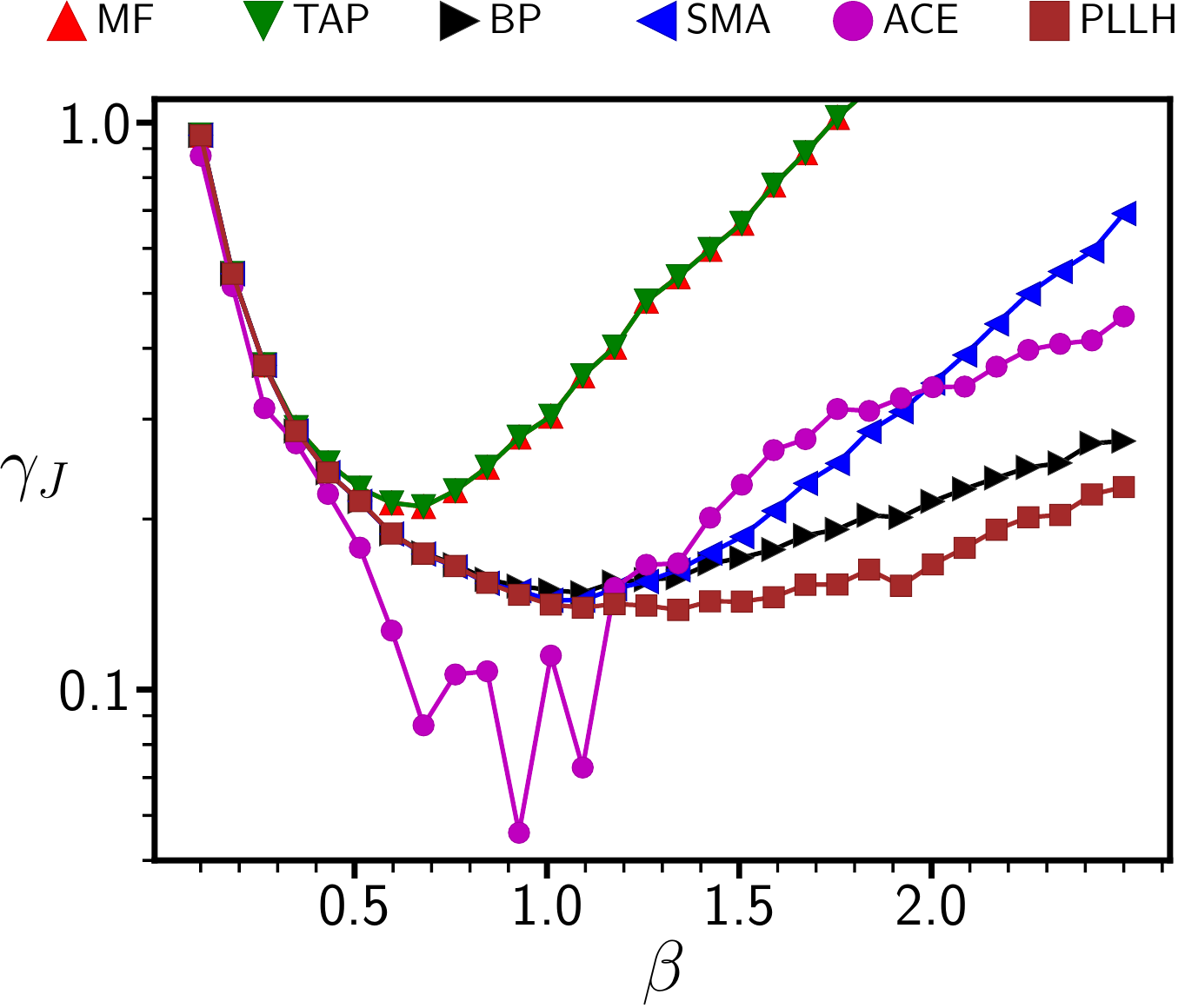} & \hspace{30pt} &
			\includegraphics[width=0.43\textwidth]{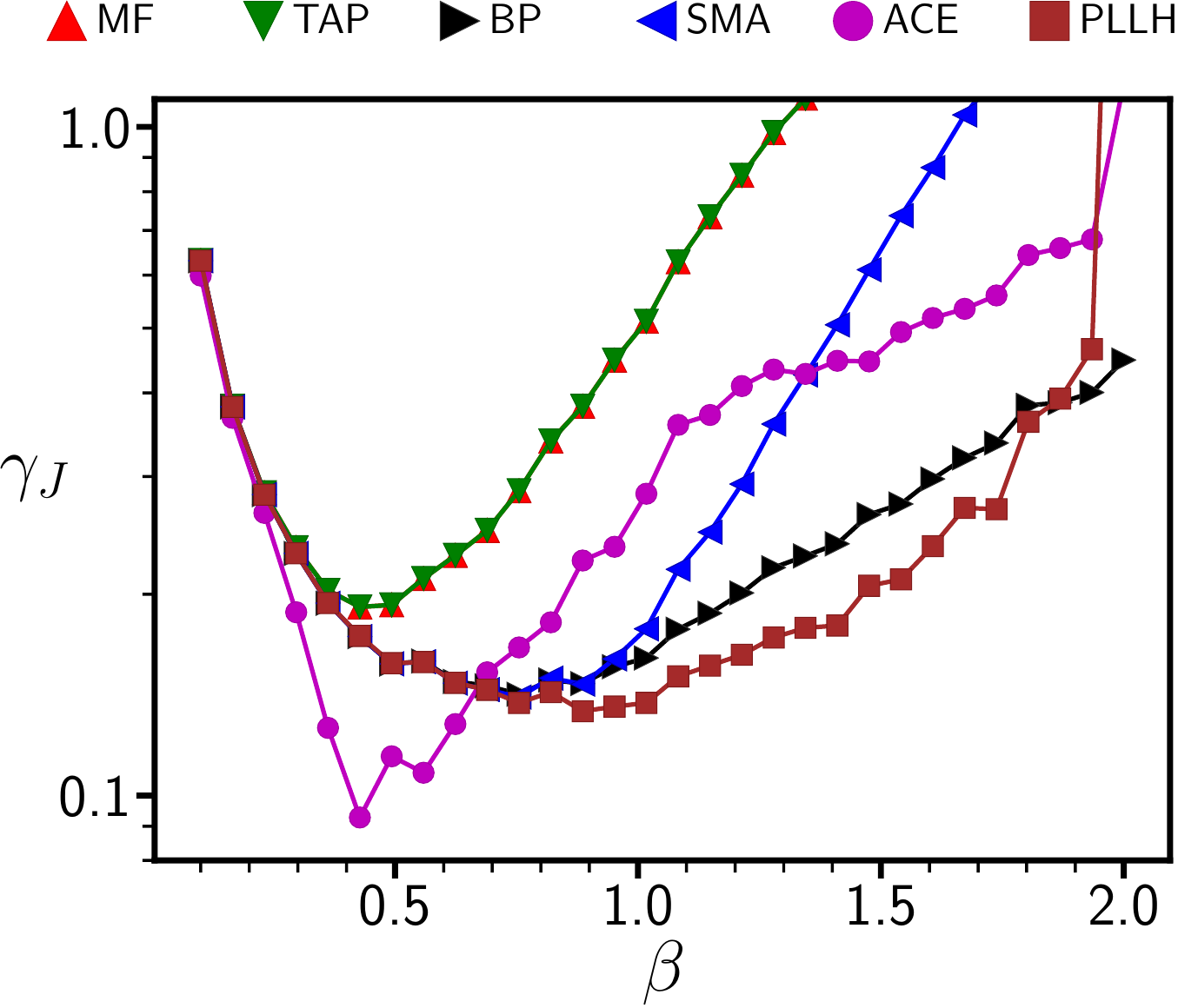} \\
			\includegraphics[width=0.43\textwidth]{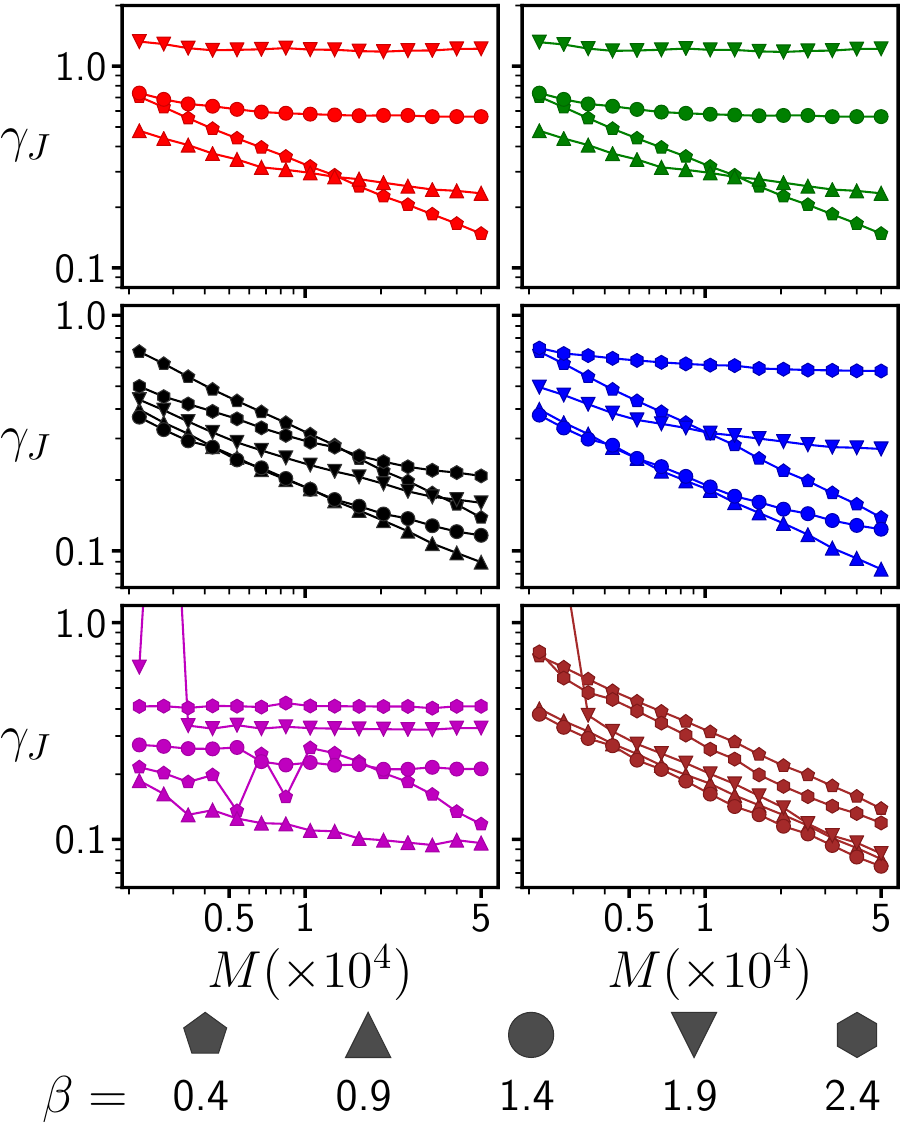} & \hspace{30pt} &
			\includegraphics[width=0.43\textwidth]{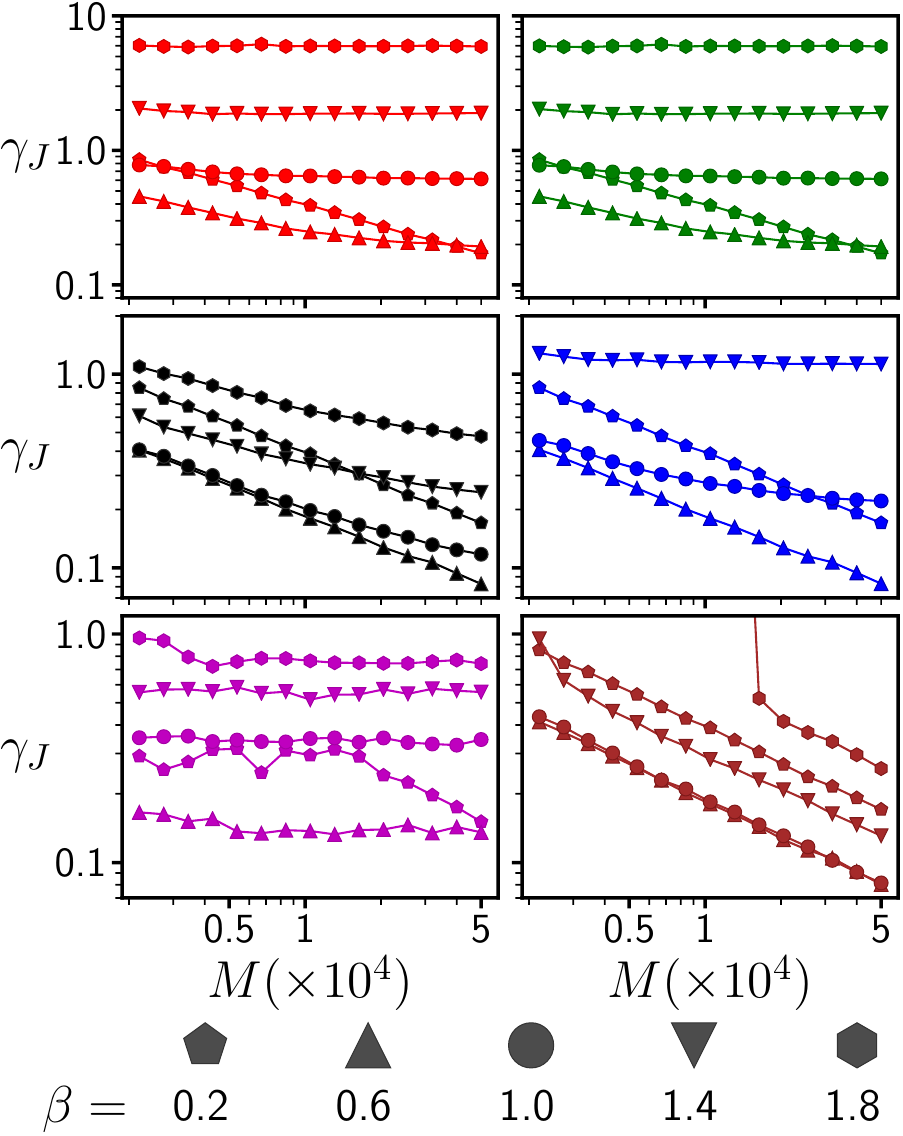}
		\end{tabular}
		\caption{\textbf{Reconstruction of the Ising model on a random
				graph of fixed degree $z=3$ (column A) and on the square lattice
				(column B) as a function of coupling strength $\beta$ and number
				of samples $M$.} The underlying couplings $J^0_{ij} $ are drawn from the uniform
			distribution on the interval $[-\beta,+\beta]$, and the external magnetic fields are set to zero
			for simplicity. The system size is $N=64$. 
			The top panel shows the reconstruction error $\gamma_J$ defined
			by~\eqref{eq:reconstruction_error} as a function of $\beta$ at
			constant $M=15000$ samples. The smaller panels show $\gamma_J$ as a function
			of the number of samples $M$ at different coupling strength $\beta$
			and for different methods (mean-field, TAP, Bethe--Peierls,
			Sessak--Monasson, ACE, and maximum pseudolikelihood).}
		\label{fig:comparison-RGZSQ}
	\end{figure}
\end{center}
\twocolumngrid

We find that the pseudolikelihood reconstruction outperforms the other
(non-exact) methods over the entire range of $\beta$, and 
correctly reconstructs the model parameters even in the glassy phase at strong couplings~\cite{Aurell2012b}. This is because the
conditional statistics of a single spin (conditioned on the other spins) is 
correctly described by~\eqref{eq:conditional_statistics_single_spin}
at any coupling strength. Although at strong couplings, the error of the
pseudolikelihood reconstruction grows with $\beta$, this can be compensated by increasing the number of samples (Figure~\ref{fig:comparison-SK}, bottom right).
This is not so for all other approximate methods (Figure~\ref{fig:comparison-SK},
small panels): At high $\beta$, the
approximation each method is based on breaks down, which cannot be
compensated by a larger number of samples. 

Figure~\ref{fig:comparison-RGZSQ} shows the reconstruction errors
of different methods for a random graph of fixed degree (column A) and the
square lattice (column B). Again, pseudolikelihood performs well in
these cases, as does the Bethe--Peierls
reconstruction. The
adaptive cluster expansion shows a remarkable behaviour: For both
graphs, it has a very small reconstruction error at weak
coupling strength, but breaks down at strong
couplings. At weak couplings, where all other methods have
similar reconstruction errors arising from
sampling noise, however the adaptive cluster expansion appears to avoid this
source of error. The adaptive
cluster expansion explicitly assumes some couplings are exactly zero, 
so the reconstruction is biased towards graphs which are not fully
connected. It thus uses \emph{extra information} beyond the
data.

Our comparison of the different methods is based on the knowledge of the
underlying couplings, so the reconstructed couplings can be compared
to the true underlying ones. In practice, the underlying couplings are
not available. However, the likelihood~\eqref{eq: static sequential
	log-likelihood} can be evaluated for different reconstructions, with
the better reconstruction resulting in a higher value of the
likelihood. Alternatively, the correlations and magnetizations of the
reconstructed model can be compared with those observed in the data as
in figure~\ref{fig:comparison-scatter}. 
Whether regularizing terms improve the reconstruction can
be decided based on statistical tests such as the Bayesian information
criterion~\cite{schwarz1978estimating} or the Akaike information
criterion~\cite{akaike1974new}. 

An aspect not discussed so far is the
inference of the interaction graph, this is, the distinction between
non-zero and zero
couplings. In practice, 
there is always some ambiguity between small non-zero couplings and
couplings which are exactly zero,
particularly at high sampling noise. 
A standard practice is to set a threshold and cut off small
couplings below the threshold. Two exceptions are the adaptive cluster
expansion and the pseudolikelihood reconstruction. The adaptive
cluster expansion has a built-in procedure to set some couplings to
zero. For
the pseudolikelihood reconstruction, the problem
is related to feature selection. Many
methods for feature selection are available from
statistics~\cite{Hastie2009a}, however there is no consensus
on the best method for the inverse Ising problem. One possibility is adding an $\ell_1$-regularization term (known
as Lasso, see section~\ref{sec:maxlikelihood}) to the
pseudolikelihood~\cite{Ravikumar2010a}. However, there 
is a critical coupling strength, below which
the $\ell_1$ regularization of the pseudolikelihood fails to recover
the interaction graph~\cite{BentoMontanari2009}.

\subsection{Reconstruction and non-ergodicity}
\label{sec:nonergodic}

At strong couplings, the dynamics of a system can become non-ergodic,
which affects the sampling of configurations, and hence the
reconstruction of parameters. Reconstruction on the basis of
non-ergodic sampling is a fundamentally difficult subject where little
is known to date. 

At low temperatures (or strong couplings), a disordered spin system can
undergo a phase transition to a state where spins are `frozen' in
different directions. This is the famous transition to the spin glass
phase~\cite{mezardparisivirasoro}. At the spin glass transition, the Gibbs free energy develops
multiple valleys with extensive barriers between them (thermodynamic
states). These free energy barriers
constrain the dynamics of the system, and the system can remain confined to one particular
valley for long times. A signature of this ergodicity breaking is the emergence of
non-trivial spin magnetisations: At the phase transition, the self-consistent
equations~\eqref{eq:c3 mean field mi} and~\eqref{eq:TAPeq}
develop multiple solutions for the
magnetisations $m_i$, corresponding to the different orientations the
spins freeze into. 
The appearance of multiple free energy minima and the resulting
ergodicity breaking has long been recognized as an obstacle to
mean-field reconstruction~\cite{marinari2010a}. In fact, all methods
based on self-consistent equations (mean-field, TAP, Bethe--Peierls
reconstruction) fail in the glassy phase~\cite{Mezard2009a}. 

The mean-field equations \eqref{eq:c3 mean field mi} describe individual thermodynamic 
states, not the mixture of many states that characterizes the
Boltzmann measure.  At large couplings (or low temperatures),
mean-field and related approaches turn out not to be limited by the validity of
self-consistent equations like~\eqref{eq:c3 mean field mi}
or~\eqref{eq:TAPeq}, but by the identification of observed
magnetisations and correlations with the corresponding quantities
calculated under one of the self-consistent equations. The former may
involve averages over multiple thermodynamic states, whereas each
solution of the self-consistent equations describes a separate
thermodynamic state.
A solution to this problem is thus to consider correlations and
magnetisations \emph{within} a single thermodynamic state, where the
mean-field result~\eqref{eq: static sequential MF LR} is 
valid within the limitations of
mean-field theory.  The different thermodynamic states (free
energy minima) can be identified from the data by searching for
clusters in the sampled spin configurations. Collecting
self-consistent equations from
different minima and jointly solving these equations using the
Moore-Penrose pseudo-inverse allows the reconstruction of couplings
and fields~\cite{Nguyen2012b}. 

The emergence of multiple states can lead to another problem: the
samples need not come from the Gibbs measure in the first place, but might be taken only
from a single thermodynamic state. In~\cite{braunstein2011a}, the reconstruction from samples
from a single thermodynamic state is studied for the concrete
case of the Hopfield model. In the so-called
memory regime, the Hopfield model has a number of 
thermodynamic states (attractor states) which scales linearly with the system
size. Samples generated from a single run of the model's dynamics
will generally come from a single state only. (The particular state the
system is `attracted' to depends on the initial conditions.) 
In this regime, the cavity-recursion equations \eqref{eq:cavity recursion_probs} typically have multiple
solutions, just like
the TAP equations~\eqref{eq:TAPeq} in the low-temperature phase. These solutions can be identified as fixed points of
belief propagation, see ~\ref{sec:bethe}. For a system where couplings
do not form a tree, each individual fixed point only approximates the marginal probability distributions
inside a single non-ergodic component of the full Gibbs
measure. However, this approximation can become exact in the thermodynamic
limit. (A necessary  condition is that loops are sufficiently long, so connected correlations measured in a
particular state decay sufficiently quickly with distance, where
distance is measured along the graph of non-zero
couplings.) 

Assuming one is able to find a fixed point of the cavity-recursion equations, it is
a straightforward  computation to express the Boltzmann
weight~\eqref{eq:thefirst} in terms of the full set of the belief propagation
marginals and of the ratio between the cavity-recursion partition function 
and the true partition function of the model. For any fixed point, labelled by the index $\alpha$, the Boltzmann weights are
\begin{equation}
p(\sss)=\frac{Z_{BP}^{\alpha}}{Z(J,h)}\prod_{i}p^{\alpha}_i(s_{i})\prod_{i<j}\frac{p_{ij}^{\alpha}(s_{i},s_{j})}{p_{i}^{\alpha}(s_{i})p_{j}^{\alpha}(s_{j})}
\ .\label{eq:non-ergodic-BP}
\end{equation}
This result also applies to more general forms of the energy
function~\cite{braunstein2011a}. It was first derived for the
zero-temperature case in the context of the so-called
tree-reweighted approximation to the 
partition function~\cite{kolmogorov2006convergent}.

The probability distribution~\eqref{eq:non-ergodic-BP} can be used
as the starting point for reconstruction in at least
two ways. The simpler one consists in replacing
$p_{ij}^{\alpha}(s_{i},s_{j})$ and $p_{i}^{\alpha}(s_{i})$ by their
sample estimates inside
state $\alpha$ and then solving the identity between \eqref{eq:non-ergodic-BP}
and \eqref{eq:thefirst} with respect to $\JJJ$
and $\hhh$: $Z(\JJJ,\hhh)$ cancels out and one is left with the
independent-pair approximation of subsection~\ref{sec:ace}. Solving
this identity can be done using belief propagation, see section
\ref{sec:BP} and~\cite{braunstein2011a}. This approach suffers from
the fact that in general there is no belief propagation fixed point
for real data sets.

A more promising approach is to guide the belief propagation equations to converge
to a fixed point corresponding to an appropriate ergodic component \emph{close}
to the empirical data. In fact, ignoring the information that the
data come from a single ergodic component results in a large
reconstruction error as one is effectively maximizing the wrong
likelihood: the (reduced) free energy $-\ln Z$  appearing in the
likelihood~\eqref{eq: static sequential log-likelihood} needs to be
replaced by a free energy restricted to a single thermodynamic state.
An
algorithmic implementation of this idea consists in restricting the spin
configurations to a subset of configuration $\Omega_{d}(\xi)$,
\textit{e.g.}, formed by a hypersphere
of diameter $d$ centered around the
centroid $\xi$ of the data samples. The system is forced to follow the measure
\begin{equation}
p_{d}(\sss)\text{\ensuremath{\propto}}I[\sss
\in{\Omega}_{d}(\xi)]e^{\beta(\sum_{i<j}J_{ij}s_{i}s_{j}+\sum_{i}h_{i}s_{i})}\label{eq:restricted}
\ ,
\end{equation}
where the indicator function $I[\sss \in{\Omega}_{d}(\xi)]$ for the
set of configurations $\Omega_{d}(\xi)$ is one for
$\sss \in \Omega_{d}(\xi)$ and zero otherwise. 
The BP approach can be used to enforce that both magnetizations $\{m_{i}^{(d)}\}$ and correlations
$\{c_{ij}^{(d)}\}$ are computed within the subspace $\Omega_{d}(\xi)$, 
see~\cite{braunstein2011a} for algorithmic details. In this way, the
reconstruction of the couplings and local fields can
be done by maximising the correct likelihood, \textit{i.e.},
replacing the partition function in the log-likelihood~\eqref{eq:
	static sequential log-likelihood} with the partition function
computed over the subset of configurations $\Omega_{d}(\xi)$.
In this way, the parameters describing all thermodynamic states can be
inferred from configurations of the system sampled from a single
state. 

Inverse problems in non-ergodic systems remain a challenging topic
with many open questions, for instance if the approach applied above
to the Hopfield model can be extended to infer generic couplings from
samples from a single thermodynamic state.



\subsection{Parameter reconstruction and criticality}
\label{sec:criticality}

Empirical evidence for critical behaviour has been reported in systems
as diverse as neural networks~\cite{bedard2006a,beggs2012a} and
financial markets~\cite{gabaix2003a,bury2013a}, leading to the
intriguing hypothesis that some information-processing systems may be
operating at a critical point~\cite{Mora2011b}. A key signature of
criticality is broad tails in some quantities, for instance the
distribution of returns in a financial market, or avalanches of neural
activity, whose sizes are distributed according to a power law. A second
sign of criticality involves an inverse statistical
problem: A statistical model like the Ising system with parameters
chosen to match empirical data (such as neural firing patterns 
or financial data) shows signs of a phase
transition~\cite{Mora2011b,mastromatteo2011a,stephens2013a,tkavcik2014a,tkacik2015,mora2015dynamical}.
Specifically, the heat capacity of an Ising model with parameters
$J_{ij}$ and $h_i$ reconstructed from data shows a peak in the heat
capacity as a function of the temperature. Temperature is introduced by changing the couplings to $\beta J_{ij}$
and $\beta h_i$. Varying the inverse temperatures $\beta$, the heat
capacity $C_h\equiv -\beta^2 \partial \langle H \rangle /\partial \beta $
shows a pronounced maximum near or at $\beta=1$, that is, at the model
parameter values inferred from the data. The implication is that the
parameters of the reconstructed model occupy a very particular area in
the space of all model parameters, namely one resulting in critical
behaviour.

\begin{figure}[bt]
	\includegraphics[width = .415\textwidth]{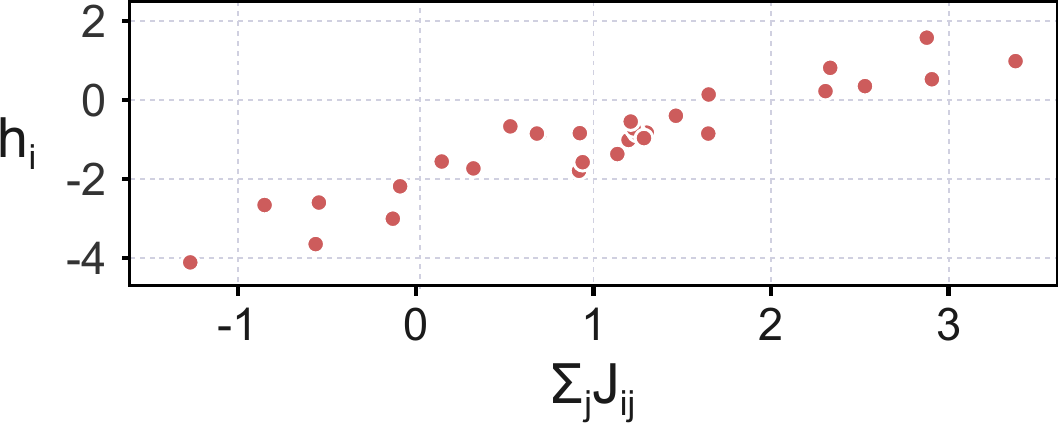}
	\includegraphics[width = .475\textwidth]{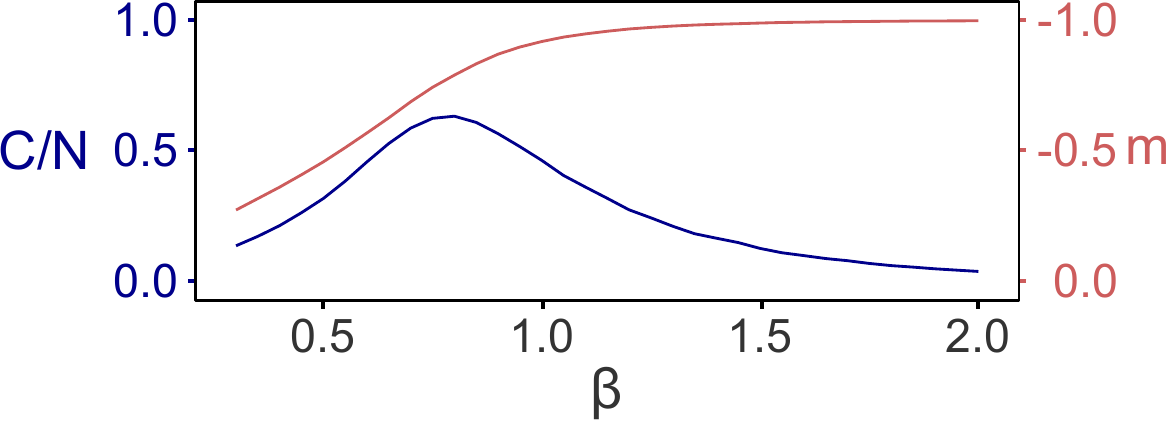}
	\caption{{\bf The Ising model with parameters matching the statistics
			of neural firing patterns.} Couplings and external fields are generated from
		neural data~\cite{tkavcik2014a,tkacik2015} as described in the
		text. The system size here is $N=30$. Left: Fields and couplings turn out not to be independent, but obey
		a linear-affine relationship, which is due to neural firing rates and
		hence magnetisations being approximately constant across
		neurons. Right: We simulate the model with reconstructed couplings at
		different temperatures using Monte Carlo simulations. The heat capacity (blue) shows a peak near $\beta=1$,
		and the magnetisation $m=\frac{1}{N} \sum_i m_i$ (red, axis
		on the right) goes from zero to minus one as the inverse temperature is increased. 
		\label{fig:critical}}
\end{figure}

An example is shown in figure~\ref{fig:critical}. It is based on
recordings of $160$ neurons in the retina of a salamander taken by the
Berry lab~\cite{tkavcik2014a,tkacik2015}. As described in
section~\ref{sec:neural}, time is discretised into intervals of $20$
ms duration, and the spin variable $s_i$ takes on the values $+1$ if
neuron $i$ fires during a particular interval, and $-1$ if it does not
fire. Correlations and magnetisations are then computed by averaging
over $297$ retinal stimulation time courses, where during each time
course, the retina is subjected to the same visual stimulation of
$953\times 20$ms duration. During most time intervals, a neuron is
typically silent. As a result, spins variables (each representing 
a different neuron) have negative magnetisations, with mean and standard deviation $-0.93\pm 0.06$ over
all spins. We note that the magnetisations are fairly homogeneous across
spins. Connected correlations between spins are small, with a slight bias towards
positive correlations; mean and standard deviation over all spin pairs
are $0.006 \pm 0.017$. Both points turn out to be important.

We use the pseudolikelihood method of
section~\ref{sec:pseudo-likelihood} to reconstruct the model parameters
from the firing patterns the first $N=30$
neurons. Similar results are found for different system sizes.
Figure~\ref{fig:critical}A shows that the reconstructed model parameters
indeed occupy a particular region of the space of model parameters:
External fields $h_i$ are linked to the sum over couplings
$\sum_j J_{ij}$ by a linear-affine relationship. If the magnetisations
$\{m_i\}$ are all equal to some $m$, such a relationship
follows directly from the mean-field equation
$h_i=\arth m_i - \sum_j J_{ij} m_j$, along with slope
$-m$ and zero-offset $\arth m$. 

Now we simulate the Ising model with the reconstructed parameters, but rescale both couplings and fields by $\beta$. At
inverse temperature $\beta=1$, magnetisations and correlations are
close to the magnetisations and correlations
found in the original
data. However, away from $\beta=1$, different fields
$h_i=\frac{1}{\beta}\arth~m_i - \frac{1}{\beta}\sum_j J_{ij}
m_j$
(in the mean-field approximation) would be needed to retain the same
magnetisation. However, as the fields remain fixed, the magnetisations change
instead, reaching $-1 (0)$ in the limit $\beta \to \infty
(0)$.
Between these limits, the magnetisations, and hence the energy changes
rapidly, leading to a peak in the heat capacity.  For small, largely
positive connected correlations $C_{ij}$ we expect couplings to scale
as $1/N$, resulting in a phase transition described by the Curie-Weiss
model. Indeed, a finite-size analysis shows the peak getting sharper and
moving closer to $\beta=1$ as the system size is
increased~\cite{Mora2011b,mastromatteo2011a,stephens2013a,tkacik2015,mora2015dynamical}.
From this, we posit that the peak in the heat capacity indeed signals
a ferromagnetic phase transition, however not necessarily one
underlying the original system. Instead, we generically expect such a
ferromagnetic transition whenever a system shows sufficient 
correlations and magnetisations that are neither plus nor minus one: The corresponding couplings
and fields then lie near (not at) a critical point, which is
characterized by the emergence of non-zero magnetisations. Changing the
temperature will drive the system to either higher or lower
magnetisations, and away from the critical point. 

This effect may not be limited to a ferromagnetic transition induced
by the empirical magnetisation. In~\cite{mastromatteo2011a}
Mastromatteo and Marsili point out a link between the criticality of
inferred models and information
geometry~\cite{balasubramanian1997a,myung2000a}: susceptibilities
such as the magnetic susceptibility can be interpreted as the entries in the
so-called Fisher information matrix used to calculated the covariances
of maximum-likelihood estimates~\cite{Cover2006a}. A high
susceptibility implies that different parameter values can be
distinguished on the basis of limited data, whereas low
susceptibilities mean that the likelihood does not differ sufficiently
between different parameter values to significantly favour one parameter value over
another. Thus, a critical point corresponds to a high density of models whose
parameters are distinguishable on the basis of the data~\cite{mastromatteo2011a}.

\section{Non-equilibrium reconstruction}
\label{sec:nonequilibrium}

In a model of interacting magnetic spins such as~\eqref{eq:thesecond},
each pair of spins $i<j$ contributes a
term $J_{ij} s_i s_j$ to the Hamiltonian. The resulting effective
fields on a spin $i$, $\sum_j J_{ij}s_j +h_i$, involve symmetric couplings
$J_{ij}=J_{ji}$; spin $i$ influences spin $j$ as much as $j$
influences $i$. A stochastic dynamics based on such local fields, such as
Monte Carlo dynamics, 
obeys detailed balance, which allows one to determine the steady-state
distribution~\cite{Gardiner1985a}. 

In applications such as neural networks or gene regulatory networks
discussed in sections~\ref{sec:neural} and~\ref{sec:geneexpression} we
have no reason to expect symmetric connections between neurons or
between genes; a synaptic connection from neuron $i$ to neuron $j$
does not imply a link in the reverse direction. Stochastic systems generally relax to a steady state at long times, where observables no
longer change with time. However, in systems with asymmetric
couplings, the resulting 
non-equilibrium steady state (NESS) violates detailed balance and
differs from the Boltzmann distribution.  As a result, none of the
results of section \ref{sec:equilibrium} on equilibrium reconstruction
apply to systems with asymmetric couplings. 

In this section we consider the inverse Ising problem in a non-equilibrium
setting. We first review different types of spin dynamics,
and then turn to the problem of reconstructing the parameters of a
spin system from either time-series data or from samples of the
non-equilibrium steady state. 

\subsection{Dynamics of the Ising model}

The Ising model lacks a prescribed dynamics, and there are many
different dynamical rules that lead to a particular steady-state
distribution. One particularly simple dynamics is the so-called
Glauber dynamics~\cite{glauber1963a}, which allows the derivation of
a number of analytical results. Other dynamical rules can be used for
parameter reconstruction in the same way at least in principle. What
dynamical rule is suitable for particular systems is however an open
question. An approach which sidesteps this issue is to apply the
maximum entropy principle to stochastic trajectories, known as the
principle of maximum caliber ~\cite{presse2013a}. In~\cite{marre2009,nassercessac2014a}
such an approach is applied to analyse neural dynamics. 

\subsubsection{Sequential Glauber dynamics}

Glauber dynamics can be based on discrete
time steps, and at each step either one or all spins have new values
assigned to them according to a stochastic rule. We first consider a sequential dynamics: In the
transition from time $t$ to $t+1$, the label of a spin
variable is picked randomly, say $i$. The value of the spin variable
$\S_i$ is then updated, with $\s_i=\pm 1$ sampled from the probability
distribution
\begin{equation}
\label{eq:glaubersq}
p(\s_i(t+1)|\sss(t))=\frac{\exp\{\s_i(t+1) \theta_i(t)\}}{2
	\ch(\theta_i(t))} \ ,
\end{equation}
where the effective local field is denoted
\begin{equation}
\label{eq:localfield}
\theta_i(t)=\sum_j J_{ij} \s_j(t) + h_i \ .
\end{equation}
One way to implement this dynamics is to set 
\begin{equation}
\label{eq:glaubersq2}
\s_i(t+1)=\textrm{sign} \left(\theta_i(t) +\xi(t) \right) \ ,
\end{equation}
where $\xi_i(t)$ is drawn independently at each step from
the distribution $p(\xi)=1-\th^2(\xi)$.  

If the couplings $J_{ij}$ between spins are symmetric and there are no
self-couplings $J_{ii}$, the sequential Glauber
dynamics~\eqref{eq:glaubersq} obeys detailed balance, and the distribution
of spin configurations at long times relaxes to a steady state
described by the Boltzmann
distribution \eqref{eq: static sequential Boltzmann}. 
However, the NESS arising at long times from Glauber dynamics with
non-symmetric couplings is generally not known. Nevertheless, 
there are several exact relations that follow
from~\eqref{eq:glaubersq}, and those can be exploited for inference.

Suppose that spin $i$ is picked for updating in
the step from time $t$ to $t+1$. Averaging over the two possible spin
configurations at time $t+1$, we obtain the magnetisation 
\begin{equation}
\label{eq:glaubersq_m}
m_i(t+1)_{\sss(t)} \equiv \langle \S_i(t+1) \rangle_{\sss(t)}=
\th\left(\theta_i(t) \right) \ , 
\end{equation}
where the average is conditioned on the configuration of all other
spins at time $t$ through the effective local field $\theta_i$. If
spin $i$
is not updated in this interval, the conditioned magnetisation is
trivially $m_i(t+1)_{\sss(t)} = \s_i(t)$. In the steady state, effective
fields 
$\Theta_i=\sum_j J_{ij} \S_j(t) + h_i$ and spins configurations $\SSS$ are 
random variables whose distribution no longer depends on
time; their averages give 
\begin{equation}
\label{eq:glaubersq_m_ness}
m_i \equiv \langle \S_i \rangle=
\langle \th(\Theta_i) \rangle=\langle \th( \sum_j J_{ij} \S_j + h_i ) \rangle\ .
\end{equation}
Similarly, the pair correlation between spins at sites $i \neq j$
at equal times obeys in the NESS 
\begin{equation}
\label{eq:glaubersq_c_ness}
\chi_{ij} \equiv \langle \S_i \S_j \rangle=\frac{1}{2}\langle \S_i\th\left(\Theta_j \right) \rangle+\frac{1}{2}\langle \S_j\th\left(\Theta_i \right) \rangle\ ,
\end{equation}
with the two terms arising from instances where the last update of $i$
was before the last update of $j$, and vice versa. Likewise, the
pair correlation of spin configurations at consecutive time intervals is in
the steady state 
\begin{equation}
\label{eq:glaubersq_d_ness}
\phi_{ij} \equiv \langle \S_i(t+1) \S_j(t) \rangle=\frac{1}{N}\langle
\th\left(\Theta_i \right)\S_j \rangle+\frac{N-1}{N}\chi_{ij} \ .
\end{equation}
These relationships are exact, but are hard to evaluate since the averages on
the right-hand sides are over the statistics of spins in the
NESS, which is generally unknown. Below, these equations will be used in different ways
for the reconstruction of the model parameters. 

The dynamics~\eqref{eq:glaubersq} defines a Markov chain of transitions
between spin configurations differing by at most one spin
flip. Equivalently, one can also define an asynchronous dynamics
described by a Master equation, where time is continuous and the time between successive spin flips is
a continuous random variable.

\subsubsection{Parallel Glauber dynamics}

Glauber dynamics can also be defined with parallel 
updates, where \emph{all} spin variables can change their configurations in
the time interval between $t$ and $t+1$ according to the stochastic
update rule
\begin{equation}
\label{eq:glauberpa}
p(\sss(t+1)|\sss(t))=\frac{\exp\{\sum_i\s_i(t+1)
	\theta_i(t)\}}{\prod_i 2 \ch(\theta_i(t))} \ .
\end{equation}
This update rule defines a Markov chain consisting of stochastic transitions
between spin configurations. 
The resulting dynamics is not realistic for biological networks, as the
synchronous update requires a central clock. 
It is however
implemented easily in technical networks, and is widely used for its simplicity. For a symmetric coupling
matrix, the steady state can still be specified in closed form using Peretto's pseudo-Hamiltonian~\cite{peretto1984a}.

Magnetisations and
correlations in the NESS obey simpler relationships for parallel
updates than for sequential updates; with the same arguments as above, 
one obtains
\begin{align}
\label{eq:glauberpa_corrs}
m_i&=\langle \th\left(\Theta_i \right) \rangle \\
\chi_{ij}&= \langle \th\left(\Theta_i \right) \th\left(\Theta_j \right)  \rangle\nonumber\\
\phi_{ij}&=\langle \th\left(\Theta_i \right) \S_j \rangle \nonumber
\ .
\end{align}

\subsection{Reconstruction from time series data}

The reconstruction of system parameters is surprisingly easy on the
basis of time series data, which specifies the state of each spin variable
at $M$ successive time points. An application where
such data is widely available is the reconstruction of neural networks from
temporal recordings of neural activity. Given a stochastic update rule
such as \eqref{eq:glaubersq} or \eqref{eq:glauberpa}, the (log-) likelihood given a time series $\D=\{\sss(t)\}$ is 
\begin{equation}
\label{eq:likelihood_timeseries}
L_{\D}(\JJJ,\hhh) = \frac{1}{M}\sum_{t=1}^{M-1} \ln
p(\sss(t+1)|\sss(t)) \ .
\end{equation}
The likelihood can be evaluated over any
time interval, both in the NESS or even before the steady state has
been reached. This approach is not limited to non-equilibrium
systems; whenever time series data are available, it can be used
equally well to reconstruct symmetric couplings. 

\subsubsection{Maximisation of the likelihood}
\label{sec:timeseries_maxlikelihood}
For Glauber dynamics with parallel updates, the likelihood is
\begin{equation}
\label{eq:likelihood_timeseries_pa}
L_{\D}(\JJJ,\hhh) = \frac{1}{M}\sum_{t=1}^{M-1} \sum_i \left[\s_i(t+1)
\theta_i(t)-\ln 2 \ch (\theta_i(t))\right] \ .
\end{equation}
Unlike the likelihood \eqref{eq: static sequential log-likelihood}
arising in equilibrium statistics, the non-equilibrium likelihood can
be evaluated easily, as the normalisation is already contained in the term
$\ch(\theta)$. Derivatives of the likelihood
\eqref{eq:likelihood_timeseries_pa} are 
\begin{align}
\frac{\partial L_{\D}}{\partial h_i} (\JJJ,\hhh)&= \frac{1}{M}\sum_{t=1}^{M-1}\left[
\s_i(t+1)-\th \theta_i(t)\right] \\
\frac{\partial L_{\D}}{\partial J_{ij}} (\JJJ,\hhh)&=\frac{1}{M}\sum_{t=1}^{M-1}\left[
\s_i(t+1) \s_j(t)-\th (\theta_i(t)) s_j(t) \right]\ . \nonumber
\end{align}
These derivatives can be evaluated in $MN^2$ computational steps and can be used to
maximize the likelihood by gradient ascent~\cite{Roudi2011a}. 

The derivatives of the likelihood can also be written as
time averages over the data, which makes a conceptual connection with
logistic regression apparent~\cite{Roudi2011a}. The derivative with respect to fields gives $\frac{\partial L_{\D}}{\partial
	h_i}=\langle \S_i(t+1)\rangle^{\D}_T-\langle \th (\sum_j J_{ij}
\S_j(t) + h_i)\rangle^{\D}_T$ and similarly for the derivative with
respect to the couplings (the subscript refers to the temporal average,
the superscript to configurations taken from the data). 
Parallel Glauber dynamics~\eqref{eq:glauberpa} defines a logistic
regression giving the statistics of $\S_{i} (t+1)$ as a function of $\{\s_{j} (t)\}$. 
In a hypothetical data set of $P$ trajectories of the system
$\D=\{\sss^{p}(t)\}_{t=1}^{M}$, each realization can be considered as
$M-1$ realisations $(\s^{p}_{i} (t+1),\{\s^{p}_{j}(t)\})$
of such regression pairs. The discussion in section~\ref{sec:pseudo-likelihood} then applies directly, giving rise to regression
equations for fields and couplings. 

For Glauber dynamics with sequential updates, the situation is not
quite so simple. In each time interval one
spin is picked for updating, but if this spin is assigned the
configuration it had before it is impossible to tell which spin was
actually chosen~\cite{zeng2013}. The solution is to sum the
likelihood~\eqref{eq:likelihood_timeseries} over all
spins (each of which might have been the one picked for updating with
probability $1/N$). It turns out that both the intervals when a spin was
actually flipped, and the intervals when no spin was flipped are
required for the inference of couplings and fields. 

\subsubsection{Mean-field theory of the non-equilibrium steady state}

As in the case of equilibrium reconstruction, mean-field theory offers
an approximation to the maximum likelihood reconstruction that can be
evaluated quickly. The speed-up is not quite as significant as it is
in equilibrium, because the
likelihood~\eqref{eq:likelihood_timeseries_pa} can already be computed
in polynomial time in $N$ and $M$.

In~\ref{sec:plefka}, the mean-field equation~\eqref{eq:c3 mean field mi} and 
the TAP equation~\eqref{eq:TAPeq} were 
derived in an expansion around a factorising ansatz for the equilibrium distribution. Kappen and Spanjers showed that,
remarkably, exactly the same equations emerge as
first- and second-order expansions around the same ansatz for the NESS
as well~\cite{Kappen2000a}.
Suppose that the (unknown) steady-state distribution of configurations $p(\sss)$ in
the NESS is `close' to another distribution with the \emph{same
	magnetisations}, $q(\sss)=\prod_i \frac{1+m_i\s_i}{2}$. According to
\eqref{eq:glaubersq_m_ness}, this distribution describes the NESS of a
different model with fields $h_i^{(q)}=\arth m_i$ and couplings $J^{(q)}_{ij}=0$.
Next, we consider a small change in these fields and couplings and ask
how the magnetisations change. To first order this change is given by the 
derivatives of~\eqref{eq:glaubersq_m_ness} evaluated at
$h_i=h_i^{(q)}$ and $J_{ij}=0$
\begin{align}
\label{eq:mf_expansionNESS}
\Delta m_i&= \sum_j \frac{\partial \langle s_i\rangle}{\partial
	h_j}\rvert_q\Delta h_j 
+\sum_{k,j} \frac{\partial \langle s_i\rangle}{\partial J_{kj}}\rvert_q\Delta J_{kj} +\ldots\\
&=(1-m_i^2)\Delta h_i + (1-m_i^2) \sum_j \Delta J_{ij} m_j +\ldots \ .\nonumber
\end{align}
Setting $\Delta h_i=h_i-h_i^{(q)}$ and $\Delta J_{ij}=J_{ij}- J_{ij}^{(q)}= J_{ij}$ and demanding that
magnetisations remain unchanged under this change of fields and
couplings gives to first order
$h_i^{(q)}=h_i+\sum_j J_{ij} m_j$ and thus
\begin{equation}
\label{eq:mf_indynamics}
m_i=\th(h_i+\sum_j J_{ij} m_j) \ .
\end{equation}
Carrying the expansion~\eqref{eq:mf_expansionNESS} to second order
in $\Delta h_j$ and $\Delta J_{jk}$ yields the TAP 
equations~\eqref{eq:TAPeq}~\cite{Kappen2000a}. As the equations for the
magnetisations ~\eqref{eq:glaubersq_m_ness} and correlations
\eqref{eq:glauberpa_corrs} are
identical for sequential and parallel dynamics, the mean-field and TAP equations apply
equally to both types of dynamics. Roudi and Hertz~\cite{roudi2011b}
extended these results to time scales before the NESS is reached 
using a generating functional approach. 

The next step is to apply the mean-field approximation to the
correlations~\eqref{eq:glauberpa_corrs} for parallel Glauber
updates~\cite{roudi2011b,zeng2011a}.  For sequential updates, 
analogous result can be derived from~\eqref{eq:glaubersq_c_ness}
and~\eqref{eq:glaubersq_d_ness}. We expand the effective local field
$\Theta_i = h_i + \sum_j J_{ij}m_j + \sum_j J_{ij}\delta \S_j$ around
the mean field $\theta^{\textrm{MF}}_i= h_i + \sum_j J_{ij}m_j $ writing
$\delta \S_i\equiv\S_i-m_i$. Expanding the $\th \Theta_i$-term of the
two-time pair correlation~\eqref{eq:glauberpa_corrs} in a formal expansion
in powers of $\delta \S$ gives
\begin{align}
\phi_{ij}&= \langle \th\left(\Theta_i \right) \S_j \rangle \\
&=\langle \th \theta^{\textrm{MF}}_i \S_j \rangle + (1- \th^2
\theta^{\textrm{MF}}_i) \sum_l J_{il} \langle \delta \S_l \S_j \rangle
+\ldots \nonumber\\
&=m_im_j+(1-m_i^2) \sum_l J_{il} (\chi_{lj}-m_lm_j) \ +\ldots \ .\nonumber
\end{align}
To first order, this equation can be read as a matrix equation in the
connected correlations functions, $D_{ij} \equiv \phi_{ij}-m_i m_j =
\sum_{m,l} A_{im}J_{ml}C_{lj}$ with $A_{im}=\delta_{im} (1-m_i^2)$ and
$C_{ij}=\chi_{ij}-m_im_j$. Inverting this relationship leads to the
mean-field reconstruction
\begin{equation}
\label{eq:Jmf_pa}
\JJJ^{\MF}= {\AAA}^{-1} \DDD \CCC^{-1} \ , 
\end{equation}
based on sample averages of magnetisations and connected correlations
in the NESS. The reconstruction based on the TAP equation can be
derived analogously~\cite{roudi2011b,zeng2011a}; the result is of the
same form as~\eqref{eq:Jmf_pa} with $A_{im}= \delta_{im} (1-m_i^2)
(1-F_i)$, where $F_i$ is the smallest root of
\begin{equation}
F_i (1-F_i)^2 = (1-m_i^2) \sum_{j} (J^{\MF})_{ij}^2
(1-m_j^2) \ .
\end{equation}

\subsubsection{The Gaussian approximation}
\label{sec:noneqGaussian}

For an asymmetric coupling matrix, with no correlation between
$J_{ij}$ and $J_{ji}$, the statistics of the effective local fields in
the NESS is remarkably simple~\cite{Mezard2011a}: In the thermodynamic
limit, 
$\eH_i$ turns out to follow a Gaussian distribution at least in some regimes. 
This distribution is characterized by a mean $\theta^{\textrm{MF}}_i$,
standard deviation $\Delta_i$ and covariance
$\epsilon_{ij}$. Using the definition of the effective  local
field~\eqref{eq:localfield}, these parameters are linked to the spin
observables by
\begin{eqnarray}
\label{eq:gaussian_parameters}
g_i  &=& h_i + \sum_{j} J_{ij} m_j   \\
\Delta_i^2  &=& \sum_{lk} J_{il}C_{lk} J_{ik}  \nonumber\\
\epsilon_{ij}&=& \sum_{lk} J_{il} C_{lk} J_{jk} \nonumber \ .
\end{eqnarray}  
The key idea is that one can transform back and forth between $\{\S_i (t)\}$ and $\{\eH_i (t)\}$ via
\begin{eqnarray}
\eHHH (t) &=& \JJJ \SSS (t) + \hhh, \\
\SSS (t) &=&  \JJJ^{-1} (\eHHH (t) - \hhh) \nonumber
\end{eqnarray}
and evaluate the correlation functions within the Gaussian theory. 
With $\theta_i=\theta^{\textrm{MF}}_i+\sqrt{\Delta_i} x$ and
$\theta_j=\theta^{\textrm{MF}}_j+\sqrt{\Delta_j} y$, where $x$ and $y$
are univariate Gaussian random variables, 
one obtains from $D_{ik}=\langle \th (\Theta_i) \S_k \rangle-\langle
\th (\Theta_i)\rangle \langle \S_k \rangle$
\begin{align}
\sum_k J_{jk}D_{ik}&=\langle \th (\Theta_i) (\Theta_j-h_j)
\rangle-\langle\th (\Theta_i)\rangle \langle \Theta_j-h_j \rangle \nonumber\\
&=\langle \th(\theta^{\textrm{MF}}_i+\sqrt{\Delta_i} x)
\sqrt{\Delta_j} y \rangle \nonumber\\
&= \epsilon_{ij}\langle 1-\th^2(\theta^{\textrm{MF}}_i+\sqrt{\Delta_i} x) \rangle\ .
\end{align}
In the last step we have used the fact that covariances between spin variables
are small~\cite{Mezard2011a}. Inserting the result for the
covariance~\eqref{eq:gaussian_parameters} gives again an equation of the
same form as~\eqref{eq:Jmf_pa}, however with $A_{im}=\delta_{im} 
\int \textrm{d}x \frac{1}{\sqrt{2 \pi}} e^{-\frac{x^2}{2}} \left[ 1-\th^2(\theta^{\textrm{MF}}_i+\sqrt{\Delta_i} x)\right]$. $ $Mean-field
theory neglects the fluctuations here and renders this term as
$1-m^2_i$, whereas the fluctuations can be captured more accurately under the
Gaussian theory. However, $A_{im}$ cannot be determined directly from 
the data alone, but also require the couplings; \cite{Mezard2011a} gives an iterative scheme to infer the
parameters of the effective local
fields~\eqref{eq:gaussian_parameters} as well as the coupling matrix
and magnetic fields. The typical-case performance of the Gaussian theory
in the thermodynamic limit
has been analysed within the framework of statistical
learning~\cite{bachschmid2015learning}, finding the Gaussian theory
breaks down at strong couplings and
a small number of samples. 

The Gaussian distribution of local fields is not limited to the
asymmetric Ising model. In fact, the asymmetric Ising model is one particular
example from a class of models called generalized linear models. In
this model class, the Gaussian approximation has been
used in the context of neural network reconstruction~\cite{toyoizumi2009mean} already prior to its application to
the asymmetric Ising model. 

\subsubsection{Method comparison}
\label{sec:noneqcomparison}

\begin{figure}[tbh]
	\includegraphics[width=.46\textwidth]{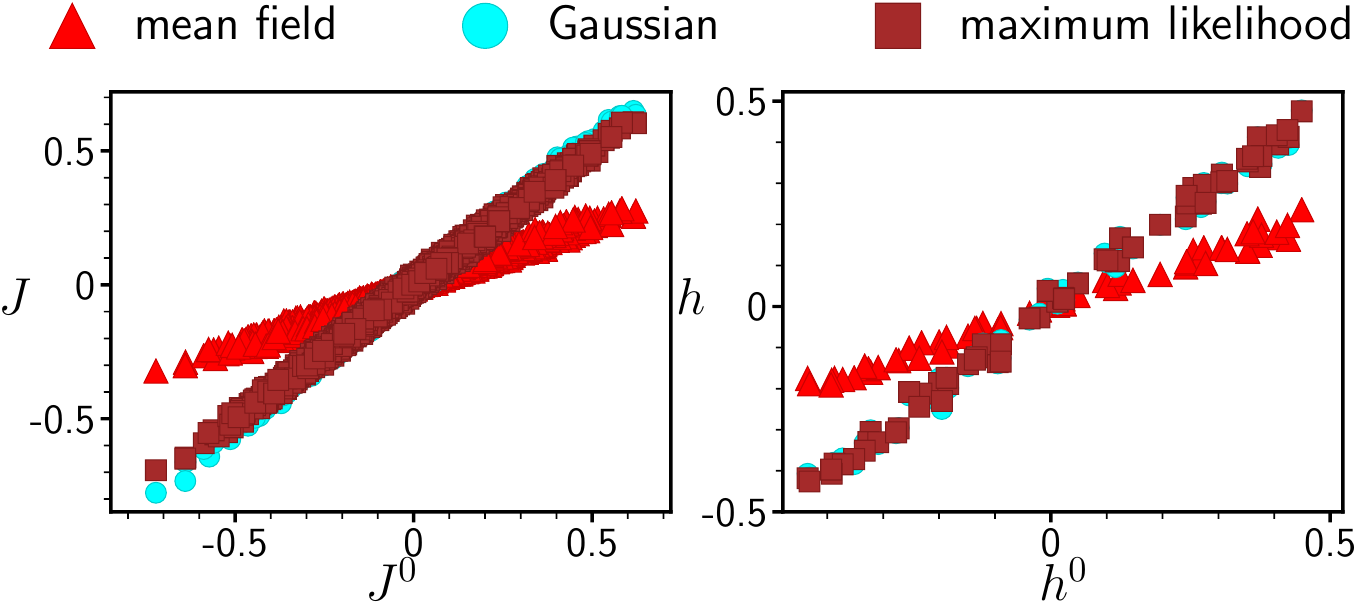}
	
	\vspace{10pt}
	
	\includegraphics[width=.46\textwidth]{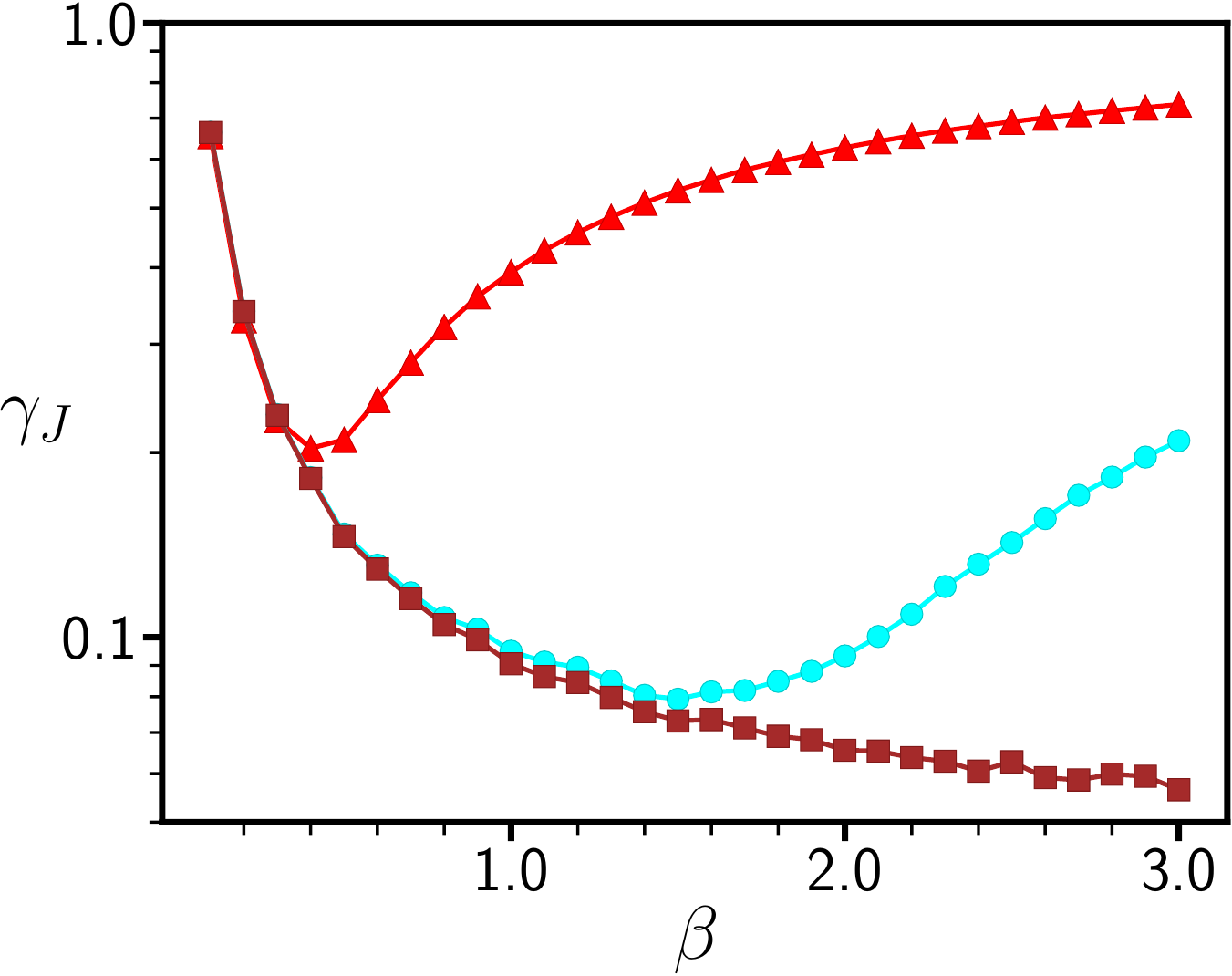}
	\caption{{\bf Reconstructing the asymmetric Ising model from time
			series. } 
		The scatter plots in the upper panels compare the reconstructed couplings and fields with the couplings and fields
		underlying the original model. The system
		size is $N=64$, the standard deviation of the original couplings is
		$\beta/\sqrt{N}$ with $\beta=1.5$, the original fields are uniformly distributed between $[-0.3\beta,0.3\beta]$, and $M=15000$ time steps are
		sampled. The bottom plot gives the
		reconstruction error as a function of $\beta$, see text. 
		\label{fig:comparison_noneq_timeseries}}
\end{figure}

We compare the results of the mean-field
approximation, the Gaussian approximation, and the maximization of the
exact likelihood~\eqref{eq:likelihood_timeseries_pa}. As in
section~\ref{sec:comparison}, we draw couplings from a Gaussian
distribution with mean zero and standard deviation $\beta/\sqrt{N}$,
but now couplings $J^0_{ij}$ are statistically independent of $J^0_{ji}$,
so the matrix of couplings is in general asymmetric.
Fields are drawn from a uniform distribution on the interval $[-0.3
\beta,0.3 \beta]$. We then sample a time series of $M=15000$ steps by parallel updates under
Glauber dynamics~\eqref{eq:glauberpa}. The
scatter plots in the top row of
figure~\ref{fig:comparison_noneq_timeseries} compares 
couplings and fields reconstructed by different methods with
the couplings and fields of the original model. It shows that couplings are
significantly underestimated by the mean-field reconstruction, a
bias which is avoided by the Gaussian
theory. The right-hand plot shows the relative reconstruction error~\eqref{eq:reconstruction_error}
against $\beta$. As in the equilibrium case shown in
figure~\ref{fig:comparison-SK}, reconstruction by any method is limited at small $\beta$
by sampling noise. At strong couplings $\beta$, mean-field theory
breaks down.
Also at strong couplings, the iterative algorithm for Gaussian
approximation reconstruction converges very slowly and stops when the
maximum number of iterations is reached (here is set $50000$ steps). 

\subsection{Outlook: Reconstruction from the steady state}

Time series data is not available in all applications. This
prompts the question how to reconstruct the parameters of a
non-equilibrium model from independent samples of the steady state. It is clear that, unlike for
equilibrium systems, pairwise correlations are insufficient to infer
couplings: The matrix of correlations is symmetric, whereas
the matrix of couplings is asymmetric for non-equilibrium
systems. Hence, there are twice as many free parameters as there are
observables. Similarly, using an equilibrium model like
\eqref{eq:thefirst} (maximum-entropy model) on data generated by a
non-equilibrium model would give parameters matching the two-spin
observables, but entirely different from the true parameters underlying the data.

One solution uses three-spin correlations to
infer the couplings~\cite{dettmernguyenberg}. A problem of
this approach is that connected three-spin correlations are small
since the effective local fields are well described by a multi-variate Gaussian
distribution (see~\ref{sec:noneqGaussian}).

A second approach uses perturbations of the non-equilibrium
steady state: We measure one set of pair correlations at
certain (unknown) model parameters, and then a second set of pair
correlations of a perturbed version of the system. Possible
perturbations include changing one or several of the couplings by a
known amount, changing the magnetic fields, or fixing particular spins to
a constant value. This generates two sets of coupled
equations specifying two symmetric correlations matrices, which can be
solved for one asymmetric coupling
matrix. Conceptually, such an
approach is well known in biology, where altering parts of a
system and checking the consequences is a standard mode of scientific
inquest. Neural stimulation can lead to a rewiring of neural connections, and the effects of
this neural plasticity can be tracked in neural
recordings~\cite{rebesco2010rewiring}. An exciting development is
optogenetic tools, which allow to
stimulate and monitor the activity of individual neurons \emph{in vivo}~\cite{tischer2014illuminating}.
In the context of inferring gene regulatory networks from
gene expression data, perturbation-based approaches have been used
both with linear~\cite{tegner2003a} and non-linear models of gene
regulation~\cite{nelander2008a,molinelli2013a}, see also
section~\ref{sec:geneexpression}. In this context, it is also fruitful to consider the
genetic variation occurring in a population of cells as a source of perturbations. Such an approach has already been used in gene
regulatory networks~\cite{rockman2008reverse,zhu2008integrating}, and with the
expansion of tools to analyze large numbers of single
cells~\cite{zeisel2015cell,cadwell2015electrophysiological}, it may
soon spread to other types of networks. 


\section{Conclusions}
\label{sec:conclusions}

At the end of this overview, we step back and summarize the 
aims and motivations behind the inverse Ising problem, discuss the
efficacy of different approaches, and outline different areas of
research that may involve the statistical mechanics of inverse
problems in the future.

\textbf{Motivation.} The inverse Ising problem arises in the context of very different
types of questions connected with the inference of model parameters.
The first and most straightforward question appears when data actually is
generated by a process obeying detailed balance and has pairwise
couplings between binary spins. The problem of inferring the
parameters of such a model can then be phrased in terms of the equilibrium
statistics~\eqref{eq: static sequential Boltzmann}, the maximisation
of the likelihood~\eqref{eq: static sequential log-likelihood}, and the
use of approximation schemes discussed in
section~\ref{sec:equilibrium}.

The second question arises when data is generated by a different, and possibly
entirely unknown type of process, and we seek a statistical
description of the data in terms of a simpler model matching only
particular aspects of the observed data. An example is models with maximum
entropy given pairwise correlations (section~\ref{sec:maxent}) used to
describe neural data in section~\ref{sec:neural}. 

The assumptions behind the Ising model (pairwise
couplings, binary spins, $\ldots$) can be relaxed. For instance, multi-valued Potts spin
variables have been used extensively in models of biological sequences
~\cite{Weigt2009a,santolini2014a}. The extension of the
Ising model to models with three- and four-spin couplings has not
been used yet in the context of inference. Nevertheless,
for many of the methods of section~\ref{sec:equilibrium}, the
extension beyond pairwise couplings would be straightforward.
A much larger rift lies between equilibrium and
non-equilibrium models. Parameter inference of a non-equilibrium
model is often based on data beyond independent samples of the steady state, specifically
time-series data. 

\textbf{Methods.} The practical question how to infer the couplings and fields that 
parametrise an Ising model has been answered using many different
approximations with different regimes of
validity. Section~\ref{sec:comparison} gives an overview. For data
sampled from the equilibrium distribution~\eqref{eq:thefirst}, the 
pseudolikelihood approach of section~\ref{sec:pseudo-likelihood} gives a reconstruction close to the
optimal one (in fact, asymptotically close in the number of
samples). However, this is paid for by a computational effort that
scales with the number of samples. For
the non-equilibrium regime and a time series of configurations sampled
from the stochastic dynamics, the likelihood of a
model can be evaluated exactly comparatively easily.

Both the pseudolikelihood and the likelihood of a time series
can be understood in the language of regression. Thus a single 
framework links two of the most successful approaches, in both the equilibrium and
non-equilibrium setting.  Regression singles out one spin variable as a dependent
variable and treats the remainder as independent variables. It then
characterizes the statistics of the dependent variable given configurations of
the independent variables.  

The application of novel concepts to the inverse Ising problem and
the development of new algorithms continues. 
An exciting new direction is interaction screening~\cite{vuffray2016interaction,lokhov2016optimal}. Vuffray, Misra, Lokhov, and
Chertkov introduce the objective function 
\begin{equation}
\label{eq:chertkov}
\frac{1}{M} \sum_{\mu=1}^M e^{H_{i}(\sss^{\mu})}=\frac{1}{M} \sum_{\mu=1}^M \exp\{-\sum_{j\ne i} J_{ij} s_i^{\mu} s_j^{\mu} -
h_i \s_i^{\mu}  \} 
\end{equation}
to be minimized with respect to the $i^{\mbox{th}}$ column of the
coupling matrix and the magnetic field on spin $i$ by convex
optimisation. $H_{i}(\sss)$ denotes the part of the
Hamiltonian containing all terms in $s_i$, see~\ref{sec:pseudo-likelihood}.
Sparsity or other properties of the model parameters can be effected
by appropriate regularisation terms. This objective function
aims to find those parameters which `screen the interactions' (and the
magnetic field) in the data, making the sum over
samples in~\eqref{eq:chertkov} as balanced as possible. 
To illustrate the appeal of this objective function, we look at the
infinite sampling limit, where the summation over samples in the
objective function~\eqref{eq:chertkov} can be replaced 
by a summation over the configurations, reweighted by the Boltzmann distribution with the true couplings $\JJJ^\ast$ and fields $\hhh^{\ast}$,
\begin{align}
\frac{1}{M} \sum_{\mu=1}^M e^{H_{i}(\sss^{\mu})} \approx &\sum_{\sss \setminus \s_i} \frac{1}{Z(\JJJ^{\ast},\hhh^{\ast})} e^{\sum_{k>j; k,j \ne i} J_{kj}^{\ast} s_k s_j + \sum_{k\ne i}h^{\ast}_k s_k} \nonumber \\ &\times \sum_{s_i} e^{-\sum_j (J_{ij}-J^{\ast}_{ij}) s_i s_j- (h_i-h^\ast_i) \s_i}.
\end{align}
Since the last sum as a function of $J_{ij}$ and $h_i$ (for a fixed
$i$) is convex and even when being reflected around 
$J^\ast_{ij}$ and $h_i^{\ast}$, so is the whole objective function. It
follows immediately that this 
objective function has a unique global minimum at $J_{ij}= J^{\ast}_{ij}$ and $h_i= h^\ast_i$.

Of course, the sample average over $e^{H_{i}(\sss)}$ is affected by
sampling fluctuations when the number of samples is small.
Nevertheless, the minimum of this objective function
nearly saturates the information theoretic bounds on the
reconstruction of sparse Ising models of Santhanam and
Wainwright~\cite{santhanam2012information} (see
section~\ref{sec:infortheo}).

The flat histogram method in Monte Carlo simulations uses a similar
rebalancing with respect to the Boltzmann measure~\cite{wang2000flat}.
Also, there may be conceptual links between interaction screening
and the fluctuations theorems such as the Jarzynski
equality~\cite{jarzynski1997nonequilibrium}, which also take sample
averages over exponentials of different thermodynamic quantities. 

Another recent development concerning the sparse inverse Ising problem
is the use of Bayesian model selection techniques by Bulso, Marsili
and Roudi~\cite{bulsomarsiliroudi2016a}.

\textbf{More to come.} Over the last two
decades, interest in inverse statistical problems has been driven by
technological progress and this progress is likely to continue opening up new applications. Beyond the
extrapolation of technological developments, there are several broad
areas at the interface of statistical mechanics, statistics, and
machine learning where
inverse statistical problems such as the inverse Ising problem might play a
role in the future.

\begin{itemize}
	\item \textbf{Stochastic control theory.} Stochastic control theory
	seeks to steer a stochastic system towards certain desired states~\cite{kappen2007a}. An inverse
	problem arises when the parameters describing the stochastic system
	are only partially known. As a result, its
	response to changes in external control parameters ('steering')  must be predicted on the
	basis of its past dynamics. A recent application is the control of
	cell populations, specifically populations of cancer cells~\cite{fischer2015value}. Such cell populations
	evolve stochastically due to random fluctuations
	of cell duplications and cell deaths. Birth and death rates of at least part of
	the population can be controlled by therapeutic drugs.
	\item \textbf{Network inference.} Like regulatory connections between
	genes discussed in section~\ref{sec:geneexpression}, 
	metabolic and signalling interactions also form intricate networks. An
	example where such a network optimizes a specific global quantity is
	flux-balance analysis~\cite{orth2010flux}, which models the flow
	of metabolites through a metabolic network. Inverting the
	relationship between metabolic rates and metabolite
        concentrations allows in principle to infer the metabolic network on
	the basis of observed metabolite concentrations~\cite{deMartino2016a}. 
	\item \textbf{Causal analysis.} The \textit{do}-calculus developed by
	Judea Pearl seeks to establish causal relationships behind
	statistical dependencies~\cite{pearl1995causal}. \textit{do}-calculus is based on
	interventions such as fixing a variable to a particular value, and then
	observing the resulting statistics of other variables. 
	Recently, Aurell and Del Ferrano have found a link
	between \textit{do}-calculus and the dynamic cavity method from
	statistical physics~\cite{aurell2015causal}.
	\item \textbf{Maximum entropy and dynamics.} The notion of a
	steady-state distribution with maximum entropy has been generalized
	to a maximum-entropy distribution over~\emph{trajectories} of a dynamical
	system~\cite{presse2013a}. Like the maximum entropy models of
	section~\ref{sec:neural}-\ref{sec:proteinstruct}, this approach can
	be used to derive simple effective models of the \emph{dynamics} of a
	system, whose parameters can be inferred from time series. In fact,
	Glauber dynamics with parallel updates gives the maximum entropy
	distribution of $\SSS(t+1)$ given $\sss(t)$. So far,
	applications have been in neural
	modelling~\cite{marre2009,nassercessac2014a}, in the effective
	dynamics of quantitative traits in
	genetics~\cite{bartondevladar2009a,bovdova2015a}, and in flocking
	dynamics~\cite{cavagna2014a}.
	\item \textbf{Hidden variables.} Even in large-scale data sets, there
	will be variables that are unobserved. Yet, those hidden variables affect the statistics of
	other variables, and hence the inference we make about interactions
	between the observed
	variables~\cite{dunn2013learning,tkacikmarremoraetal2013a,battistin2015belief,rouditaylor2015a}. 
	This can lead to a signature of critical
	behaviour, even when the original system is not critical~\cite{marsilimastromatteoroudi2013a,schwabnemenmanMehta2014a}.
	\item \textbf{High-dimensional statistics and inference.}  In many
	applications, the number of systems parameters to be inferred is of the same order of
	magnitude or exceeds the sample size. The field of high-dimensional
	statistics deals with this regime, and a fruitful interaction with statistical physics has emerged over
	the last decade~\cite{ZdeborovaKrzakala2015}, spurred by
	applications such as compressed sensing. Vuffray et al. 
	~\cite{vuffray2016interaction}
	propose the objective function~\eqref{eq:chertkov} for the inverse Ising problem in the
	high-dimensional regime. Other objective functions might perform even
	better, and the optimal objective function may depend on the number
	of samples and the statistics of the underlying couplings. 
	~\cite{berg2016,bachschmid2017statistical} use the statistical mechanics of disordered
	systems to find the objective function which minimizes the difference
	between the reconstructed and the underlying couplings for Gaussian
	distributed couplings. 
	\item \textbf{Restricted Boltzmann machines and deep learning.} Recently, (deep) feed-forward neural networks have
	re-established themselves as powerful learning architectures,
	leading to spectacular applications in the area of computer vision,
	speech recognition, data visualisation, and game
	playing~\cite{lecun2015deep}. This progress has also demonstrated that the
	most challenging step in data analysis is the extraction of features
	from unlabelled data. There is a wide range of methods for feature
	extraction, the most well-known ones being convolutional networks
	for images~\cite{lecun1998gradient,lecun2015deep}, or
	auto-encoders~\cite{hinton2006a,vincent2010stacked}, 
	and restricted Boltzmann machines
	(RBMs)~\cite{hinton2006a} for less structured data. RBMs are possibly one of the most
	general methods, although the algorithms used for finding
	the optimal parameters are  heuristic and approximate in many
	respects. For instance, in the case of multiple layers, RMBs correspond to generic Boltzmann 
	machines (with feedback loops), for which learning is a 
	hard computational  problem. Usually, the sub-optimal solution which
	is adopted is used to train each layer independently.
	These observations are  not surprising since RBMs are nothing but an
	inverse Ising problem with a layer of visible spins connected to a
	layer of hidden spins. Empirical data is available only for the
	visible spins. The role of the hidden variables is to compress
	information and identify structural features in the data. Learning
	consists of finding the visible-to-hidden couplings $J_{ij}=J_{ji}$
	and the local fields such that summing over to
	the hidden variables gives back a (marginalized) probability distribution over the
	visible variables which is maximally consistent with the data,
	\textit{i.e.} has 
	minimal KL divergence~\eqref{eq:def_kullbachleibler} from the
	empirical distribution of the data. 
	\item \textbf{Learning phases of matter.} One aspect of the emerging
	applications of machine learning in quantum physics is the identification of
	non-trivial quantum states from data. This is of particular interest
	in phases of matter where the order parameter is either unknown or
	hard to compute (such as the so-called entanglement
	entropy~\cite{islam2015measuring,levin2006detecting}).
	Techniques from machine learning, specifically deep learning with
	neural networks, have recently been used to classify quantum states
	without knowledge of the underlying
	Hamiltonian~\cite{carrasquilla2017machine}.
\end{itemize}

\section*{Acknowledgements} Discussion with our colleagues and students
have inspired and shaped this work. In particular we would like to
thank Erik Aurell, Michael Berry, Andreas Beyer, Simona Cocco, Simon Dettmer,  David
Gross, Jiang Yijing, David R. Jones, Bert Kappen, Alessia Marruzzo, Matteo Marsili,
Marc M\'ezard, R{\'e}mi Monasson, Thierry Mora,
Manfred Opper, Andrea Pagnani, Federico Ricci-Tersenghi, Yasser Roudi,
Ga{\v{s}}per Tka{\v{c}}ik, Aleksandra Walczak,
Martin Weigt and Pieter Rein ten Wolde. Many thanks to Ga{\v{s}}per Tka{\v{c}}ik and  Michael Berry for
making the neural recordings from~\cite{tkavcik2014a,tkacik2015}
available. This work was supported by the DFG under Grant SFB
680; BMBF under Grants emed:SMOOSE and SYBACOL.

\newpage
\bibliographystyle{abbrv}
\bibliography{inv-ising}

\end{document}